\documentclass[a4paper,11pt]{article}
\usepackage{amsmath, amsthm, amssymb,amsbsy, amsfonts}
\usepackage{graphicx, color}
 \usepackage{jheppub} 
\usepackage{tikz}
\usetikzlibrary{plotmarks}
\usepackage{adjustbox}

\usepackage{mathtools}\usepackage{framed}
\usepackage{float}
\newcommand{\nn}{\nonumber}

\def\cN{{\mathcal N}}

\def\ni{{\noindent}}

\title{\boldmath 5d $E_n$ Seiberg-Witten curve via toric-like diagram}

\author{Sung-Soo Kim,}
\author{Futoshi Yagi}

\affiliation{School of Physics,\\ Korea Institute for Advanced Study,\\
85 Hoegiro, Dongdaemun-gu\\
Seoul 130-722, Korea}
\emailAdd{sungsoo.kim@kias.re.kr}
\emailAdd{fyagi@kias.re.kr}

\abstract{We consider 5d $Sp(1)$ gauge theory with $E_{N_f+1}$ global symmetries based on toric(-like) diagram constructed from $(p,q)$-web with 7-branes. We propose a systematic procedure to compute the Seiberg-Witten curve for generic toric-like diagram. For $N_f=6,7$ flavors, we explicitly compute the Seiberg-Witten curves for 5d $Sp(1)$ gauge theory, and show that these Seiberg-Witten curves agree with already known $E_{7,8}$ results. We also discuss a generalization of the Seiberg-Witten curve to rank-$N$ cases.}
\begin{document} 
\begin{flushright} %
\hfill {\tt KIAS-P14064}\\
\end{flushright}

\maketitle
\flushbottom
%%%%%%%%%%%%%%%%%%%%%%%%%%%%%%%%%%%%%%%%%%%%%%%
\section{Introduction}
Various aspects of supersymmetric gauge theories have been studied
via branes, and rich physics associated with branes has been revealed.
Wide class of four-dimensional $\cN=2$ supersymmetric gauge theories
are given by D4-NS5 brane setup,
whose asymptotic distance of the NS5 branes gives 1-loop correction
to the gauge coupling constant.
The condition that NS5 branes do not intersect 
gives the asymptotic free/conformal condition of the gauge theory.
Furthermore, the M-theory uplift of the corresponding
brane setup gives Seiberg-Witten (SW) curve \cite{Witten:1997sc}.

Five-dimensional uplift of this derivation of SW curve from branes
was studied in \cite{Brandhuber:1997ua} for the case with simple Lie groups like $SU, SO, Sp$,
and extended to more general setup by using $(p,q)$ 5-brane web in \cite{Aharony:1997bh}.
The method with $(p,q)$ 5-brane web gives a systematic and simple procedure
to derive the SW curve, using the dual graph of the 5-brane web
whose vertices correspond to the non-vanishing coefficients of the polynomial describing the curve. 
 In \cite{Aharony:1997bh}, it was discussed that five-dimensional $Sp(1)~ (\simeq SU(2))$ gauge theories
could be studied up to four flavors by this $(p,q)$ 5-brane web%
\footnote{In this paper, we do not consider O-planes. If one introduces it, 5-brane web can describes up to six flavors \cite{Brunner:1997gk}.}.
As five-dimensional theories are given as natural uplift of four-dimensional gauge theories,
studying the case with more than four flavors via the $(p,q)$ 5-branes %seems to have difficulty.
seemed not so straightforward. 
As discussed, for example, in \cite{Aharony:1997ju}, more flavors lead to the
intersection of NS5-branes analogously to the four-dimensional case,
which makes it hard to interpret. 

On the other hand, 
five-dimensional $Sp(1)$ gauge theory with $N_f$ flavors is expected to have
nontrivial UV fixed point up to \footnote{UV fixed point of the five-dimensional $Sp(1)$ theory with eight flavors is believed to be six-dimensional. Although this case is also interesting, we do not study in this paper.} $N_f \le 7$ \cite{Seiberg:1996bd},
where the global symmetry enhancement 
to $E_{N_f+1}$ group is realized.
Such class of isolated conformal field theories (CFT) has been studied in various different methods.
Recently, the global symmetry enhancement was explicitly checked in
\cite{Kim:2012gu} by computing the superconformal index (up to $N_f=5$), and confirmed later up to $N_f=6, 7$ in \cite{Hwang:2014uwa}.
It was also shown that the Nekrasov partition function is invariant under the enhanced $E_{N_f+1}$
symmetry when one expands it in terms of the properly redefined Coulomb moduli parameter \cite{Mitev:2014jza},
where fiber-base duality \cite{Aharony:1997bh,Katz:1997eq,Bao:2011rc, Bao:2013wqa} plays an important role.

The SW curves for corresponding four-dimensional theory with $E_{6, 7, 8}$ global symmetries were studied in \cite{Minahan:1996fg, Minahan:1996cj}. These curves were obtained from dimensional analysis and symmetry argument without referring to specific field theory description at UV. Uplift of the SW curves to five dimensions \cite{Minahan:1997ch} and to six dimensions \cite{Eguchi:2002fc,Eguchi:2002nx} was performed via the effective action of the E-string theory, which is obtained by
compactification to the corresponding local del Pezzo surface or half K3 manifold.
When the geometry is toric, the corresponding SW curve can be computed also from the toric diagram which can be 
%The toric diagram of this Calabi-Yau geometry can be
 reinterpreted as the dual graph of the $(p,q)$ 5-brane web \cite{Leung:1997tw}.
It followed that the SW curve for five-dimensional $Sp(1)$ gauge theory with $N_f \le 4$ flavors was reproduced as mirror curve of the corresponding Calabi-Yau geometry \cite{Eguchi:2000fv}. While it can be toric for $N_f \le 5$, %\footnote{For $N_f=3,4,5$, the geometry becomes toric only when we choose the specific complex structure.For $N_f=0,1,2$, the geometry is toric for arbitrary complex structure.}
the corresponding local del Pezzo surfaces for $N_f = 6,7$ are non-toric, and thus there have been difficulties finding corresponding $(p,q)$ 5-brane web diagrams.

Such obstacles seem avoidable in the brane setup introduced in \cite{Benini:2009gi} which thus opens up a possibility to analyze the $Sp(1)$ gauge theory with more than four flavors in the $(p,q)$ 5-brane setup. It is $[p,q]$ 7-branes at infinity that resolve the non-toric nature of dual diagrams, each of which binds an arbitrary number of $(p,q)$ 5-branes. Some of the bound 5-branes ``jump over'' other 5-branes in such a way not to break the s-rule. The brane setup in \cite{Benini:2009gi} was originally introduced as a five-dimensional uplift of isolated CFTs discussed in \cite{Gaiotto:2009we}. It includes five-dimensional CFT of $E_6$, $E_7$ and $E_8$ global symmetries, which are expected to be identified as the UV fixed point of the $Sp(1)$ gauge theories with five, six and seven flavors, respectively.

It was known that the five-dimensional $Sp(1)$ gauge theory with $N_f$ flavors can be realized by adding $N_f$ D7-branes inside the 5-brane loop of the pure $Sp(1)$ brane setup, and that the global symmetry enhancement to $E_{N_f+1}$ can be shown from the monodromy properties of 7-branes \cite{DeWolfe:1999hj,Taki:2014pba}. More intuitive and relevant explanation connecting D7-branes inside the 5-brane loop to 7-branes at infinity leading to $E_6$ symmetry was presented in \cite{Bao:2013pwa} for $N_f=5$ where the brane setup with $E_{N_f+1}$ symmetry studied in \cite{Benini:2009gi} can be derived by properly pulling out all the 7-branes outside by the Hanany-Witten effect.

With this brane setup, the SW curve for $Sp(1)$ with five flavors
was computed in \cite{Benini:2009gi,Bao:2013pwa} and is in agreement with
the aforementioned result \cite{Minahan:1997ch, Eguchi:2002fc, Eguchi:2002nx}.
Moreover, by generalizing the relation between toric $(p,q)$ 5-brane web and toric diagram discussed in \cite{Leung:1997tw},
the superconformal index/Nekrasov partition function in \cite{Kim:2012gu, Hwang:2014uwa}
were reproduced from the computations of topological string partition function
\cite{Bao:2013pwa, Hayashi:2013qwa, Hayashi:2014wfa}.
Throughout this paper, such dual graph of the original $(p,q)$ 5-brane web diagram,
introduced in \cite{Benini:2009gi}, we call ``toric-like diagram'' as the counterpart of the toric diagram. In this paper, we develop a systematic way of computing the SW curve for the five-dimensional theory with $N_f=6,7$ flavors based on the toric-like diagram.

%Along the direction that these recent developments allow some of the theories obtained bycompactifying non-toric Calabi-Yau geometry to be analyzed with the technology developed for toric Calabi-Yau geometry or $(p,q)$ 5-brane web, %in this paper, we would like to further develop the $(p,q)$ 5-brane configuration by computing the SW curve for the five-dimensional theory with $N_f=6,7$ flavors based on the toric-like diagram.

The rest of this paper is organized as follows:
In section 2, we review the derivation of SW curve of five-dimensional $Sp(1)$
gauge theory with $N_f=5$ flavors \cite{Bao:2013pwa},
which is identified as five-dimensional $T_3$ theory.
Throughout this review, we clarify generic procedure to
compute SW curve from toric-like diagram.
In section 3 and 4, we compute the SW curve for
the theory with $N_f=6$ and $N_f=7$ flavors, respectively.
We present the SW curves obtained based on three representative web diagrams which are related by the Hanany-Witten transitions, and then show that each of these curves is obtained from another one simply by taking a proper coordinate transformation, thus demonstrating that the Hanany-Witten transition is realized in the SW curve as a coordinate transformation. 
%which turns out to 
We also show that the obtained curves agree with the known results \cite{Minahan:1997ch, Eguchi:2002fc, Eguchi:2002nx}.
In section 5, we study mass decoupling limit to reproduce
the SW curve for lower flavors from the curve for higher flavors, especially from $N_f=7$ to $N_f=6$.
In section 6, we consider the SW curve
obtained from the toric-like web diagram
corresponding to higher rank $E_n$ theory given in \cite{Benini:2009gi}
and shows that the rank-2 curve actually factorizes into the two copies of
the SW curve for the rank-1 $E_n$ theory.
We then conclude and discuss the observed relation%
\footnote{Although this relation should be closely related to \cite{Bergman:2014kza},
we give slightly different interpretation. Our observation is closer to the one in \cite{Bao:2013pwa}.}
 analogous to the special case of $\cN=2$ dualities \cite{Gaiotto:2009we}.
In Appendices, we give various complimentary computation and results
of the SW curves for $Sp(1)$ theory with $N_f$ flavors.

%%%%%%%%%%%%%%%%%%%%%%%%%%%%%%%%%%%%%%%%%%%%%%%
%%%%%%%%%%%%%%%%%%%%%%%%%%%%%%%%%%%%%%%%%%%%%%%
%%%%%%%%%%%%%%%%%%%%%%%%%%%%%%%%%%%%%%%%%%%%%%%
\section{5d Seiberg-Witten curve from toric-like diagram}
In this section, after reviewing the SW curve of the 5d $T_3$
theory which is identified as 5d $Sp(1)$ theory with five flavors,
we propose a procedure to derive the SW curve from generic toric-like diagram.

\subsection{5d $T_3$ theory}\label{sec:T3}
5d version of 4d $T_N$ theories was studied based on the web or dual toric diagrams \cite{Benini:2009gi} where it describes M-theory compactified on a non-compact CY threefold. The dual toric diagram is obtained by associating a vertex to each face of 5-brane junctions. For a single junction, $T_1$, it corresponds to $\mathbb{C}^3$ and for multi-junctions, $T_N$, it corresponds to $\mathbb{C}^3/(\mathbb{Z}_N\times\mathbb{Z}_N)$. Upon compactification on $S^1$, it gives rise to 4d $T_N$ constructed in \cite{Gaiotto:2009we}.

We now briefly review 5d $T_3$ theory in relation with its SW curve. A detail analysis for $T_3$ theory has been done in \cite{Bao:2013pwa}. Rather than summarizing the result of \cite{Bao:2013pwa}, we here point out salient features of the analysis and then use this $T_3$ theory to give an intuitive idea on how the SW curves for $E_7$ and $E_8$ can be derived.

As shown in \cite{Benini:2009gi}, the 5d uplift of rank-1 4d $T_N$ theories well fits into the multi-junction of the 5d $(p,q)$ web. 
4d Minahan-Nemenschansky's isolated superconformal theories with the exceptional $E_n$ symmetry can be uplifted and studied in this framework. For instance, in 5d, the $N_f=5$ superconformal theory with $E_6$ global symmetry at the UV fixed point corresponds to $T_3$ theory; the $N_f=6,7$ superconformal theory with $E_7, E_8$ global symmetry corresponds to 
 a Higgsed $T_{4},T_{6}$ (to keep one Coulomb modulus) theory, respectively.

\begin{figure}[H]
\centering
\begin{adjustbox}{width=0.2\textwidth}
\begin{tikzpicture}
  [inner sep=0.5mm, line width=0.5mm,
 dot/.style={fill=black,draw,circle,minimum size=1pt},
 whitedot/.style={fill=white,draw,circle,minimum size=1pt},
 mark size=4pt, mark options={fill=white} ]
    
    \draw (-1,0) -- (0,0) -- (0,-1);
    \draw (0,0) -- (1,1);
    \draw (1,1) -- (2,1) -- (3, 2) -- (3,3) -- (2,3) -- (1,2) -- (1,1);
    \draw (1,2) -- (0,2); \draw (2,3) -- (2, 4) -- (1,4);
    \draw (2,4) -- (4,6);
    \draw (3,3) -- (4,4); \draw (4,5) -- (4,4) -- (5,4); \draw[dashed] (4,5) -- (6,5);
    \draw (3, 2) -- (5, 2);
    \draw (2,1) -- (2,0) ;
    \draw[<-, very thick, color=red] (4.4,4.4) --  (4.4,4.8);

  \foreach \plm in {otimes*}
  \foreach \plm in {otimes*} \draw plot[mark=\plm] coordinates {(-1,0)} ;
  \foreach \plm in {otimes*} \draw plot[mark=\plm] coordinates {(0,2)} ;
  \foreach \plm in {otimes*} \draw plot[mark=\plm] coordinates {(1,4)} ;

  \foreach \plm in {otimes*} \draw plot[mark=\plm] coordinates {(0,-1)} ;
  \foreach \plm in {otimes*} \draw plot[mark=\plm] coordinates {(2,0)} ;
  \foreach \plm in {otimes*} \draw plot[mark=\plm] coordinates {(5,2)} ;

  \foreach \plm in {otimes*} \draw plot[mark=\plm] coordinates {(4,6)} ;
  \foreach \plm in {otimes*} \draw[mark options={fill=red}] plot[mark=\plm] coordinates {(4,5)} ;
  \foreach \plm in {otimes*} \draw plot[mark=\plm] coordinates {(5,4)} ;
\end{tikzpicture}
\end{adjustbox}
\quad
\begin{tikzpicture}
 \draw[->,thick] (-1.5,1) -- (-0.5, 1);
   \node[circle, fill=white] at (-0.5,-0.5) {};
\end{tikzpicture}
\begin{adjustbox}{width=0.2\textwidth}
\begin{tikzpicture}
  [inner sep=0.5mm, line width=0.5mm,
 dot/.style={fill=black,draw,circle,minimum size=1pt},
 whitedot/.style={fill=white,draw,circle,minimum size=1pt},
 mark size=4pt, mark options={fill=white} ]

    \draw (-1,0) -- (0,0) -- (0,-1);
    \draw (0,0) -- (1,1);
    \draw (1,1) -- (2,1) -- (3, 2) -- (3,3) -- (2,3) -- (1,2) -- (1,1);
    \draw (1,2) -- (0,2); \draw (2,3) -- (2, 4) -- (1,4);
    \draw (2,4) -- (4,6);
    \draw (3,3) -- (5,5);  \draw[dashed] (4,3) -- (6,3);
    \draw (3, 2) -- (5, 2);
    \draw (2,1) -- (2,0) ;
    \draw[<-, very thick, color=red] (4.4,2.4) --  (4.4,2.8);

  \foreach \plm in {otimes*}
  \foreach \plm in {otimes*} \draw plot[mark=\plm] coordinates {(-1,0)} ;
  \foreach \plm in {otimes*} \draw plot[mark=\plm] coordinates {(0,2)} ;
  \foreach \plm in {otimes*} \draw plot[mark=\plm] coordinates {(1,4)} ;

  \foreach \plm in {otimes*} \draw plot[mark=\plm] coordinates {(0,-1)} ;
  \foreach \plm in {otimes*} \draw plot[mark=\plm] coordinates {(2,0)} ;
  \foreach \plm in {otimes*} \draw plot[mark=\plm] coordinates {(5,2)} ;

  \foreach \plm in {otimes*} \draw plot[mark=\plm] coordinates {(4,6)} ;
  \foreach \plm in {otimes*} \draw[mark options={fill=red}] plot[mark=\plm] coordinates {(4,3)} ;
  \foreach \plm in {otimes*} \draw plot[mark=\plm] coordinates {(5,5)} ;
\end{tikzpicture}
\end{adjustbox}
\quad
\begin{tikzpicture}
 \draw[->,thick] (-1.5,1) -- (-0.5, 1);
   \node[circle, fill=white] at (-0.5,-0.5) {};
\end{tikzpicture}
\begin{adjustbox}{width=0.2\textwidth}
\begin{tikzpicture}
  [inner sep=0.5mm, line width=0.5mm,
 dot/.style={fill=black,draw,circle,minimum size=1pt},
 whitedot/.style={fill=white,draw,circle,minimum size=1pt},
 mark size=4pt, mark options={fill=white} ]
    \draw (-1,0) -- (0,0) -- (0,-1);
    \draw (0,0) -- (1,1);
    \draw (1,1) -- (2,1) -- (3, 2) -- (3,3) -- (2,3) -- (1,2) -- (1,1);
    \draw (1,2) -- (0,2); \draw (2,3) -- (2, 4) -- (1,4);
    \draw (2,4) -- (3,5);
    \draw (3,3) -- (4,4); \draw (3, 2) -- (4, 2);
    \draw (4,1) -- (4,2) -- (5,3);  \draw[dashed] (4,1) -- (6,1);
    \draw (2,1) -- (2,0) ;

  \foreach \plm in {otimes*}
  \foreach \plm in {otimes*} \draw plot[mark=\plm] coordinates {(-1,0)} ;
  \foreach \plm in {otimes*} \draw plot[mark=\plm] coordinates {(0,2)} ;
  \foreach \plm in {otimes*} \draw plot[mark=\plm] coordinates {(1,4)} ;

  \foreach \plm in {otimes*} \draw plot[mark=\plm] coordinates {(0,-1)} ;
  \foreach \plm in {otimes*} \draw plot[mark=\plm] coordinates {(2,0)} ;
  \foreach \plm in {otimes*} \draw [mark options={fill=red}] plot[mark=\plm] coordinates {(4,1)} ;

  \foreach \plm in {otimes*} \draw plot[mark=\plm] coordinates {(3,5)} ;
  \foreach \plm in {otimes*} \draw plot[mark=\plm] coordinates {(4,4)} ;
  \foreach \plm in {otimes*} \draw plot[mark=\plm] coordinates {(5,3)} ;
\end{tikzpicture}
\end{adjustbox}
\caption{A brane configuration with five flavors leading to $T_3$-diagram.}\label{webNf5}
\end{figure}
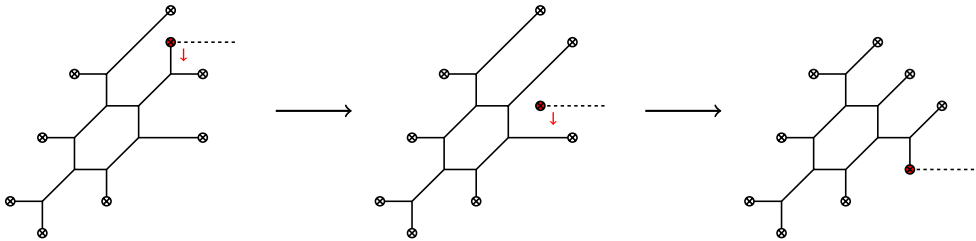

$E_n$ symmetries realized in the framework is not manifest, only subgroup of $E_n$ is manifest.
As an instructive example, we consider $T_3$ theory.
It has the following web diagram or corresponding dual toric diagram:
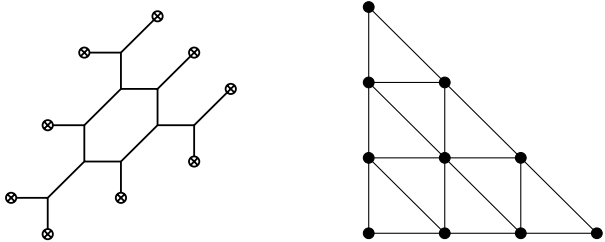
\begin{figure}[H]
\centering
\begin{adjustbox}{width=0.2\textwidth}
\begin{tikzpicture}
  [inner sep=0.5mm, line width=0.5mm,
 dot/.style={fill=black,draw,circle,minimum size=1pt},
 whitedot/.style={fill=white,draw,circle,minimum size=1pt},
 mark size=4pt, mark options={fill=white} ]
    \draw (-1,0) -- (0,0) -- (0,-1);
    \draw (0,0) -- (1,1);
    \draw (1,1) -- (2,1) -- (3, 2) -- (3,3) -- (2,3) -- (1,2) -- (1,1);
    \draw (1,2) -- (0,2); \draw (2,3) -- (2, 4) -- (1,4);
    \draw (2,4) -- (3,5);
    \draw (3,3) -- (4,4); \draw (3, 2) -- (4, 2);
    \draw (4,1) -- (4,2) -- (5,3);
    \draw (2,1) -- (2,0) ;

  \foreach \plm in {otimes*}
  \foreach \plm in {otimes*} \draw plot[mark=\plm] coordinates {(-1,0)} ;
  \foreach \plm in {otimes*} \draw plot[mark=\plm] coordinates {(0,2)} ;
  \foreach \plm in {otimes*} \draw plot[mark=\plm] coordinates {(1,4)} ;

  \foreach \plm in {otimes*} \draw plot[mark=\plm] coordinates {(0,-1)} ;
  \foreach \plm in {otimes*} \draw plot[mark=\plm] coordinates {(2,0)} ;
  \foreach \plm in {otimes*} \draw plot[mark=\plm] coordinates {(4,1)} ;

  \foreach \plm in {otimes*} \draw plot[mark=\plm] coordinates {(3,5)} ;
  \foreach \plm in {otimes*} \draw plot[mark=\plm] coordinates {(4,4)} ;
  \foreach \plm in {otimes*} \draw plot[mark=\plm] coordinates {(5,3)} ;
\end{tikzpicture}
\end{adjustbox}
\qquad \qquad
\begin{tikzpicture}
  [inner sep=0.5mm,
 dot/.style={fill=black,draw,circle,minimum size=1pt},
 whitedot/.style={fill=white,draw,circle,minimum size=1pt}]
    \draw (1,0) -- (1,3) -- (4,0) -- (1,0);
    \draw (2,0) -- (2,2);
    \draw (1,1) -- (3,1) -- (3,0);
    \draw (1,2) -- (2,2);
    \draw (1,2) -- (3,0);
    \draw (1,1) -- (2,0);
    \node[dot] at (1,0) {}; \node[dot] at (1,1) {}; \node[dot] at (1,2) {}; \node[dot] at (1,3) {};
    \node[dot] at (2,0) {}; \node[dot] at (2,1) {}; \node[dot] at (2,2) {};
    \node[dot] at (3,0) {}; \node[dot] at (3,1) {};
    \node[dot] at (4,0) {};
\end{tikzpicture}
\caption{A toric diagram for $E_6$}\label{webe6}
\end{figure}
\noindent As the diagram indicates, it has manifest $SU(3)^3$ global symmetry which is a maximal compact subgroup of $E_6$
\begin{align}
E_6 \supset SU(3)\times SU(3) \times SU(3).
\end{align}
Using the Hanany-Witten transition (as well as monodromy of 7-brane branch cut), we can explain that this diagram is related to the diagram of five flavors ($N_f=5$). In the $(p,q)$ 5-brane web configuration, the matters are represented by semi-infinite $(1,0)$ 5-branes.  We introduce $[p,q]$ 7-branes such that $(p,q)$ 5-branes can end without breaking supersymmetry. See Figure \ref{webNf5}.
\noindent Let us imagine a web diagram with five $(1,0)$ 5-branes ending on 7-branes (denoted by $\otimes$). 
For convenience, we put three on the left and two on the right. 
This is the leftmost web diagram configuration in Figure \ref{webNf5}. In order to avoid colliding of 7-branes, we bring down the $[0,1]$ 7-brane filled in red. Recall that this $[0,1]$ 7-brane has a branch cut denoted by the dashed line. When the $[0,1]$ 7-brane passes through a $(p,q)$ 5-brane, the charge of the 5-brane changes as it experiences monodromy due to the $[0,1]$ 7-brane. For instance, as depicted in the middle of Figure \ref{webNf5}, $(1,0)$ 5-brane charge is altered to $(1,1)$ as the $[0,1]$ 7-brane passes through it. As it is brought to further down, the web configuration becomes the web diagram of $T_3$, which is the rightmost Figure \ref{webNf5}.

\begin{figure}[H]
\centering
\begin{adjustbox}{width=0.2\textwidth}
\begin{tikzpicture}
  [inner sep=0.5mm, line width=0.5mm,
 dot/.style={fill=black,draw,circle,minimum size=1pt},
 whitedot/.style={fill=white,draw,circle,minimum size=1pt},
 mark size=4pt, mark options={fill=white} ]
    \draw (-1,0) -- (0,0) -- (0,-1);
    \draw (0,0) -- (1,1);
    \draw (1,1) -- (2,1) -- (3, 2) -- (3,3) -- (2,3) -- (1,2) -- (1,1);
    \draw (1,2) -- (0,2); \draw (2,3) -- (2, 4) -- (1,4);
    \draw (2,4) -- (5,7);
    \draw (3,3) -- (4,4); \draw (4,5) -- (4,4) -- (5,4);
    \draw (3, 2) -- (5, 2);
    \draw (2,1) -- (2,0) ;
    \draw[->, very thick, color=red] (4.4,5) --  (4.4, 5.5);

  \foreach \plm in {otimes*}
  \foreach \plm in {otimes*} \draw plot[mark=\plm] coordinates {(-1,0)} ;
  \foreach \plm in {otimes*} \draw plot[mark=\plm] coordinates {(0,2)} ;
  \foreach \plm in {otimes*} \draw plot[mark=\plm] coordinates {(1,4)} ;

  \foreach \plm in {otimes*} \draw plot[mark=\plm] coordinates {(0,-1)} ;
  \foreach \plm in {otimes*} \draw plot[mark=\plm] coordinates {(2,0)} ;
  \foreach \plm in {otimes*} \draw plot[mark=\plm] coordinates {(5,2)} ;

  \foreach \plm in {otimes*} \draw plot[mark=\plm] coordinates {(5,7)} ;
  \foreach \plm in {otimes*} \draw[mark options={fill=red}] plot[mark=\plm] coordinates {(4,5)} ;
  \foreach \plm in {otimes*} \draw plot[mark=\plm] coordinates {(5,4)} ;
  \draw[dashed] (4.1,5) -- (5.5,5);
\end{tikzpicture}
\end{adjustbox}
\quad
\begin{tikzpicture}
 \draw[->,thick] (-1.5,1.5) -- (-0.5, 1.5);
   \node[circle, fill=white] at (-0.5,-0.5) {};
\end{tikzpicture}
\begin{adjustbox}{width=0.2\textwidth}
\begin{tikzpicture}
  [inner sep=0.5mm, line width=0.5mm,
 dot/.style={fill=black,draw,circle,minimum size=1pt},
 whitedot/.style={fill=white,draw,circle,minimum size=1pt},
 mark size=4pt, mark options={fill=white} ]
    
    \draw (-1,0) -- (0,0) -- (0,-1);
    \draw (0,0) -- (1,1);
    \draw (1,1) -- (2,1) -- (3, 2) -- (3,3) -- (2,3) -- (1,2) -- (1,1);
    \draw (1,2) -- (0,2); \draw (2,3) -- (2, 4) -- (1,4);
    \draw (2,4) -- (4,6) -- (5,6);
    \draw (3,3) -- (4.1,4.1); \draw (4.1,5.8) -- (4.1,4.1) -- (5,4.1);
    \draw (3, 2) -- (5, 2);
    \draw (2,1) -- (2,0) ;
    \draw (4,6) -- (4,7) ; \draw (4.1, 6.2) -- (4.1,7); \draw (4.1,5.8) arc (-90:90:0.2cm);

  \foreach \plm in {otimes*}
  \foreach \plm in {otimes*} \draw plot[mark=\plm] coordinates {(-1,0)} ;
  \foreach \plm in {otimes*} \draw plot[mark=\plm] coordinates {(0,2)} ;
  \foreach \plm in {otimes*} \draw plot[mark=\plm] coordinates {(1,4)} ;

  \foreach \plm in {otimes*} \draw plot[mark=\plm] coordinates {(0,-1)} ;
  \foreach \plm in {otimes*} \draw plot[mark=\plm] coordinates {(2,0)} ;
  \foreach \plm in {otimes*} \draw plot[mark=\plm] coordinates {(5,2)} ;

  \foreach \plm in {otimes*} \draw plot[mark=\plm] coordinates {(5,6)} ;
  \foreach \plm in {otimes*} \draw[mark options={fill=red}] plot[mark=\plm] coordinates {(4.05,7)} ;
  \foreach \plm in {otimes*} \draw plot[mark=\plm] coordinates {(5,4.1)} ;
	\draw[dashed] (4.15,7) -- (5.55,7);
\end{tikzpicture}
\end{adjustbox}
\quad
\begin{tikzpicture}
  [inner sep=0.5mm,
  dot/.style={fill=black,draw,circle,minimum size=1pt},
 whitedot/.style={fill=white,draw,circle,minimum size=1pt}]
  \draw[<->,thick] (-1.5,1.5) -- (-0.5, 1.5);

    \draw (0,0) -- (0,3) -- (2,3) -- (2,0) -- (0,0);
    \draw (0,1) -- (2,1);
    \draw (0,2) -- (2,2);
    \draw (0,3) -- (2,3);
    \draw (0,1) -- (1,0);
    \draw (0,2) -- (2,0);
    \draw (0,3) -- (2,1);
    \draw (1,0) -- (1,2);
    \node[dot] at (0,0) {}; \node[dot] at (0,1) {}; \node[dot] at (0,2) {}; \node[dot] at (0,3) {};
    \node[dot] at (1,0) {}; \node[dot] at (1,1) {}; \node[dot] at (1,2) {};
    \node[dot] at (2,0) {}; \node[dot] at (2,1) {};
    \node[dot] at (2,2) {}; \node[dot] at (2,3) {}; \node[whitedot] at (1,3) {};
    \node[circle, fill=white] at (-0.5,-0.5) {};
\end{tikzpicture}
\caption{A brane configuration with five flavors leading to a toric-like diagram.}\label{webNf5-1}
\end{figure}
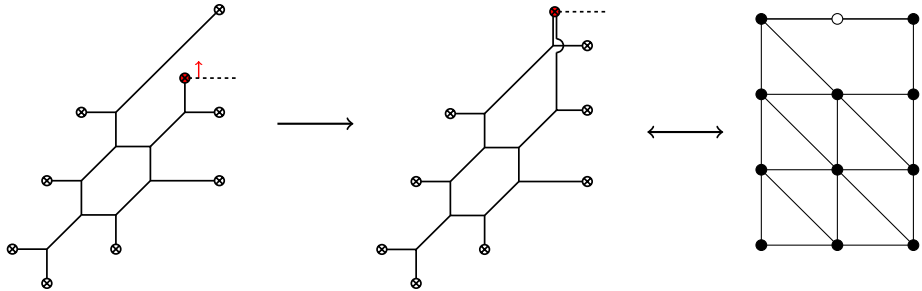

It is interesting to see what happens to the web diagram if one pushes up the $[0,1]$ 7-brane rather than bringing it down. See Figure \ref{webNf5-1}.
As explained in \cite{Benini:2009gi}, the $[0,1]$ 7-brane jumps over the $(1,1)$ 5-brane when crossing, and another $(0,1)$ 5-brane is created to attach to the $[0,1]$ 7-brane (the Hanany-Witten effect). The resultant dual toric-like diagram involves a new kind of dot (white dot) to indicate this jumping phenomenon associated with binding multiple 5-branes attached to a single 7-brane. In this way, the number of the Coulomb moduli remains unaltered. 
 This dual toric-like diagram was called a dot diagram in \cite{Benini:2009gi}.

If the white dot above were a black dot, it would increase the number of both the dimension of the Coulomb moduli and triangulation, hence it would be a toric diagram for $SU(3)$ gauge theory. Notice that in this specially tuned web diagram or toric-like diagram in Figure \ref{webNf5-1}, it appear to have six flavors coming from six $(1,0)$ 5-branes. This implies that  the  manifest global symmetry is no longer $SU(3)^3$ but it is $S[U(3)\times U(3)]\times SU(2)$. Turning a black dot into a white dot can be interpreted as a procedure of Higgsing. But we call it a special tuning, as the it can be understood as a procedure keeping dimension of the Coulomb moduli to be one.  We then say that it provides a dual picture of the $SU(2)$ theory of $E_6$ symmetry as a special tuning of $SU(3)$ gauge theory with six flavors.

It is clear that the toric diagram for the $T_3$ theory can be given in a different way through the Hanany-Witten transition as explained. This means that as one writes the SW curve based on a toric diagram, two SW curves obtained from two different toric diagrams should be related by the Hanany-Witten effect.
We emphasize that the way that the Hanany-Witten effect
is realized in the SW curve is a coordinate transformation.
It is also worth noting that as a toric diagram shows manifest global symmetry,
one can find different manifest symmetry by the Hanany-Witten transition. 
As we will show later, we find that the Hanany-Witten transition on a given (tuned) 
$T_N$ theory gives rise to different compact subgroups of $E_n$ symmetry.

We now consider the construction of the SW curve. For this, we compactify the theory on a circle, and T-dualize it to become IIA theory, and then we uplift it to M-theory. The curve then describes M5 brane configuration embedded in $\mathbb{R}^2\times T^2$. Given a toric diagram, say Figure. \ref{webe6}, the SW curve takes the form as
\begin{align}
\sum_{ij}c_{ij}\,t^i \,w^j = 0,
\label{SW curve gen}
\end{align}
with the SW one-form
\begin{align}
\lambda_{SW}= \log t~ {\rm d}(\log w).
\end{align}
Here, $c_{ij}$ in (\ref{SW curve gen}) is the non-vanishing coefficient that corresponds to the ($i,j$) dot in the toric diagram.

\begin{figure}[H]
\centering
\begin{tikzpicture}
  [inner sep=0.5mm,
 dot/.style={fill=black,draw,circle,minimum size=1pt},
 whitedot/.style={fill=white,draw,circle,minimum size=1pt}]
    \draw[->] (-1,-1) -- (-1, 4);
    \draw[->] (-1, 4) -- (-1,-1) -- (4, -1);
    \draw[red] (0,0) -- (0,3) -- (3,0) -- (0,0);
    \draw (0, 2.5) -- (-0.9, 2.5); \draw (0, 1.5) -- (-0.9, 1.5); \draw (0, 0.5) -- (-0.9, 0.5);
    \draw (0.5, 2.5) -- (1.5, 3.5); \draw (1.5, 1.5) -- (2.5, 2.5); \draw (2.5, 0.5) -- (3.5, 1.5);
    \draw (0.5, 0) -- (0.5, -0.9); \draw (1.5, 0) -- (1.5, -0.9); \draw (2.5, 0) -- (2.5, -0.9);

    \node[dot] at (0,0) {}; \node[dot] at (0,1) {}; \node[dot] at (0,2) {}; \node[dot] at (0,3) {};
    \node[dot] at (1,0) {}; \node[dot] at (1,1) {}; \node[dot] at (1,2) {};
    \node[dot] at (2,0) {}; \node[dot] at (2,1) {};
    \node[dot] at (3,0) {};

    \node[] at (0.5,-1.6) [label=above:$N_1$] {};
    \node[] at (1.5,-1.6) [label=above:$N_2$] {};
    \node[] at (2.5,-1.6) [label=above:$N_3$] {};
    
    \node[] at (-1.5,0.2) [label=above:$M_1$] {};
    \node[] at (-1.5,1.2) [label=above:$M_2$] {};
    \node[] at (-1.5,2.2) [label=above:$M_3$] {};
    
    \node[] at (-0.5,-0.5) [label=above:$c_{00}$] {};
    \node[] at (0,3.2) [label=above:$c_{03}$] {};
    \node[] at (3.5,-0.3) [label=above:$c_{30}$] {};

    \node[] at (1.5,3.5) [label=above:$w\approx-L_{3}t$] {};
    \node[] at (2.5,2.5) [label=above:$w\approx-L_{2}t$] {};
    \node[] at (3.5,1.5) [label=above:$w\approx-L_{1}t$] {};

    \node[] at (4.3,-1.3) [label=above:$t$] {};
    \node[] at (-1.2,4.0) [label=above:$w$] {};

    \node[] at (1,1) [label=left:$U$] {};

\end{tikzpicture}
\caption{The configuration for the SW curve of $T_3$ toric diagram}\label{torict4}
\end{figure}
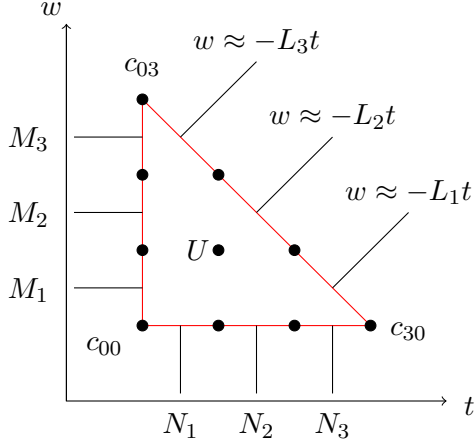
For instance, consider the configuration for the $T_3$ toric diagram given in Figure \ref{torict4}.
There are 10 coefficients $c_{ij}$ which can be determined from the boundary conditions.
Here, we impose the following boundary condition
\begin{align}
&t = N_1,N_2,N_3 \quad {\rm as } \quad w\to0 \cr
&w = M_1, M_2, M_3 \quad {\rm as } \quad t\to0 \cr
&w = - L_1 t,  - L_2 t, - L_3 t \quad {\rm as } \quad |t| \sim |w| \to \infty,
\label{bcLMN}
\end{align}
where the first, second and third lines in Figure \ref{torict4} correspond to
three NS5-branes, three D5-branes and three (1,1) 5-branes, respectively,
in the original type IIB picture.
%See Figure \ref{torict4}.
These conditions yield the constraints to the coefficients:
\begin{align}
\sum_{i=0}^3 c_{i0}t^i &= c_{30} (t-N_1)(t-N_2)(t-N_3), \cr
\sum_{j=0}^3 c_{0j}w^j &= c_{03} (w-M_1)(w-M_2)(w-M_3), \cr
\sum_{j=0}^3 c_{i,3-i}t^iw^{3-i} &= c_{03} (w+L_1t)(w+L_2t)(w+L_3t),
\label{bcE6const}
\end{align}
which enables us to express the coefficients $c_{ij}$ in terms of $L_i, M_i$, and $N_i$.
In order for these conditions to be consistent,
we find that the following compatibility conditions are necessary
\begin{align}
M_1M_2M_3= N_1N_2N_3\cdot L_1L_2L_3.
\end{align}
This means that one of the constraint equations in \eqref{bcE6const}
%out of nine equations in (\ref{bcE6const}) 
is used not to determine the coefficient but to determine this compatibility condition.
Since the SW curve does not change when we multiply an identical constant to all the
coefficients, this degree of freedom can be also used to determine one coefficient as we like.
Including this, we can determine 9 coefficients out of 10 in (\ref{SW curve gen}).
The undetermined coefficient $c_{11}$ corresponds to the internal dot
in the toric diagram in Figure \ref{torict4} and is not affected by the boundary condition.
This coefficient $c_{11}$ is interpreted as the Coulomb moduli parameter and we denote it as $U$.

Using the degrees of freedom of rescaling $t$ and $w$,
we can further impose the conditions on $N_i$ and $M_i$.
Together with the compatibility condition, it is convenient to impose
\begin{align}
M_1M_2M_3 = N_1N_2N_3 = L_1L_2L_3.
\label{LMNconv}
\end{align}
It is then straightforward to find the curve corresponding to the toric diagram, Figure \ref{torict4}:
\begin{align}
&w^3-\sum_i M_i w^2 + \sum_i L_i  w^2 t+ \sum_i M^{-1}_i w + U wt\cr
&\qquad \qquad + \sum_i L^{-1}_i wt^2 -1 +\sum_i N_i^{-1} \,t - \sum_i N_i \,t^2 + t^3=0.
\label{SWE6-tri}
\end{align}

The procedure to move from the diagram in Figure \ref{torict4} to the one in Figure \ref{webNf5-1}
by moving the [0,1] 7-brane in Figure \ref{torict4} at $t=N_3$ upward by using Hanany-Witten effect
can be realized by the coordinate transformation
\begin{align}
w=W(t-N_3).
\label{HWN3}
\end{align}
With this transformation, the SW curve (\ref{SWE6-tri}) can be expressed as
\begin{align}
&(t-N_3)^2 W^3 + (t-N_3)  \left( \sum_i L_i t  - \sum_i M_i \right) W^2 \cr
&\qquad\qquad + \left( \sum_i L^{-1}_i t^2 + U t + \sum_i M^{-1}_i \right) W
+ (t-N_1)(t-N_2)=0,
\label{SWE6-rect}
\end{align}
where we have divided entire equation by the factor $(t-N_3)$.
Although the corresponding diagram in Figure \ref{webNf5-1} includes a white dot, every dots correspond to non-vanishing coefficients of the SW curve.
%we observe that the non-vanishing coefficients still
%corresponds to the dots of the toric-like diagram also in this case,
%which is the same as the rule for usual toric diagrams.
However, compared to the SW curve corresponding to the usual toric diagram,
we find that the coefficients are tuned to be specific values.
That is, some extra conditions are imposed to this SW curve due to the white dot.
Writing the left hand side of (\ref{SWE6-rect}) as $\sum_{i,j} c_{ij} t^i W^j$,
it is straightforward to see that this SW curve satisfies the following relation:
\begin{align}
&
\sum_{i=0}^2 c_{i3} t^i \propto (t-N_3)^2, \qquad
\sum_{i=0}^2 c_{i2} t^i \propto (t-N_3), \cr
&
\sum_{i=0}^2 c_{i0} t^i \propto (t-N_1)(t-N_2), \cr
&
\sum_{j=0}^3 c_{2j} W^j \propto (W-L_1)( W-L_2)( W-L_3), \cr
&
\sum_{j=0}^3 c_{0j} W^j \propto
\left( W-\frac{M_1}{N_3} \right)
\left( W-\frac{M_2}{N_3} \right)
\left( W-\frac{M_3}{N_3} \right),
\label{bcE6}
\end{align}
\begin{figure}
\centering
\begin{adjustbox}{width=0.5\textwidth}
\begin{tikzpicture}
  [inner sep=0.5mm,
  dot/.style={fill=black,draw,circle,minimum size=1pt},
 whitedot/.style={fill=white,draw,circle,minimum size=1pt}]

    \draw (0,0) -- (0,3) -- (2,3) -- (2,0) -- (0,0);
    \draw (0,1) -- (2,1);
    \draw (0,2) -- (2,2);
    \draw (0,3) -- (2,3);
    \draw (0,1) -- (1,0);
    \draw (0,2) -- (2,0);
    \draw (0,3) -- (2,1);
    \draw (1,0) -- (1,2);
    \node[dot] at (0,0) [label=left:$c_{00}$] {}; 
    \node[dot] at (0,1) [label=left:$c_{01}$] {}; 
    \node[dot] at (0,2) [label=left:$c_{02}$] {}; 
    \node[dot] at (0,3) [label=left:$c_{03}$] {};
    \node[dot] at (1,0) [label=below:$c_{10}$] {}; 
    \node[dot] at (1,1) [label={[label distance=0.03cm]45:$c_{11}$}] {}; 
    \node[dot] at (1,2) [label={[label distance=0.03cm]45:$c_{12}$}]{};
    \node[whitedot] at (1,3) [label=above:$c_{13}$] {};
    \node[dot] at (2,0) [label=right:$c_{20}$] {}; 
    \node[dot] at (2,1) [label=right:$c_{21}$] {};
    \node[dot] at (2,2) [label=right:$c_{22}$] {}; 
    \node[dot] at (2,3) [label=right:$c_{23}$] {}; 

\draw [red] (1,0) circle [x radius=2cm, y radius=0.45cm]; 
\draw [red] (1,2) circle [x radius=2cm, y radius=0.45cm]; 
\draw [red] (1,3) circle [x radius=2cm, y radius=0.45cm]; 

\draw [blue] (0,1.5) circle [x radius=0.5cm, y radius=2.3cm]; 
\draw [blue] (2,1.5) circle [x radius=0.5cm, y radius=2.3cm]; 

\node[] at (3,3) [label=right:$ {\color{red}\propto (t-N_3)^2}$] {}; 
\node[] at (3,2) [label=right:$ {\color{red}\propto (t-N_3)}$] {}; 
\node[] at (3,0) [label=right:$ {\color{red}\propto (t-N_1)(t-N_2)}$] {}; 

\node[] at (2.5,-2.5) [label=above:$ {\color{blue}\propto (W-\frac{M_1}{N_3})(W-\frac{M_2}{N_3})(W-\frac{M_3}{N_3})}$] {};
\node[] at (4.5,-0.8) [label=below:${\color{blue} \propto (W-L_1)(W-L_2)(W-L_3)}$] {}; 
\draw [blue] [->] (0,-1.8)--(0,-1);

\end{tikzpicture}
\end{adjustbox}
\caption{The boundary conditions toric-like diagram for $E_6$.}\label{bcfore6wpwhites}
\end{figure}
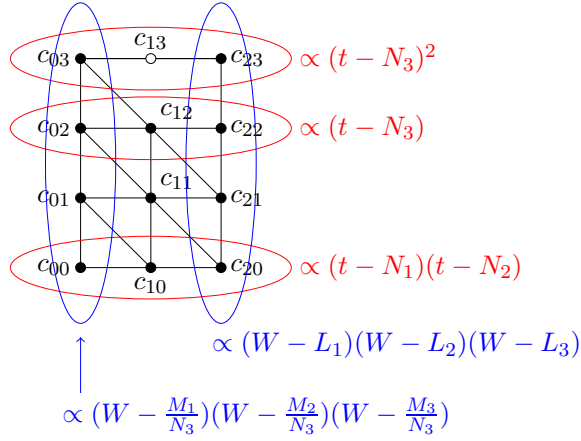
which we interpret as the constraints coming from the boundary conditions.

As depicted in Figure \ref{bcfore6wpwhites}, the white dots in the toric-like diagram is associated with the first line of (\ref{bcE6}).
The first relation, which gives the leading behavior at $W \to \infty$,
is consistent with the two NS5-branes are coincident, bound by one [0,1] 7-brane.
Furthermore, as for the second relation, which gives the subleading contribution
in the region $W \to \infty$,
we would like to interpret that this is the consequence that
one out of two coincident NS5-branes jumps over the other 5-brane.
The remaining three relations from the third to the fifth are more straightforward to give the interpret.
They correspond to the boundary conditions for two NS5-branes at $W \to 0$,
three external D5-branes at $t \to \infty$, and
three more external D5-branes at $t \to 0$.

It is worth noting that these boundary conditions (\ref{bcE6})
together with our convention (\ref{LMNconv})
are enough to reproduce the SW curve (\ref{SWE6-rect}).
This example gives us an intuition of what kind of boundary condition we should impose,
if we have white dots in a generic toric-like diagram. In the following, we discuss how the boundary conditions yields the SW curve in more details.

\subsection{General procedure}\label{generalprocedure}
Based on the computation of the SW curve for $E_6$ theory,
we propose a systematic procedure of deriving the SW curve from
any given toric-like diagram.

If all the vertices are black dots, it is a usual toric diagram
and thus the SW curve is given by
\begin{eqnarray}
\sum_{(i,j) \in {\rm vertices}} c_{ij} t^i w^j = 0,
\label{SW-gen}
\end{eqnarray}
where $(i,j)$ are summed over all the vertices in the diagram.
The coefficients $c_{ij}$ corresponding to the dots
on the boundary edge of the toric diagram
are determined by the boundary condition for the external $(p,q)$ 5-branes.
These boundary conditions are given in the form
\begin{eqnarray}
w^{p} t^{-q} \sim \tilde{m}_n^{(p,q)} \qquad {\rm at} \qquad |w|^q \sim |t|^p \to \infty,
\end{eqnarray}
where this $\tilde{m}_n^{(p,q)}$ corresponds to the ``mass parameter''\footnote{The ``mass parameter'' here is not physical mass of the gauge theory. It is typically the exponential of some linear combination of masses and inverse gauge coupling with some shift.}.
For example, the first line in (\ref{bcLMN}) is obtained by identifying $p=0$, $q=-1$ and $\tilde{m}_n^{(-1,0)}=N_n$.
These boundary conditions give the constraints
\begin{eqnarray}
\sum_{(ij) \atop {ip+jq=N_{i,j}}} c_{i,j} t^i q^j
\propto \prod_{n} (w^p - \tilde{m}^{(p,q)}_n t^q),
\end{eqnarray}
where $N_{i,j}$ is the maximum
of the value $ip+jq$ among all the combination of $(i,j)$
corresponding to the vertex in the toric diagram.
In other words, the $(i,j)$ is summed over the vertex on the boundary edge
which is perpendicular to the considered $(p,q)$ 5-brane.
We impose this boundary condition corresponding to all the external $(p,q)$ 5-branes
and solve for $c_{i,j}$.
Note that in order for a solution to exist, there must be one
constraint among all the mass parameters.

Generic toric-like diagram is  obtained by converting some of black dots
into white dots in toric diagram.
This procedure corresponds to the tuning of the coefficients $c_{ij}$.
Therefore, the SW curve is still the same form as (\ref{SW-gen})
but more conditions are added to the coefficients.
Suppose that $n$-th $(p,q)$ 7-brane binds $k_n$ external $(p,q)$ 5-branes.
In this case, the boundary conditions are of the following sets of constraint
\begin{eqnarray}
\sum_{(ij) \atop {ip+jq=N_{i,j}}} c_{i,j} t^i q^j
&\propto& \prod_{n} (w^p - \tilde{m}^{(p,q)}_n t^q)^{k_n},
\nonumber \\
\sum_{(ij) \atop ip+jq=N_{i,j} - 1 } c_{i,j} t^i q^j
&\propto& \prod_{n} (w^p - \tilde{m}^{(p,q)}_n t^q)^{{\rm max}(k_n-1,0)},
\nonumber \\
\cdots
\nonumber \\
\sum_{(ij) \atop {ip+jq=N_{i,j}} - \ell} c_{i,j} t^i q^j
&\propto& \prod_{n} (w^p - \tilde{m}^{(p,q)}_n t^q)^{{\rm max}(k_n-\ell,0)},
\nonumber \\
\cdots
\label{proposal}
\end{eqnarray}
These sets of constraints are understood as natural generalization of the first two
conditions in (\ref{bcE6}).

\subsection{Comment on general procedure}

In the previous subsection, we proposed the procedure to derive
SW curve from generic toric-like diagram.
In this subsection, we discuss that this procedure
can be understood as a natural 5d uplift of
the SW curve for the following D4-NS5 brane setup with flavor D6-branes \cite{Witten:1997sc}.

To begin with, we review the related result in \cite{Witten:1997sc}.
Suppose that we have $n+1$ NS5-branes labeled by $\alpha=0,1,\cdots,n$.
Between $\alpha$-th NS5-brane and $(\alpha-1)$-th NS5-brane,
$k_{\alpha}$ color D4 branes are suspended ($\alpha=1,\cdots, n$).
Moreover, $(i_{\alpha}-i_{\alpha-1})$ D6-branes exist
between $\alpha$-th NS5-brane and $(\alpha-1)$-th NS5-brane
at the place $v = e_a$ with $a=i_{\alpha-1}+1 ,\cdots, i_{\alpha}$,
where we set $i_0=0$.
In this setup, no D4-branes are attached to any of the D6-branes.
The D6-branes are uplifted to the Taub-NUT space in the M-theory,
which is defined by
\begin{align}
\frac{y}{z} = \prod_{a=1}^{i_{n}} (v-e_a) = \prod_{s=1}^n J_s(v),
\label{TaubNUT}
\end{align}
embedded in $\mathbb{C}^3$ with three complex coordinates%
\footnote{Our $z$ corresponds to $z^{-1}$ in \cite{Witten:1997sc}.}
 $y$, $z$, and $v$,
where we put
\begin{align}
J_s(v) = \prod_{a=i_{s-1}+1}^{i_s} (v-e_a).
\end{align}
It is known that the SW curve is given in the following form:
\begin{align}
&
y^{n+1} + g_1(v) y^n + g_2(v) J_1(v) y^{n-1} + g_3(v) J_1(v)^2 J_2(v) y^{n-2}\cr
& \qquad \qquad
+ \cdots + g_{\alpha} (v) \prod_{s=1}^{\alpha-1} J_s(v)^{\alpha-s} \cdot y^{n+1-\alpha}
+ \cdots
+ f \prod_{s=1}^n J_s(v)^{n+1-s} = 0,
\label{yeq}
\end{align}
where $g_{\alpha}(v)$ are polynomials of degree $k_{\alpha}$.
Or, if we change the coordinate from $y$ to $z$ by using (\ref{TaubNUT}), it reads
\begin{align}
&
\prod_{s=1}^n J_s^{s} \cdot z^{n+1}
+ g_1(v) \prod_{s=2}^{n} J_s^{s-1} \cdot z^n
+ g_2(v) \prod_{s=3}^{n} J_s^{s-2} \cdot z^{n-1} \cr
& \qquad \qquad
+ \cdots + g_{\alpha} (v) \prod_{s=\alpha+1}^{n} J_s^{s-\alpha} \cdot z^{n+1-\alpha}
+ \cdots + g_{n-1}(v) J_n(v) z + f  = 0.
\label{zeq}
\end{align}

In order to connect this results to our proposal,
we reinterpret this in a slightly different way.
Instead of placing the D6-branes between the NS5-branes without any D4 branes attached,
we can move these D6-branes horizontally to infinity using the Hanany-Witten effect.
Suppose that we move all the D6-branes to the direction of the outside of the $0$-th NS5-brane.
Through this process, the $(i_{\alpha}-i_{\alpha-1})$ D6-branes originally placed
between $\alpha$-th NS5-brane and $(\alpha-1)$-th NS5-brane
pass through NS5-branes $\alpha$ times, and thus
each D6 brane binds $\alpha$ D4-branes.
The bound D4-branes jump over NS5-branes properly in such a way to avoid breaking the s-rule.
In this setup, since D6-branes are placed at infinity,
the space is not Taub NUT space anymore but just flat $\mathbb{C}^2$.

We reinterpret (\ref{zeq}) as the M5-brane configuration for this situation,
where the space is now flat $\mathbb{C}^2$ spanned by the two coordinates $v$ and $z$.
All the D6-branes now exist at the region $z \to \infty$, and
the first term in (\ref{zeq}) is consistent with the situation that
$(i_{s}-i_{s-1})$ D6-branes bind $s$ D4-branes each.
The second term has also the factor $J_s$ but with one power less.
As we decrease the power of $z$, the power of $J_s$ also reduces one by one.
We claim that this is the counterpart of our proposal (\ref{proposal}).
What is generalized in our proposal is that we consider not only for
$D7$-branes but also for arbitrary $[p,q]$ 7-branes.
This generalization appears only when we uplift to five dimensions.

We note that with this reinterpretation, we find that (\ref{TaubNUT}),
which originally defines the multi Taub NUT space,
can be seen as the coordinate transformation to move
the external D4-branes from one side to the other side by
the Hanany-Witten transition, which is the analogue of (\ref{HWN3}).
Then, the curve (\ref{yeq}) is also consistent with
this interpretation, where
the $(i_{\alpha}-i_{\alpha-1})$ D6-branes originally placed
between $\alpha$-th NS5-brane and $(\alpha-1)$-th NS5-brane
bind $n+1-\alpha$ D4-branes each.
An analogous consistency check is also possible for our examples dealt in this paper.
That is, even after such coordinate transformation,
our proposal (\ref{proposal}) is still satisfied.

%%%%%%%%%%%%%%%%%%%%%%%%%%%%%%%%%%%%%%%%%%%%%%
%%%%%%%%%%%%%%%%%%%%%%%%%%%%%%%%%%%%%%%%%%%%%%
%%%%%%%%%%%%%%%%%%%%%%%%%%%%%%%%%%%%%%%%%%%%%%
%%%%%%%%%%%%%%%%%%%%%%%%%%%%%%%%%%%%%%%%%%%%%%
\section{$E_7$ Seiberg-Witten curve}

In this section, we compute the Seiberg-Witten curve for 5d Sp(1) theory $N_f=6$ flavor based on toric-like diagram.
After constructing the corresponding toric-like diagram, we compute the SW curve
using the technique developed in the previous section.
Then, we check that it is consistent with the known expression 
written in $E_7$ invariant manner \cite{Eguchi:2002fc, Eguchi:2002nx}.
We also study the 4d limit of the 5d SW curve and show that it reproduces the expected curve \cite{Gaiotto:2009we}.
\begin{figure}[t]
\centering
\begin{adjustbox}{width=0.2\textwidth}
\begin{tikzpicture}
  [inner sep=0.5mm, line width=0.5mm,
 dot/.style={fill=black,draw,circle,minimum size=1pt},
 whitedot/.style={fill=white,draw,circle,minimum size=1pt},
 mark size=4pt, mark options={fill=white} ]
    
    \draw (-1,0) -- (0,0) -- (0,-1);  \draw[dashed] (0,-1) -- (-2,-1);
    \draw (0,0) -- (1,1);
    \draw (1,1) -- (2,1) -- (3, 2) -- (3,3) -- (2,3) -- (1,2) -- (1,1);
    \draw (1,2) -- (0,2); \draw (2,3) -- (2, 4) -- (1,4);
    \draw (2,4) -- (4,6);
    \draw (3,3) -- (4,4); \draw (4,5) -- (4,4) -- (5,4); \draw[dashed] (4,5) -- (6,5);
    \draw (3, 2) -- (4, 2);
    \draw (2,1) -- (2,0) ;
    \draw (0,-2) -- (2,0) -- (3,0);
    \draw[<-, very thick, color=red] (4.4,4.4) --  (4.4,4.8);
    \draw[->, very thick, color=red] (-0.4,-1.2) --  (-0.4,-1.6);

  \foreach \plm in {otimes*} \draw plot[mark=\plm] coordinates {(0,-2)} ;
  \foreach \plm in {otimes*} \draw plot[mark=\plm] coordinates {(3,0)} ;

  \foreach \plm in {otimes*} \draw plot[mark=\plm] coordinates {(-1,0)} ;
  \foreach \plm in {otimes*} \draw plot[mark=\plm] coordinates {(0,2)} ;
  \foreach \plm in {otimes*} \draw plot[mark=\plm] coordinates {(1,4)} ;

  \foreach \plm in {otimes*} \draw plot[mark=\plm] coordinates {(0,-1)} ;
  \foreach \plm in {otimes*} \draw plot[mark=\plm] coordinates {(4,2)} ;

  \foreach \plm in {otimes*} \draw plot[mark=\plm] coordinates {(4,6)} ;
  \foreach \plm in {otimes*} \draw plot[mark=\plm] coordinates {(4,5)} ;
    \foreach \plm in {otimes*} \draw plot[mark=\plm] coordinates {(5,4)} ;
   \node[circle, fill=white] at (0,-3.5) {};
\end{tikzpicture}
\end{adjustbox}
\quad
\begin{tikzpicture}
 \draw[->,thick] (-1.5,1) -- (-0.5, 1);
   \node[circle, fill=white] at (-0.5,-1.18) {};
\end{tikzpicture}
\begin{adjustbox}{width=0.25\textwidth}
\begin{tikzpicture}
  [inner sep=0.5mm, line width=0.5mm,
 dot/.style={fill=black,draw,circle,minimum size=1pt},
 whitedot/.style={fill=white,draw,circle,minimum size=1pt},
 mark size=4pt, mark options={fill=white} ]
    \draw (-1,0) -- (0,0) -- (0,-1.82);\draw (0,-1.8) arc (90:270:0.2cm); \draw (0,-2.18) -- (0,-3);
    \draw (0,0) -- (1,1);
    \draw (1,1) -- (2,1) -- (3, 2) -- (3,3) -- (2,3) -- (1,2) -- (1,1);
    \draw (1,2) -- (-1,2); \draw (2,3) -- (2, 4) -- (-1,4);
    \draw (2,4) -- (4,6);
    \draw (3,3) -- (5,5); \draw (3, 2) -- (4, 2);
    \draw (4,0.18) -- (4,2) -- (6,4);  \draw[dashed] (4,-3) -- (6,-3);
    \draw (4,-0.18) -- (4,-3);
    \draw (4,0.2) arc (90:270:0.2cm);

    \draw (2,1) -- (2,0);
    \draw (2,0) -- (0.1,-2) -- (-1,-2);\draw (0.1,-2) --(0.1, -3) ;\draw[dashed] (0,-3) -- (-2,-3);
    \draw (2,0) -- (4.1,0) -- (7.1,3);\draw (4.1,0) -- (4.1,-3);

  \foreach \plm in {otimes*}
 \foreach \plm in {otimes*} \draw plot[mark=\plm] coordinates {(-1,0)} ;
  \foreach \plm in {otimes*} \draw plot[mark=\plm] coordinates {(-1,2)} ;
  \foreach \plm in {otimes*} \draw plot[mark=\plm] coordinates {(-1,4)} ;

  \foreach \plm in {otimes*} \draw plot[mark=\plm] coordinates {(0.05,-3)} ;
  \foreach \plm in {otimes*} \draw plot[mark=\plm] coordinates {(-1,-2)} ;
  \foreach \plm in {otimes*} \draw plot[mark=\plm] coordinates {(4.05,-3)} ;

  \foreach \plm in {otimes*} \draw plot[mark=\plm] coordinates {(4,6)} ;
  \foreach \plm in {otimes*} \draw plot[mark=\plm] coordinates {(5,5)} ;
  \foreach \plm in {otimes*} \draw plot[mark=\plm] coordinates {(6,4)} ;
  \foreach \plm in {otimes*} \draw plot[mark=\plm] coordinates {(7.1,3)} ;
\end{tikzpicture}
\end{adjustbox}
\caption{$N_f=6$ brane configuration (left) and a tuned $T_4$ diagram after Hanany-Witten transitions (right).}\label{webNf6andt4}
\end{figure}
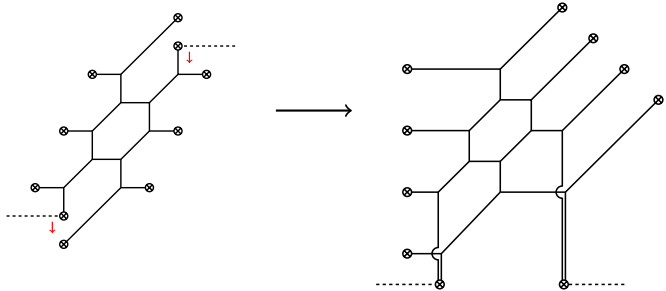

\subsection{Construction of the toric-like diagram}
Let us first start by adding one more flavor to the brane configuration for $N_f=5$, for instance, the first diagram of Figure \ref{webNf6andt4}. As shown for the $N_f=5$ case leading to $T_3$ diagram, it is then straightforward to obtain a tuned $T_4$ diagram, via successive use of the Hanany-Witten transition.
\begin{figure}[H]
\centering

\begin{tikzpicture}
  [inner sep=0.5mm,
 dot/.style={fill=black,draw,circle,minimum size=1pt},
 whitedot/.style={fill=white,draw,circle,minimum size=1pt}]
    \draw (0,0) -- (4,0) -- (0,4) -- (0,0);
    \draw (0,1) -- (3,1);
    \draw (0,2) -- (2,2);
    \draw (2,0) -- (2,2);
    \draw (1,1) -- (1,3) -- (0,3);
    \draw (0,2) -- (2,0);
    \draw (2,1) -- (0,3);

    \node[dot] at (0,0) {}; \node[dot] at (0,1) {}; \node[dot] at (0,2) {}; \node[dot] at (0,3) {}; \node[dot] at (0,4) {};
    \node[whitedot] at (1,0) {}; \node[dot] at (1,1) {}; \node[dot] at (1,2) {}; \node[dot] at (1,3) {};
    \node[dot] at (2,0) {}; \node[dot] at (2,1) {}; \node[dot] at (2,2) {};
    \node[whitedot] at (3,0) {}; \node[dot] at (3,1) {};
    \node[dot] at (4,0) {};

\end{tikzpicture}
\caption{A toric-like diagram for the $SU(2)$ theory with $N_f=6$ flavors. It can be viewed as a tuned $T_4$ diagram. It has manifest $SU(4)\times SU(4)\times SU(2)$ symmetry, but it is invariant under which $E_7$ symmetry}\label{webe7}
\end{figure}
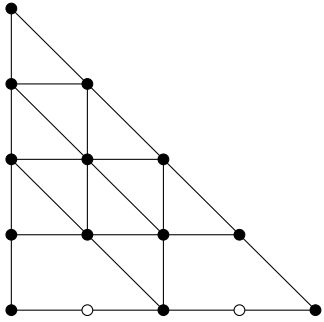
\noindent The corresponding toric-like diagram is given in Figure \ref{webe7}. 
This is a $T_4$ diagram that is specially tuned to have only one Coulomb modulus. It has manifest global symmetry $SU(4)\times SU(4)\times SU(2)$, which is a maximal compact subgroup of $E_7$.

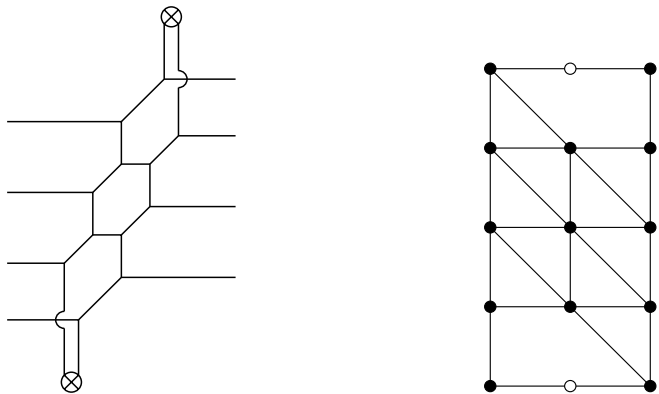
\begin{figure}[H]
\centering
\begin{adjustbox}{width=0.2\textwidth}
\begin{tikzpicture}
  [inner sep=0.5mm, line width=1.5pt,
 dot/.style={fill=black,draw,circle,minimum size=1pt},
 whitedot/.style={fill=white,draw,circle,minimum size=1pt},
 mark size=10pt, mark options={fill=white} ]
	\draw (-2,0)  -- (0.5,0) -- (0.5, -2);
    \draw (0.5,0)  -- (2,1.5) -- (6, 1.5);
    \draw (2,1.5)  -- (2,3) -- (1,3);
    \draw (2,3)  -- (3,4) -- (6,4);
    \draw (3,4)  -- (3,5.5) -- (4,6.5) -- (6, 6.5);
    \draw (4,6.5)  -- (4,8.2);
    \draw (4,8.2) arc (-90:90:0.3cm);
    \draw (4,8.8) -- (4,10.5);

    \draw (0,-2) -- (0,-0.3);
    \draw (0,0.3) arc (90:270:0.3cm);
    \draw (0,0.3) -- (0,2) -- (-2,2);
    \draw (0,2) -- (1,3) -- (1, 4.5) -- (-2, 4.5);
    \draw (1,4.5) -- (2, 5.5) -- (3, 5.5);
    \draw (2,5.5) -- (2, 7) -- (-2, 7);
    \draw (2,7) -- (3.5,8.5) -- (3.5,10.5);
    \draw (3.5,8.5) --(6,8.5);

  	\foreach \plm in {otimes*} \draw plot[mark=\plm] coordinates {(0.25,-2.2)} ;
  	\foreach \plm in {otimes*} \draw plot[mark=\plm] coordinates {(3.75,10.7)} ;
  	\end{tikzpicture}
\end{adjustbox}
\hspace{3cm}
%%%%%%%%%%%%%%%%%%%%%%%%%%%%%%%%%%%%%%%
\begin{adjustbox}{width=0.15\textwidth}
\begin{tikzpicture}
  [inner sep=0.5mm,
 dot/.style={fill=black,draw,circle,minimum size=1pt},
 whitedot/.style={fill=white,draw,circle,minimum size=1pt}]
    \draw (0,0)  -- (2,0)  -- (2,4) -- (0,4) -- (0,0);
    \draw (0,1)  -- (2,1);
    \draw (0,2)  -- (2,2);
    \draw (0,3)  -- (2,3);
    \draw (2,0)  -- (0,2);
    \draw (2,1)  -- (0,3);
    \draw (2,2)  -- (0,4);
    \draw (1,1)  -- (1,3);

    \node[dot] at (2,0) {};% [label=above:$c_{20}$] {};
    \node[dot] at (2,1) {};%[label=above:$c_{21}$] {};
    \node[dot] at (2,2) {};%[label=above:$c_{22}$] {};
    \node[dot] at (2,3) {};%[label=above:$c_{23}$] {};
    \node[dot] at (2,4) {};%[label=above:$c_{24}$] {};
    \node[whitedot] at (1,0) {};%[label=right:$c_{10}$] {};
    \node[dot] at (1,1) {};%[label=right:$c_{11}$] {};
    \node[dot] at (1,2) {};%[label=right:$c_{12}$] {};
    \node[dot] at (1,3) {};%[label=right:$c_{13}$] {};
    \node[whitedot] at (1,4) {};%[label=right:$c_{14}$] {};
    \node[dot] at (0,0) {};%[label=below:$c_{00}$] {};
    \node[dot] at (0,1) {};%[label=below:$c_{01}$] {};
    \node[dot] at (0,2) {};%[label=below:$c_{02}$] {};
    \node[dot] at (0,3) {};%[label=below:$c_{03}$] {};
    \node[dot] at (0,4) {};%[label=below:$c_{04}$] {};
\end{tikzpicture}
\end{adjustbox}

\caption{
Left:  Another web diagram for the $SU(2)$ theory with $N_f=7$ flavors. We use $\otimes$ to denote the 7-brane that combine 5-branes.
Right:  Another toric-like diagram for $E_7$. 
It has manifest $S[U(4)\times U(4)]$ symmetry. 
This can be viewed as a tuned toric diagram for $SU(4)$ gauge theory with eight flavors. 
}\label{e7dot2}
\end{figure}

As discussed in section \ref{sec:T3}, the curve can be seen in various ways through the Hanany-Witten transition. 
For instance, one can push a 7-brane upward instead of bringing it downward. This gives a rectangular shape toric-like diagram, which has manifest $S[U(4)\times U(4)]$ symmetry.
See Figure \ref{e7dot2}.
It follow from the figure that the $SU(2)$ theory with $N_f=6$ flavors can be viewed as a a special case of toric diagram for $SU(4)$ gauge theory with $N_f=8$ flavors.

%
% %We note that as it is obtained via Hanany-Witten transition, very the same diagram can also be obtained from the tuned $T_4$ diagram above, by taking upward one of the white dot in the bottom. 
% We can also take upward one of the white dots of this tune $T_4$ diagram. The resultant diagram is given in Figure \ref{dote7su8}. 
% This toric-like diagram for $E_7$ has manifest $S[U(4)\times U(4)]$ symmetry. 

% %%%%%%%%%%%%%%%%%%%%%%%%%%%%%%%%%%%%<<<<<<I am here.
% It is then natural to see $N_f=6$, $E_7$ theory can be viewed as a a special case of toric diagram for $SU(4)$ gauge theory with $N_f=8$ flavors. 

\begin{figure}[t]
\centering
\begin{adjustbox}{width=0.10\textwidth}
\begin{tikzpicture}
  [inner sep=0.5mm,
 dot/.style={fill=black,draw,circle,minimum size=1pt},
 whitedot/.style={fill=white,draw,circle,minimum size=1pt}]
    \draw (0,0) -- (2,0) -- (0,8) -- (0,0);
    \draw (0,1) -- (1,1) -- (1,3);
    \draw (2,0) -- (0,2);     \draw (2,0) -- (1,2);

    \draw (2,0) -- (1,3) -- (0,6);
    \draw (1,3) -- (0,7);     \draw (1,3) -- (0,5);
    \draw (1,2) -- (0,5);     \draw (1,2) -- (0,4);    \draw (1,2) -- (0,3);
    \draw (1,1) -- (0,3);

    \node[dot] at (0,0) {}; 
    \node[whitedot] at (1,0) {}; 
    \node[dot] at (2,0) {};
    \node[dot] at (0,1) {}; 
    \node[dot] at (1,1) {};
    \node[dot] at (0,2) {}; 
    \node[dot] at (1,2) {};
    \node[dot] at (0,3) {}; 
    \node[dot] at (1,3) {};
    \node[dot] at (0,4) {}; 
    \node[whitedot] at (1,4) {};
    \node[dot] at (0,5) {};
    \node[dot] at (0,6) {};
    \node[dot] at (0,7) {};
    \node[dot] at (0,8) {};

%    \draw[->] (-0.3,-0.3) -- (-0.3,0.7); \node[] at (-0.7,0.4) [label=above:$w$] {};
%    \draw[->] (-0.3,-0.3) -- (0.7,-0.3); \node[] at (0.4,-0.9) [label=above:$T$] {};
\end{tikzpicture}
\end{adjustbox}
\caption{A toric-like diagram for the $SU(2)$ theory with $N_f=6$ flavors. It has manifest $SU(8)$ symmetry}\label{dote7su8}
\end{figure}
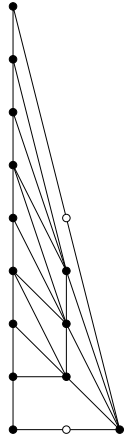

We can further apply the Hanany-Witten to other 7-branes. For instance, let us move all the $[1,0]$ 7-branes on the right hand sides of Figure \ref{e7dot2} to the left.
We then obtain the toric-like diagram given in Figure \ref{dote7su8}.
This toric-like diagram has manifest $SU(8)$ symmetry, which is again a maximal subgroup of $E_7$.

%%%%%%%%%%%%

\subsection{$N_f=6$ Seiberg-Witten curve from M-theory}\label{subsec:E7curve}
We now have at least three toric-like diagrams describing the $SU(2)$ theory with $N_f=6$ flavors. 
Let us call the diagrams as the tuned $T_4$ digram for Figure \ref{webe7}, the rectangular diagram for Figure \ref{e7dot2}, and the long triangle diagram for Figure \ref{dote7su8}.
In this subsection, we compute the SW curve for each diagram following 
%We compute the SW curve by using 
the procedure discussed in section \ref{generalprocedure}. We then explicitly show that 
the curves are related by a proper coordinate transformation confirming that the Hanany-Witten transition is realized as a coordinate transformation in the SW curve.

% The SW curve can be computed from either of the three toric diagrams in
% Figure \ref{webe7}, Figure \ref{e7dot2}, or Figure \ref{dote7su8}.
% Since these diagrams are related by the Hanany-Witten transition,
% the curves computed from the different diagrams should be equivalent.

% In the following, we first compute the SW curve from Figure \ref{webe7}.
% After a coordinate transformation, which we interpret as Hanany-Witten transition,
% we obtain the SW curve which is expected to correspond to Figure \ref{e7dot2}.
% We check that this curve can be also reproduced by computing the SW curve directly from Figure \ref{e7dot2}.
% Analogously, after a further coordinate transformation, we obtain the curve 
% corresponding to Figure \ref{dote7su8}.
% We check that this can be also reproduced directly from Figure \ref{dote7su8}.

\subsubsection{Seiberg-Witten curve from the tuned $T_4$ diagram} %based on Figure \ref{webe7}}
\begin{figure}[H]
\centering
\begin{adjustbox}{width=0.35\textwidth}
\begin{tikzpicture}
  [inner sep=0.5mm,
 dot/.style={fill=black,draw,circle,minimum size=1pt},
 whitedot/.style={fill=white,draw,circle,minimum size=1pt}, 
 mark size=5pt, mark options={draw=blue, fill=white}
 ]
    \draw (0,0) -- (4,0) -- (0,4) -- (0,0);
    \draw (0,1) -- (3,1);
    \draw (0,2) -- (2,2);
    \draw (0,3) -- (1,3);
    \draw (2,0) -- (0,2);
    \draw (2,1) -- (0,3);
    \draw (2,2) -- (0,4);
    \draw (1,1) -- (1,3);
    \draw (2,0) -- (2,2);

    \node[dot] at (4,0) [label=right:$c_{40}$] {};
    \node[whitedot] at (3,0) [label=below:$c_{30}$] {};
    \node[dot] at (3,1) [label=right:$c_{31}$] {};
    \node[dot] at (2,0) [label=below:$c_{20}$] {};
    \node[dot] at (2,1) [label={[label distance=0.03cm]-45:$c_{21}$}] {};
    \node[dot] at (2,2) [label=right:$c_{22}$] {};
    \node[whitedot] at (1,0) [label=below:$c_{10}$] {};
    \node[dot] at (1,1) [label={[label distance=0.03cm]-135:$c_{11}$}] {};
    \node[dot] at (1,2) [label={[label distance=0.03cm]-135:$c_{12}$}]  {};
    \node[dot] at (1,3) [label=right:$c_{13}$] {};
    \node[dot] at (0,0) [label=left:$c_{00}$] {};
    \node[dot] at (0,1) [label=left:$c_{01}$] {};
    \node[dot] at (0,2) [label=left:$c_{02}$] {};
    \node[dot] at (0,3) [label=left:$c_{03}$] {};
    \node[dot] at (0,4) [label=left:$c_{04}$] {};

    \draw [color=blue, densely dashed] (0.9,-0.4) -- (0.9,-1.2);
    \draw [color=blue, densely dashed] (1.1,-0.4) -- (1.1,-1.2);
	\foreach \plm in {otimes*} \draw plot[mark=\plm] coordinates {(1,-1.3)} ;

    \draw [color=blue, densely dashed] (2.9,-0.4) -- (2.9,-1.2);
    \draw [color=blue, densely dashed] (3.1,-0.4) -- (3.1,-1.2);
	\foreach \plm in {otimes*} \draw plot[mark=\plm] coordinates {(3,-1.3)} ;

	\node[] at (1.2,-1.3) [label=right:$N_2$] {};
	\node[] at (3.2,-1.3) [label=right:$N_1$] {};

	\draw [color=blue, densely dashed] (-0.2,0.5) -- (-1,0.5);
	\node[] at (-1,0.5) [label=left:$M_1$] {};

    \draw [color=blue, densely dashed] (-0.2,1.5) -- (-1,1.5);
    \node[] at (-1,1.5) [label=left:$M_2$] {};
    
    \draw [color=blue, densely dashed] (-0.2,2.5) -- (-1,2.5);
    \node[] at (-1,2.5) [label=left:$M_3$] {};

    \draw [color=blue, densely dashed] (-0.2,3.5) -- (-1,3.5);
    \node[] at (-1,3.5) [label=left:$M_4$] {};

	\draw [color=blue, densely dashed] (0.7,3.7) -- (1.5,4.5);
	\node[] at (2.2,4.8) [label=left:$L_1$] {};
	
    \draw [color=blue, densely dashed] (1.7,2.7) -- (2.5,3.5);
    \node[] at (3.2,3.8) [label=left:$L_2$] {};

    \draw [color=blue, densely dashed] (2.7,1.7) -- (3.5,2.5);
    \node[] at (4.2,2.8) [label=left:$L_3$] {};
    
    \draw [color=blue, densely dashed] (3.7,0.7) -- (4.5,1.5);
    \node[] at (5.2,1.8) [label=left:$L_4$] {};

\end{tikzpicture}
\end{adjustbox}
\caption{Coefficients in the tuned $T_4$ diagram}\label{T4e7dots}
\end{figure}
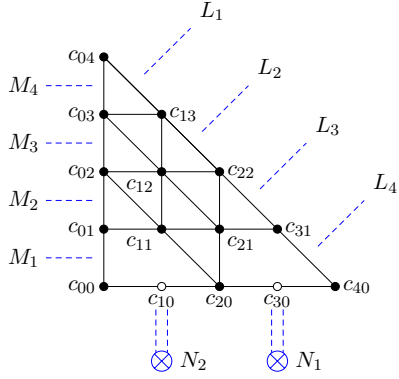

Before computing the SW curve % for $E_7$ theory 
from Figure \ref{webe7}, 
we first briefly mention the curve for $T_4$ theory,
which corresponds to the diagram obtained by replacing all the white dots into black dots in Figure \ref{T4e7dots}.
The SW curve is written in the form.
\begin{eqnarray}\label{su4swm}
\sum_{p \ge 0, q\ge 0 \atop p+q \le 4} c_{pq} t^p w^q = 0.
\label{SW}
\end{eqnarray}
We then impose the following conditions.
The polynomial at asymptotic region behaves
\begin{align}
t \sim w \to \infty \quad: &\quad
\sum_{p=0}^4 c_{p,4-p} t^p w^{4-p} = c_{04} \prod_{i=1}^4 (w+L_i t),
\cr
t \to 0 \quad: &\quad
\sum_{q=0}^4 c_{0q} w^q = c_{04} \prod_{i=1}^4 (w-M_i),
\cr
w \to 0 \quad: &\quad
\sum_{p=0}^4 c_{p0} t^p = c_{40} \prod_{i=1}^4 (t-N_i),
\label{condT4}
\end{align}
from which we can obtain the $T_4$ SW curve.
The coefficients $c_{11}$, $c_{12}$, and $c_{21}$ 
are not determined from the above conditions 
and are treated as the Coulomb moduli parameters.
This $T_4$ theory corresponds to the 5d uplift of 
sphere with three full punctures in 4d setup \cite{Gaiotto:2009we}.

Now, we move on to the SW curve for the $SU(2)$ theory with $N_f=6$ flavors.
In 4d, $E_7$ theory is obtained by replacing one of the full punctures with the degenerate one.
In 5d, it is equivalent to replacing some of the black dots into the white dots in the toric-like diagram as in Figure \ref{T4e7dots}.
Let us apply the generic procedure discussed in section \ref{generalprocedure}.
Structurally, the SW curve is still of the form (\ref{SW}), but
the conditions (\ref{condT4}) are replaced by the following ones: 
\begin{align}
t, w \to \infty \quad: &\quad
\sum_{p=0}^4 c_{p,4-p} t^p w^{4-p} = c_{0,4} \prod_{i=1}^4 (w+L_i t),
\cr
t \to 0 \quad: &\quad
\sum_{q=0}^4 c_{0,q} w^q = c_{0,4} \prod_{i=1}^4 (w-M_i),
\cr
w \to 0 \quad: &\quad
\sum_{p=0}^4 c_{p,0} t^p = c_{4,0} \prod_{i=1}^2 (t-N_i)^2,
\qquad
\sum_{p=0}^4 c_{p,1} t^p \propto \prod_{i=1}^2 (t-N_i).
\label{condE7}
\end{align}
This amounts not only to put $N_3=N_1$, $N_4=N_2$
but also gives rise to to one more condition which is the last condition in (\ref{condE7})
%further condition coming from .
This extra condition enables us to determine $c_{11}$ and $c_{21}$ in terms of physical parameters. Recall that these coefficients are the Coulomb moduli parameters in $T_4$ case, and tuning the $T_4$ keeping only one modulus,  these coefficients are no longer Coulomb moduli parameters.   This is the effect of ``degenerating puncture''. 

The next thing we do is to determine the coefficients $c_{ij}$ from the boundary conditions \eqref{condE7}.
There are 15 dots in the toric-like diagram in Figure \ref{T4e7dots},
which means that we have 15 non-zero coefficients in the SW curve \eqref{su4swm}. 
Naive counting says that the conditions \eqref{condE7} give 14 relations
between the coefficients $c_{ij}$ and the parameters $L_i$, $M_i$, $N_i$.
However, not all the relations are independent.
%available to determine these coefficients in terms of these parameters.
For instance, three relations including $c_{00}, c_{04}, c_{40}$ are given as
\begin{align}
c_{40}  =  c_{04} \prod_i^{4} L_i,
\qquad
c_{00}  =  c_{04} \prod_i^{4} M_i,
\qquad
c_{00} = c_{40} \prod_i^{2} N_i{}^2,
\end{align}
which lead to one compatibility condition on the parameters $L_i$, $M_i$, $N_i$
\begin{align}\label{relationtm}
\prod_{i=1}^4 M_i = \prod_{j=1}^4 L_j \cdot \prod_{k=1}^2 N_k{}^2.
\end{align}
That is, one out of  the 14 relations gives this compatibility condition
rather than determining the coefficients.
It means out of the 15 coefficients, one can determine 13 coefficients.
Two remaining undetermined coefficients can be identified as follows:
one can be identified as an overall constant and the other plays a role
of the Coulomb modulus $U$ of the theory which is the black dot in
the middle of the toric diagram, $c_{11}$.

As shifting along the $t$- and $w$-axes is irrelevant, one can say that there are three rescaling degrees of freedom one can freely choose (overall constant, shifts in $t$ and $w$ coordinates).  For an overall constant, we choose
\begin{align}
c_{04}=1.
\end{align}
With the rescaling of $t$ and $w$, we can choose,
$
N_1 N_2 =1
$
and 
$
\prod_{i=1}^4 M_i = 1,
$
respectively.
Together with the compatibility condition \eqref{relationtm}, we obtain
\begin{eqnarray}
\prod_{i=1}^4 L_i = \prod_{i=1}^4 M_i = \prod_{i=1}^2 N_i = 1.
\end{eqnarray}
The resulting Seiberg-Witten curve is then  written as 
\begin{eqnarray}
&& w^4
+ \left( \chi_{1}(L) t  - \chi_{1} (M) \right) w^3
+ \left( 
\chi_{2} (L) t^2 
+ c_{12} t 
+ \chi_{2} (M) 
\right)
w^2
\nonumber \\
&& \qquad
+ ( t - N_1 ) ( t - N_2 )
\left( \chi_3(L) t - \chi_3(M) \right) w
+ ( t - N_1 )^2 ( t - N_2 )^2
 = 0,
\label{E7SW}
\end{eqnarray}
where, $\chi_i(X)$ is the character of $SU(4)$ defined as 
\begin{eqnarray}
\chi_n (X) = \sum_{1 \le i_1 \le i_2 \le \cdots \le i_n \le 4} X_{i_1}  X_{i_2} \cdots X_{i_n}.
\end{eqnarray}
As the SW curve is obtained from the tuned $T_4$ diagram in Figure \ref{T4e7dots},
%toric-like diagram in Figure \ref{webe7}, 
manifest symmetry is $SU(4)\times SU(4)\times SU(2)$ and the curve is 
the curve is expressed in terms of the characters of $SU(4)\times SU(4)\times SU(2)$. 
%In addition, the curve is supposed to be invariant under the discrete Weyl group of $E_7$.

\subsubsection{Seiberg-Witten curve from the rectangular diagram}%based on Figure \ref{e7dot2}}
We now discuss the SW curve corresponding to the diagram in Figure \ref{e7dot2}.
This rectangular diagram is obtained by moving one of the $[0,1]$ 7-brane in Figure \ref{webe7} to upward by Hanany-Witten transition.
As mentioned in section \ref{generalprocedure}, 
such Hanany-Witten transition can be realized by the coordinate transformation
\begin{eqnarray}
w \to  w (t-N_1),
\label{coord trans}
\end{eqnarray}
which moves $[0,1]$ 7-brane at $t=N_1$ upward.
By substituting (\ref{coord trans}) to (\ref{E7SW}) 
and by multiplying the factor $(t-N_1)^{-2}$,
we obtain
\begin{eqnarray}\label{eq:E7square}
&& ( t - N_1 )^2  w^4
+( t - N_1 )  \left( \chi_{1}(L) t  - \chi_{1} (M) \right) w^3
+ \left( 
\chi_{2} (L) t^2 
+ C_{12} t 
+ \chi_{2} (M) 
\right)
w^2
\nonumber \\
&& \qquad
 +( t - N_2 )
\left( \chi_3(L) t - \chi_3(M) \right) w
+ ( t - N_2 )^2
 = 0.
\end{eqnarray}

\begin{figure}[t]
\centering
\begin{adjustbox}{width=0.30\textwidth}
\begin{tikzpicture}
  [inner sep=0.5mm,
 dot/.style={fill=black,draw,circle,minimum size=1pt},
 whitedot/.style={fill=white,draw,circle,minimum size=1pt}, 
 mark size=5pt, mark options={draw=blue, fill=white}
 ]
    \draw (0,0) -- (2,0)  -- (2,4) -- (0,4) -- (0,0);
    \draw (0,1) -- (2,1);
    \draw (0,2) -- (2,2);
    \draw (0,3) -- (2,3);
    \draw (2,0) -- (0,2);
    \draw (2,1) -- (0,3);
    \draw (2,2) -- (0,4);
    \draw (1,1) -- (1,3);

    \node[dot] at (2,0) [label=right:$c_{20}$] {};
    \node[dot] at (2,1) [label=right:$c_{21}$] {};
    \node[dot] at (2,2) [label=right:$c_{22}$] {};
    \node[dot] at (2,3) [label=right:$c_{23}$] {};
    \node[dot] at (2,4) [label=right:$c_{24}$] {};
    \node[whitedot] at (1,0) [label=below:$c_{10}$] {};
    \node[dot] at (1,1) [label={[label distance=0.03cm]-135:$c_{11}$}] {};
    \node[dot] at (1,2) [label={[label distance=0.03cm]-135:$c_{12}$}]  {};
    \node[dot] at (1,3) [label={[label distance=0.03cm]-135:$c_{13}$}] {};
    \node[whitedot] at (1,4) [label=above:$c_{14}$] {};
    \node[dot] at (0,0) [label=left:$c_{00}$] {};
    \node[dot] at (0,1) [label=left:$c_{01}$] {};
    \node[dot] at (0,2) [label=left:$c_{02}$] {};
    \node[dot] at (0,3) [label=left:$c_{03}$] {};
    \node[dot] at (0,4) [label=left:$c_{04}$] {};

    \draw [color=blue, densely dashed] (0.9,-0.4) -- (0.9,-1.2);
    \draw [color=blue, densely dashed] (1.1,-0.4) -- (1.1,-1.2);
	\foreach \plm in {otimes*} \draw plot[mark=\plm] coordinates {(1,-1.3)} ;

	\draw [color=blue, densely dashed] (0.9,4.4) -- (0.9,5.2);
   	\draw [color=blue, densely dashed] (1.1,4.3) -- (1.1,5.2);
	\foreach \plm in {otimes*} \draw plot[mark=\plm] coordinates {(1,5.3)} ;

	\node[] at (1.2,-1.3) [label=right:$t_2$] {};
	\node[] at (1.2,5.3) [label=right:$t_1$] {};

	\draw [color=blue, densely dashed] (-0.2,0.5) -- (-1,0.5);
	\node[] at (-1,0.5) [label=left:$\widetilde m_5$] {};

    \draw [color=blue, densely dashed] (-0.2,1.5) -- (-1,1.5);
    \node[] at (-1,1.5) [label=left:$\widetilde m_6$] {};

    \draw [color=blue, densely dashed] (-0.2,2.5) -- (-1,2.5);
    \node[] at (-1,2.5) [label=left:$\widetilde m_7$] {};

    \draw [color=blue, densely dashed] (-0.2,3.5) -- (-1,3.5);
    \node[] at (-1,3.5) [label=left:$\widetilde m_8$] {};

	\draw [color=blue, densely dashed] (2.2,0.5) -- (3,0.5);
	\node[] at (3,0.5) [label=right:$\widetilde m_1$] {};

	\draw [color=blue, densely dashed] (2.2,1.5) -- (3,1.5);
	\node[] at (3,1.5) [label=right:$\widetilde m_2$] {};

	\draw [color=blue, densely dashed] (2.2,2.5) -- (3,2.5);
	\node[] at (3,2.5) [label=right:$\widetilde m_3$] {};

	\draw [color=blue, densely dashed] (2.2,3.5) -- (3,3.5);
	\node[] at (3,3.5) [label=right:$\widetilde m_4$] {};

\end{tikzpicture}
\end{adjustbox}
\caption{Coefficients in the rectangular diagram}\label{e7dots}
\end{figure}
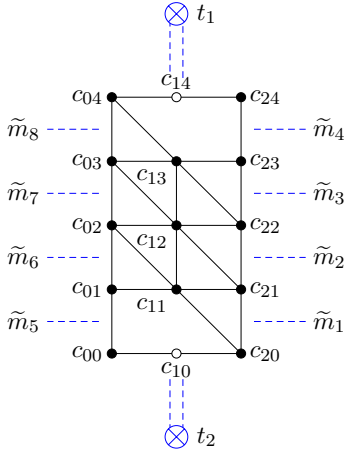

It is also possible to obtain the SW curve from the rectangular diagram in Figure \ref{e7dot2} or 
Figure \ref{e7dots} directly.
The curve should take the form of the polynomial
\begin{align}
\sum^{2}_{i=0}\sum^{4}_{j=0} c_{ij}t^iw^j=0,
\end{align}
with the boundary conditions given as follows
\begin{align}
t\to 0\quad &:\quad  \sum_{j=0}^4 c_{0j}w^j =c_{04}\prod^{8}_{j=5}(w-\widetilde{m}_j),&\cr
t\to \infty\quad &:\quad \sum_{j=0}^4 c_{2j}t^2w^j =c_{24}t^2\prod^{4}_{j=1}(w-\widetilde{m}_j),&
\cr
w\to 0 \quad &:\quad \sum_{i=0}^2c_{i0}t^i=c_{20}(t-t_2)^2;\qquad
\sum_{i=0}^2c_{i1}t^i\propto (t-t_2),
&\cr
w\to \infty\quad &:\quad  \sum_{i=0}^2c_{i4}t^iw^4 =w^4 c_{24}(t-t_1)^2
;\qquad
\sum_{i=0}^2c_{i3}t^i\propto (t-t_1)
.& \label{wbcnf6}
\end{align}
By using the rescaling of $t$, we set $t_1=1$.
With the rescaling of $w$, we set the center of mass position of $m_i$'s to be unity
 \begin{align}\label{E7wres}
 \prod_{i=1}^{8}\widetilde{m}_i=1.
 \end{align}
The compatibility condition corresponding to \eqref{relationtm} is given as
\begin{align}
\frac{t^2_2}{t^2_1}
= \frac{\prod^{8}_{i=5} \widetilde{m}_i}{\prod^{4}_{j=1} \widetilde{m}_j},
\end{align}
which leads to $t_2=\prod^{8}_{i=5}\widetilde{m}_i$.
With a little calculation, one finds that the SW curve is given by
\begin{align}\label{E7curvegeneral}
\prod^{4}_{i=1}(w-\widetilde{m}_i)\, t^2 + k(w)\, t+ \prod^{8}_{i=5}(w-\widetilde{m}_i)=0,
\end{align}
where
\begin{align}
k(w)
&= -2 w^4+\chi^{SU(8)}_{\mu_1}\, w^3 + U \,w^2 +\chi^{SU(8)}_{\mu_7}\,w -2.
\end{align}
Up to the rescaling of the coordinates
\begin{eqnarray}
t \to N_1^{-1} t, \qquad
w \to -N_1{}^{\frac{1}{2}} w,
\end{eqnarray}
we find that this curve is exactly the same as \eqref{eq:E7square}, 
where the parameters are related by
\begin{eqnarray}
\widetilde{m}_{i} = N_2{}^{\frac{1}{2}} L_{i},
 \qquad
\widetilde{m}_{i+4} =  N_1{}^{\frac{1}{2}} M_i, 
\quad (i=1,2,3,4),
\qquad 
U = c_{12} .
\label{E7paramrel}
\end{eqnarray}
Therefore, we find that the SW curve computed from the tuned $T_4$ diagram %Figure \ref{webe7}
and the one computed from the rectangular diagram %Figure \ref{e7dot2} 
is indeed related by a simple coordinate transformation \eqref{coord trans}.

\subsubsection{Seiberg-Witten curve from the long triangle diagram} %based on Figure \ref{dote7su8}}

Finally, we move to the SW curve corresponding to Figure \ref{dote7su8}.
Here we used the choice \eqref{E7wres} enabling us to express in terms of $\chi_{\mu_i}$,  the characters of the fundamental weights $\mu_i$ of $SU(8)$,\footnote{Dynkin diagram for $SU(8)$ is given by
\begin{flalign*}
 \underset{\mathclap{\mu_1}}{\circ} -\!\!\!-
 \underset{\mathclap{\mu_2}}{\circ}-\!\!\!-
  \underset{\mathclap{\mu_3}}{\circ}-\!\!\!-
 \underset{\mathclap{\mu_4}}{\circ}-\!\!\!-
\underset{\mathclap{\mu_5}}{\circ}-\!\!\!-
\underset{\mathclap{\mu_6}}{\circ} -\!\!\!-
\underset{\mathclap{\mu_7}}{\circ}%-\!\!\!-
\end{flalign*}
and $\mu_i$ are the fundamental weight corresponding to the Dynkin label. }
\begin{align}
\chi_{\mu_i}^{SU(8)} \equiv \sum_{k_1<k_2<\cdots<k_i}^{8}\widetilde{m}_{k_1}\widetilde{m}_{k_2}\cdots \widetilde{m}_{k_i}.%\qquad (\chi_{\mu_8}=1).
\end{align}
By performing the coordinate transformation
\begin{align}\label{HWforE7}
t\to\frac{T}{\prod^{4}_{i=1}(w-\widetilde{m}_i)},
\end{align}
we can write the SW curve in an $SU(8)$ manifest way as
\begin{align}
T^2 +k(w)\, T+ \prod^{8}_{i=1}(w-\widetilde{m}_i)=0,
\end{align}
or
\begin{align}\label{su8manicurveTW}
&T^2+(-2w^4+\chi_{\mu_1}\, w^3 +U \,w^2 +\chi_{\mu_7}\,w -2)T\cr
&+w^8 -\chi_{\mu_1}w^7
+\chi_{\mu_2}w^6 -\chi_{\mu_3}w^5+\chi_{\mu_4} w^4 -\chi_{\mu_5}w^3+\chi_{\mu_6} w^2-\chi_{\mu_7} w+ 1=0,
\end{align}
where we have dropped the superscript of the characters, $\chi_{\mu_i}\equiv\chi_{\mu_i}^{SU(8)}$.
The coordinate transformation \eqref{HWforE7} is interpreted as the Hanany-Witten  
to obtain the diagram in Figure \ref{dote7su8} from that in Figure \ref{e7dot2}.

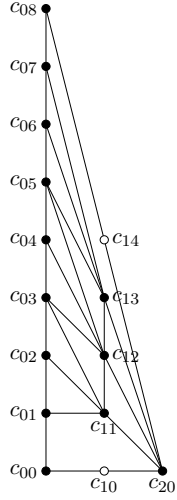
\begin{figure}[t]
\centering
\begin{adjustbox}{width=0.15\textwidth}
\begin{tikzpicture}
  [inner sep=0.5mm,
 dot/.style={fill=black,draw,circle,minimum size=1pt},
 whitedot/.style={fill=white,draw,circle,minimum size=1pt}]
    \draw (0,0) -- (2,0) -- (0,8) -- (0,0);
    \draw (0,1) -- (1,1) -- (1,3);
    \draw (2,0) -- (0,2);     \draw (2,0) -- (1,2);

    \draw (2,0) -- (1,3) -- (0,6);
    \draw (1,3) -- (0,7);     \draw (1,3) -- (0,5);
    \draw (1,2) -- (0,5);     \draw (1,2) -- (0,4);    \draw (1,2) -- (0,3);
    \draw (1,1) -- (0,3);

    \node[dot] at (0,0) [label=left:$c_{00}$]{}; 
    \node[whitedot] at (1,0) [label=below:$c_{10}$]{}; 
    \node[dot] at (2,0)[label=below:$c_{20}$] {};
    \node[dot] at (0,1)[label=left:$c_{01}$] {}; 
    \node[dot] at (1,1)[label=below:$c_{11}$] {};
    \node[dot] at (0,2)[label=left:$c_{02}$] {}; 
    \node[dot] at (1,2) [label=right:$c_{12}$]{};
    \node[dot] at (0,3) [label=left:$c_{03}$]{}; 
    \node[dot] at (1,3) [label=right:$c_{13}$]{};
    \node[dot] at (0,4)[label=left:$c_{04}$] {}; 
    \node[whitedot] at (1,4)[label=right:$c_{14}$] {};
    \node[dot] at (0,5)[label=left:$c_{05}$] {};
    \node[dot] at (0,6) [label=left:$c_{06}$]{};
    \node[dot] at (0,7) [label=left:$c_{07}$]{};
    \node[dot] at (0,8)[label=left:$c_{08}$] {};

%    \draw[->] (-0.3,-0.3) -- (-0.3,0.7); \node[] at (-0.7,0.4) [label=above:$w$] {};
%    \draw[->] (-0.3,-0.3) -- (0.7,-0.3); \node[] at (0.4,-0.9) [label=above:$T$] {};
\end{tikzpicture}
\end{adjustbox}
\caption{Coefficients in the long triangle diagram} \label{dote7su8coef}
\end{figure}
Again, we can obtain \eqref{su8manicurveTW} directly from the diagram in Figure \ref{dote7su8}
or Figure \ref{dote7su8coef}
by considering the polynomial
\begin{align}
\sum_{p \ge 0, q \ge 0 \atop 4p+q \le 8.} c_{pq} T^p w^q=0
\end{align}
with the boundary conditions given as follows
\begin{align}
T \to 0\quad &:
\quad  \sum_{q=0}^8 c_{0q}w^q = c_{08}\prod^{8}_{q=1}(w-\widetilde{m}_q),&\cr
w\to 0 \quad &:
\quad \sum_{p=0}^2c_{p0}t^p=c_{20}(T-1)^2;\qquad
\sum_{p=0}^2c_{p1}t^p \propto (T-1),
&\cr
w^4 \sim T \to \infty\quad &:
\quad  \sum_{p=0}^2c_{p, 8-4p} T^p w^{8-4p} = c_{20}(T-w^4)^2
;\qquad 
&\cr
& \qquad  \sum_{p=0}^1 c_{p, 7-4p} T^p w^{7-4p}  \propto (T-w^4)
.& 
 \label{wbcnf6} 
\end{align}

\subsection{$E_7$ invariance}
In this subsection, we discuss $E_7$ symmetry of the SW curve with six flavors.
The curve \eqref{su8manicurveTW} that we obtained for the $N_f=6$ case is manifestly $SU(8)$ invariant, which is a maximal compact subgroup of $E_7$, and is expected to be $E_7$ invariant. One way to check the $E_7$ invariance is to check whether the curve \eqref{su8manicurveTW} is invariant under the Weyl invariance of $E_7$. This can be done, but in practice it is not so straightforward because it involves mixing of coordinate transformations. 
Another way, which is more direct, is to compare \eqref{su8manicurveTW} to the curve which is written in a $E_7$ manifest way \cite{Minahan:1997ch,Eguchi:2002fc,Eguchi:2002nx} by examining the modular function called $j$-invariant of the elliptic curve.

To this end, we rewrite the $SU(8)$ invariant curve \eqref{su8manicurveTW} in a form that is easier to extract the $j$-invariant
\begin{align} \label{su8manicurve2}
{y}^2&=
\left(-4U+\chi_{\mu_1}^2-4 \chi_{\mu_2}\right)\,w^4
+ (2 U \chi_{\mu_1}+4 \chi_{\mu_3}-4\chi_{\mu_7})\,w^3\\
&
+\left(U^2+2 \chi_{\mu_1}\chi_{\mu_7}-4 \chi_{\mu_4}+8\right)\,w^2
+ (2 U \chi_{\mu_7}-4 \chi_{\mu_1}+4 \chi_{\mu_5})\,w
-4 U-4\chi_{\mu_6}+\chi_{\mu_7}^2\nn
\end{align}
where
\begin{align}
y = T - w^4+\frac12\chi_{\mu_1}\, w^3 +\frac12 U \,w^2 +\frac12\chi_{\mu_7}\,w -1.
\end{align}

The $E_7$ manifest curve \cite{Eguchi:2002fc,Eguchi:2002nx}, that we want to compare, is of the following form \begin{align}\label{E7manicurve}
y^2=&~ 4x^3+ ( -u^2 + 4\chi^{E_7}_{\mu_1}-100)x^2
+\Big((2\chi^{E_7}_{\mu_2}-12\chi^{E_7}_{\mu_7}\big)u +4\chi^{E_7}_{\mu_3}-4\chi^{E_7}_{\mu_6}-64\chi^{E_7}_{\mu_1} +824\Big)x \cr
&+ 4u^4 + 4\chi^{E_7}_{\mu_7}u^3 + (4\chi^{E_7}_{\mu_6}-8\chi^{E_7}_{\mu_1} + 92)u^2
+(4\chi^{E_7}_{\mu_5} -4\chi^{E_7}_{\mu_1}\chi^{E_7}_{\mu_7} -20\chi^{E_7}_{\mu_2} +116\chi^{E_7}_{\mu_7})u \cr
&+4\chi^{E_7}_{\mu_4} -\chi^{E_7}_{\mu_2}\chi^{E_7}_{\mu_2} +4\chi^{E_7}_{\mu_1}\chi^{E_7}_{\mu_1} -40\chi^{E_7}_{\mu_3} +36\chi^{E_7}_{\mu_6} +248\chi^{E_7}_{\mu_1}-2232 ,
\end{align}
where $\chi^{E_7}_{\mu_i}$ is the character of the fundamental weight $\mu_i$ of $E_7$ which is associated to the node of the $E_7$ Dynkin diagram as
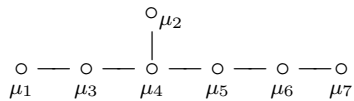
\begin{flalign*}
 \underset{\mathclap{\mu_1}}{\circ} -\!\!\!-
 \underset{\mathclap{\mu_3}}{\circ}-\!\!\!-
\underset{\mathclap{\mu_4}}{\overset{\overset{\textstyle\circ_{\mathrlap{\mu_2}}}{\textstyle\vert}}{\circ}} -\!\!\!-
\underset{\mathclap{\mu_5}}{\circ}-\!\!\!-
\underset{\mathclap{\mu_6}}{\circ} -\!\!\!-
\underset{\mathclap{\mu_7}}{\circ}
\end{flalign*}

Expressed in the standard Weierstrass form, this $E_7$ manifest curve is of degree three polynomial in $x$. On the other hand, the our $SU(8)$ manifest curve is quartic in $w$. In order to compare both, we use the $j$-invariant of elliptic curve 
\begin{align}\label{defjinvariant}
j = \frac{g_2^3}{g_2^3-27g_3^2}.
\end{align}
For generic cubic and quartic polynomials, the forms of $g_2$ and $g_3$ are given in Appendix \ref{app:jinv}

We first compare the massless case by taking all the mass parameters $\widetilde m_i$ to unity ($\widetilde m_i= e^{-\beta m_i}\to 1$). This means that
the character of the fundamental weights becomes the dimension of the corresponding fundamental weights: For $E_7$,
\begin{align}
&\chi^{E_7}_{\mu_1} \to 133, \quad
\chi^{E_7}_{\mu_2} \to 912, \quad
\chi^{E_7}_{\mu_3} \to 8645, \cr
&\chi^{E_7}_{\mu_4} \to 365750,\quad
\chi^{E_7}_{\mu_5} \to 27664,\quad
\chi^{E_7}_{\mu_6} \to 1539,\quad
\chi^{E_7}_{\mu_7} \to 56.
\end{align}
and the curve \eqref{E7manicurve} is written as
\begin{align}
y^2= 4 x^3+\left(432-u^2\right) x^2+(1152 u+20736) x+ 4 u^4+224 u^3+5184 u^2+69120 u+442368.
\end{align}
The corresponding $j$-invariants is given by
\begin{align}
\frac{(u-36 )^3}{1728 (u-52)}.
\end{align}
For $SU(8)$, the characters again become the dimensions of the representations
\begin{align}
&\chi_{\mu_1}=\chi_{\mu_7} \to 8, \quad
\chi_{\mu_2}=\chi_{\mu_6} \to 28, \quad
\chi_{\mu_3}=\chi_{\mu_5} \to 56, \quad
\chi_{\mu_4} \to 70,
\end{align}
and the curve \eqref{su8manicurve2} is written as
\begin{align}
y^2=(U+12)\Big[-4 w^4+16 w^3+(U-12) w^2+16 w-4\Big].
\end{align}
The corresponding $j$-invariants for this is given by
\begin{align}
\frac{(U-36 )^3}{1728 (U-52)},
\end{align}
which coincide with that of $E_7$ manifest curve. From this, we can identify the Coulomb moduli parameter $U$ with $u$ used in \cite{Eguchi:2002nx}. 

With this identification of the Coulomb modulus and agreement of the $j$-invariant for massless case, one can check a generic massive case. Although it is tedious, it is straightforward to see that the $j$-invariant
for the $E_7$ manifest curve \eqref{E7manicurve} exactly coincides with the $j$-invariant for the $SU(8)$ manifest curve \eqref{su8manicurve2} by implementing the decomposition of the $E_7$ fundamental weights into the $SU(8)$ fundamental weights listed in Appendix \ref{che7tosu8}. (We list the form of $g_2$ and $g_3$ for the $SU(8)$ manifest curve \eqref{su8manicurve2} in Appendix \ref{g2gesu8}.)
Therefore, our expression \eqref{su8manicurve2} of the SW curve for $N_f=6$ flavors describes the $E_7$ curve, although it is not manifestly $E_7$ invariant. 

\subsection{4d limit of 5d $E_7$ Seiberg-Witten curve}
From the 5d curve \eqref{E7curvegeneral}
\begin{align}\label{SWE7for4dlimit}
\prod^{4}_{i=1}(w-\widetilde{m}_i)\, t^2 +
\left(-2 w^4+\chi^{SU(8)}_{\mu_1}\, w^3 + U \,w^2 +\chi^{SU(8)}_{\mu_7}\,w -2\right) t+ \prod^{8}_{i=5}(w-\widetilde{m}_i)=0,
\end{align}
whose corresponding toric-like diagram is of rectangular shape given in Figure \ref{e7dot2},
we discuss 4d limit of 5d theory which is to take zero radius limit of the compactified circle. We associate the radius of the circle $\beta$ and 
then the 5d coordinate $w$ and mass parameters $\widetilde m_i$ are related to the 4d coordinate $v$ and masses $m_i$ as
\begin{align}
w =e^{-\beta v}, \quad {\rm and} \quad \widetilde m_i =e^{-\beta m_i}.
\end{align}
To take zero size limit of the radius $\beta$, we expand the Coulomb moduli parameter $U$ in five dimensions as
\begin{align}\label{Uexpand}
U=\sum^{\infty}_{k=0}u_k\beta^k.
\end{align}
Expansion of the curve \eqref{SWE7for4dlimit} leads to consistent conditions determining the expansion coefficient of the 5d Coulomb moduli parameter $U$, and non-trivial relation occurs at order $\beta^4$ which gives rise to 4d SW curve
\begin{align}\label{4dE7curve1}
t^2 \prod^{4}_{i=1}(v- m_i) + t\Big(-2 v^4 - D_2\, v^2 + D_3 \,v + u \Big) + \prod^{8}_{i=5}(v- m_i)=0,
\end{align}
where $D_n\equiv\sum^{8}_{i_1 < \cdots <i_n} m_{i_1} \cdots m_{i_n}$ are the symmetric product and 
the 4d Coulomb moduli parameter $u$ appears at order $\beta^4$ in the expansion of $U$ 
\begin{align}
U= -12 +2D_2 \,\beta^2+ \left( u + \frac{1}{6}(2D_2-D_2) \right) \,\beta^4 +\mathcal{O}(\beta^5).
\end{align}

\begin{figure}[H]
\centering
\begin{adjustbox}{width=0.25\textwidth}
\begin{tikzpicture}
  [inner sep=0.5mm,
 dot/.style={fill=black,draw,circle,minimum size=1pt},
 whitedot/.style={fill=white,draw,circle,minimum size=1pt}]
\draw (0,0.3) circle [radius=3cm];

\draw (-2,-0.5) -- (0,-0.5) -- (0,0) -- (-2, 0) -- (-2,-0.5);
\draw (-1.5,-0.5) -- (-1.5,0);
\draw (-1,-0.5) -- (-1,0);
\draw (-0.5,-0.5) -- (-0.5,0);
\node at (-1,- 1) [below] {$t=0$};

\draw (1,-0.75)--(2,-0.75)--(2,0.25)--(1,0.25)--(1,-0.75);
\draw (1.5,-0.75)--(1.5,0.25);
\draw (1,-0.25) -- (2,-0.25);
\node at (1.5,-1) [below] {$t=1$};

\draw (-1,1.5) -- (1,1.5) -- (1,2) -- (-1,2) -- (-1,1.5);
\draw (-0.5,1.5) -- (-0.5,2);
\draw (0,1.5) -- (0,2);
\draw (0.5,1.5) -- (0.5,2);
\node at (0,1.3) [below] {$t=\infty$};
\end{tikzpicture}
\end{adjustbox}
\caption{Sphere with three punctures which corresponds to $E_7$ CFT.}
\label{surface E7}
\end{figure}
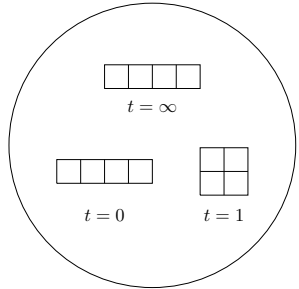

In the following, we check that the 4d SW curve (\ref{4dE7curve1}) is exactly the SW curve for the 4d $E_7$ CFT found in \cite{Gaiotto:2009we, Benini:2009gi}, which is given by the quadruple cover of the sphere with three punctures with specific type. See Figure \ref{surface E7}.
 For later convenience, we reparametrize the mass parameters as
\begin{align}
&m = \frac{1}{4}\sum_{i=1}^4 m_i = -  \frac{1}{4}\sum_{i=5}^8 m_i, \cr
&\hat{m}_i = m_i - m \quad (i=1,2,3,4), \qquad
\hat{m}_i = m_i + m \quad (i=5,6,7,8)
\end{align}
By changing the coordinate as
\begin{align}
v = xt + m \frac{t+1}{t-1},
\end{align}
we can write the curve in the way
\begin{align}
x^4 + \sum_{n=2}^4 \phi_n(t) x^{4-n} = 0,
\end{align}
with SW one-form $\lambda=x dt$.
Here, $\phi_n(t)$ has poles at $t=0,1,\infty$ where the three punctures exist.
The residues at each pole are given by
\begin{align}
& \{ \hat{m}_5,\hat{m}_6,\hat{m}_7,\hat{m}_8 \} \quad \text{at} \quad t=0, \cr
& \{ -2m, -2m , 2m, 2m \} \quad \text{at} \quad t=1, \cr
& \{ \hat{m}_1,\hat{m}_2,\hat{m}_3,\hat{m}_4 \} \quad \text{at} \quad t=\infty,
\end{align}
which we identify as mass parameters.
This is consistent with the type of each puncture.
The type of each puncture can be further checked by looking at the order of the pole
of $\phi_n$ at each puncture when we turn off the mass parameters associated with the corresponding puncture.
The expected order of $\phi_n$ are given by ``$n-$(height of the $n$-th boxes)'',
where we label the boxes in the Young diagram in such a way that the height of the box does not decrease.
See \cite{Gaiotto:2009we} for detail.
Denoting the order of the pole of $\phi_n$ as $p_n$, we can explicitly check
\begin{align}
&(p_2, p_3, p_4 ) = (1,2,3) \quad \text{at} \quad t=0 \quad \text{when} \quad \hat{m}_5 = \hat{m}_6 = \hat{m}_7 = \hat{m}_8 = 0, \cr
&(p_2, p_3, p_4 ) = (1,1,2) \quad \text{at} \quad t=1 \quad \text{when} \quad m = 0, \cr
&(p_2, p_3, p_4 ) = (1,2,3) \quad \text{at} \quad t=\infty \quad \text{when} \quad \hat{m}_1 = \hat{m}_2 = \hat{m}_3 = \hat{m}_4 = 0.
\end{align}
This is again consistent with the type of punctures.
Thus, we have checked that our 4d curve (\ref{4dE7curve1}) agree with
that of the 4d $E_7$ CFT.

%%%%%%%%%%%%%%%%%%%%%%%%%%%%%%%%%%%%%%%%%%%%%%%%%
%%%%%%%%%%%%%%%%%%%%%%%%%%%%%%%%%%%%%%%%%%%%%%%%%
%%%%%%%%%%%%%%%%%%%%%%%%%%%%%%%%%%%%%%%%%%%%%%%%%
%%%%%%%%%%%%%%%%%%%%%%%%%%%%%%%%%%%%%%%%%%%%%%%%%
\section{$E_8$ Seiberg-Witten curve}
In this section, we consider the SW curve for $Sp(1)$ gauge theory with $N_f=7$ flavors.
The construction of the toric-like diagram, the computation of the corresponding SW curve,
 the comparison with the known curve, and the 4d limit can be studied analogously to the previous section.

\subsection{Constructing toric-like diagram}

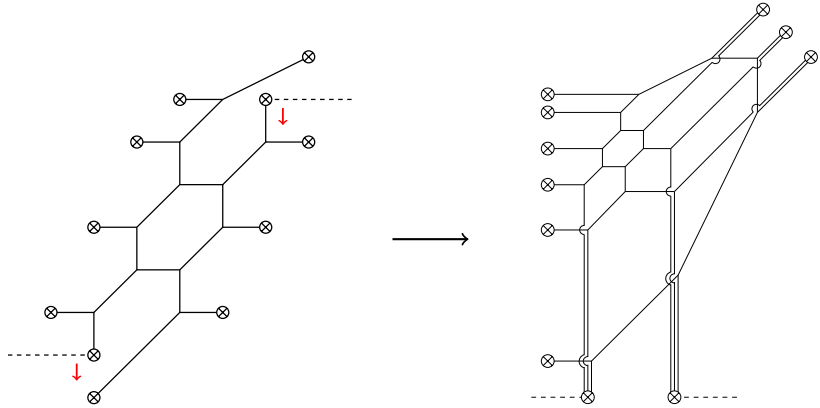
\begin{figure}[t]
\centering
\begin{adjustbox}{width=0.3\textwidth}
\begin{tikzpicture}
  [inner sep=0.5mm, line width=0.3mm,
 dot/.style={fill=black,draw,circle,minimum size=1pt},
 whitedot/.style={fill=white,draw,circle,minimum size=1pt},
 mark size=4pt, mark options={fill=white} ]
    \draw (-1,0) -- (0,0) -- (0,-1);  \draw[dashed] (0,-1) -- (-2,-1);
    \draw (0,0) -- (1,1);
    \draw (1,1) -- (2,1) -- (3, 2) -- (3,3) -- (2,3) -- (1,2) -- (1,1);
    \draw (1,2) -- (0,2); \draw (2,3) -- (2, 4) -- (1,4);
    \draw (2,4) -- (3,5) -- (5,6);\draw (2,5) -- (3,5);
    \draw (3,3) -- (4,4); \draw (4,5) -- (4,4) -- (5,4); \draw[dashed] (4,5) -- (6,5);
    \draw (3, 2) -- (4, 2);
    \draw (2,1) -- (2,0) ;
    \draw (0,-2) -- (2,0) -- (3,0);
    \draw[<-, very thick, color=red] (4.4,4.4) --  (4.4,4.8);
    \draw[->, very thick, color=red] (-0.4,-1.2) --  (-0.4,-1.6);

  \foreach \plm in {otimes*} \draw plot[mark=\plm] coordinates {(0,-2)} ;
  \foreach \plm in {otimes*} \draw plot[mark=\plm] coordinates {(3,0)} ;

  \foreach \plm in {otimes*} \draw plot[mark=\plm] coordinates {(-1,0)} ;
  \foreach \plm in {otimes*} \draw plot[mark=\plm] coordinates {(0,2)} ;
  \foreach \plm in {otimes*} \draw plot[mark=\plm] coordinates {(1,4)} ;

  \foreach \plm in {otimes*} \draw plot[mark=\plm] coordinates {(0,-1)} ;
  %\foreach \plm in {otimes*} \draw plot[mark=\plm] coordinates {(2,0)} ;
  \foreach \plm in {otimes*} \draw plot[mark=\plm] coordinates {(4,2)} ;

  \foreach \plm in {otimes*} \draw plot[mark=\plm] coordinates {(5,6)} ;
  \foreach \plm in {otimes*} \draw plot[mark=\plm] coordinates {(2,5)} ;
  \foreach \plm in {otimes*} \draw plot[mark=\plm] coordinates {(4,5)} ;
  \foreach \plm in {otimes*} \draw plot[mark=\plm] coordinates {(5,4)} ;
\end{tikzpicture}
\end{adjustbox}
\quad
\begin{tikzpicture}
 \draw[->,thick] (-2,1.5) -- (-1, 1.5);
   \node[circle, fill=white] at (-0.5,-0.5) {};
\end{tikzpicture}
\begin{adjustbox}{width=0.25\textwidth}
\begin{tikzpicture}
  [inner sep=0.5mm, line width=0.2mm,
 dot/.style={fill=black,draw,circle,minimum size=1pt},
 whitedot/.style={fill=white,draw,circle,minimum size=1pt},
 mark size=4pt, mark options={fill=white} ]

    \draw (3.5,3.1) -- (2.4,2) -- (0.8,1.2) -- (0.4,0.8) -- (0.4,0.4) -- (0,0)
           -- (0,-0.4) -- (-0.4,-0.8) -- (-0.4,-1.7) arc (90:270:0.1cm);
    \draw (-0.4,-1.9) -- (-0.4, -4.56) arc (105:255:0.15cm);
    \draw (-0.4, -4.83) -- (-0.4, -5.5);
    %-------------
    \draw (3.55,3.05) -- (2.55,2.05) arc (45:-135:0.1cm);
    \draw (2.41,1.91) -- (0.9,0.4) -- (0.9,0) -- (0.5,-0.4)
    -- (0.5,-0.95) -- (-0.32,-1.8) -- (-0.32, -4.6) arc (90:270:0.1cm);
    \draw (-0.32, -4.78)-- (-0.32,-5.5);
    %-----------------
    \draw (4,2.6) -- (3.4,2) -- (3.4,0.8) -- (1.66,-2.8) -- (-0.24,-4.7) -- (-0.24, -5.5);
    %---------------
    \draw (1.5,0) -- (1.5,-0.85) arc (90:270:0.1cm);
    \draw (1.5, -1.05) -- (1.5, -2.73) arc (105:255:0.15cm);
    \draw (1.5, -3) -- (1.5, -5.5);
    %-------
    \draw (1.66,-2.8) -- (1.66,-5.5); %%
    %-------
    \draw (1.58,-0.95) -- (1.58,-2.77) arc (90:270:0.1cm);
    \draw (1.58,-2.95) -- (1.58,-5.5);
    %---------------
    \draw (-1.2, -4.7) -- (-0.24,-4.7);
    \draw (-1.2, -1.8) -- (-0.32, -1.8);
    \draw (-1.2, -0.8) -- (-0.4,-0.8);
    \draw (-1.2, 0) -- (0,0);
    \draw (-1.2, 0.8) -- (0.4,0.8);
    \draw (-1.2, 1.2) -- (0.8,1.2);
    %---------------
    \draw (0.4,0.4) -- (0.9,0.4);
    \draw (0,-0.4) -- (0.5,-0.4);
    %---------------
    \draw (4.05,2.55) -- (3.45,1.95) arc (45:-135:0.1cm);
    \draw (3.31,1.81) -- (1.5,0) ;
     %-----------------
    \draw (4.55, 2.05) -- (3.45, 0.95) arc (45:225:0.1cm);
    \draw (3.31, 0.81) -- (1.58,-0.95);
    %-----------------
    \draw (2.4,2) -- (3.4,2);
    \draw (0.9,0) -- (1.5,0);
    \draw (3.4,0.8) -- (4.6,2);
    %-----------------
    \draw (0.5,-0.95) -- (1.58,-0.95);
    %-----------------
    \draw [dashed] (-0.4, -5.5) -- (-1.58,-5.5);
    \draw [dashed] (1.58,-5.5) -- (3,-5.5);

  \foreach \plm in {otimes*}
  %\draw[ mark size=4pt, mark options={fill=white}]
  \foreach \plm in {otimes*} \draw plot[mark=\plm] coordinates {(-1.2, -4.7)} ;
  \foreach \plm in {otimes*} \draw plot[mark=\plm] coordinates {(-1.2, -1.8)} ;
  \foreach \plm in {otimes*} \draw plot[mark=\plm] coordinates {(-1.2, -0.8)} ;
  \foreach \plm in {otimes*} \draw plot[mark=\plm] coordinates {(-1.2, 0)} ;
  \foreach \plm in {otimes*} \draw plot[mark=\plm] coordinates {(-1.2, 0.8)} ;
  \foreach \plm in {otimes*} \draw plot[mark=\plm] coordinates {(-1.2, 1.2)} ;

  \foreach \plm in {otimes*} \draw plot[mark=\plm] coordinates {(1.58,-5.5)} ;
  \foreach \plm in {otimes*} \draw plot[mark=\plm] coordinates {(-0.32,-5.5)} ;

  \foreach \plm in {otimes*} \draw plot[mark=\plm] coordinates {(3.525,3.075)} ;
  \foreach \plm in {otimes*} \draw plot[mark=\plm] coordinates {(4.025,2.575)} ;
  \foreach \plm in {otimes*} \draw plot[mark=\plm] coordinates {(4.575,2.025)} ;
\end{tikzpicture}
\end{adjustbox}
\caption{$N_f=7$ brane configuration (left) and tuned $T_6$ diagram after Hanany-Witten transitions (right)}\label{webNf7andt6}
\end{figure}

\begin{figure}[H]
\centering
\begin{adjustbox}{width=0.3\textwidth}
\begin{tikzpicture}
  [inner sep=0.5mm,
 dot/.style={fill=black,draw,circle,minimum size=1pt},
 whitedot/.style={fill=white,draw,circle,minimum size=1pt}]

    \draw (0,0) -- (6,0) -- (0,6) -- (0,0);
    \draw (0,6) -- (1,4) ;
    \draw (0,3) -- (3,0);
    \draw (0,4) -- (2,2);
    \draw (0,4) -- (1,4);
    \draw (0,5) -- (3,2);
    \draw (0,3) -- (2,3);
    \draw (0,2) -- (4,2);
    \draw (1,2) -- (1,4);
    \draw (0,1) -- (4,1);
    \draw (2,2) -- (2,4);
    \draw (4,1) -- (3,2);
    \draw (4,1) -- (6,0);
    \draw (2,1) -- (2,2);

    \node[dot] at (0,0) {}; \node[dot] at (0,1) {}; \node[dot] at (0,2) {}; \node[dot] at (0,3) {}; \node[dot] at (0,4) {};\node[dot] at (0,5) {};\node[dot] at (0,6) {};

    \node[whitedot] at (1,0) {}; 
    \node[whitedot] at (1,1) {}; 
    \node[dot] at (1,2) {}; 
    \node[dot] at (1,3) {};
    \node[dot] at (1,4) {};
    \node[whitedot] at (1,5) {};

    \node[whitedot] at (2,0) {}; 
    \node[dot] at (2,1) {}; 
    \node[dot] at (2,2) {}; 
    \node[dot] at (2,3) {}; 
    \node[dot] at (2,4) {};

    \node[dot] at (3,0) {}; \node[whitedot] at (3,1) {}; \node[dot] at (3,2) {}; \node[whitedot] at (3,3) {};

    \node[whitedot] at (4,0) {};\node[dot] at (4,1) {};\node[dot] at (4,2) {};
    \node[whitedot] at (5,0) {};\node[whitedot] at (5,1) {};
    \node[dot] at (6,0) {};

\end{tikzpicture}
\end{adjustbox}
\caption{A toric-like diagram with $E_8$ symmetry. It can be viewed as a tuned $T_6$ diagram.}\label{T6e8toric}
\end{figure}
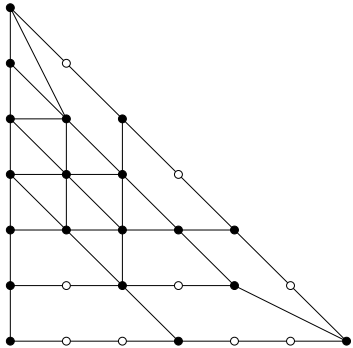

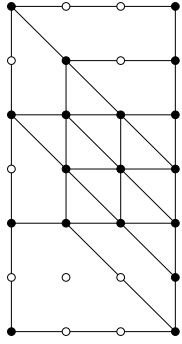
\begin{figure}[H]
\centering
\begin{adjustbox}{width=0.15\textwidth}
\begin{tikzpicture}
  [inner sep=0.5mm,
 dot/.style={fill=black,draw,circle,minimum size=1pt},
 whitedot/.style={fill=white,draw,circle,minimum size=1pt}]

    \draw (0,0) -- (3,0) -- (3,6) -- (0,6) -- (0,0);
    \draw (1,5) -- (3,5); \draw (0,4) -- (3,4); \draw (1,3) -- (3,3); \draw (0,2) -- (3,2);
    \draw (0,6) -- (3,3); \draw (1,4) -- (3,2); \draw (0,4) -- (3,1); \draw (1,2) -- (3,0);
    \draw (1,2) -- (1,5); \draw (2,2) -- (2,4);

    \node[dot] at (0,0){}; 
    \node[whitedot] at (0,1){}; 
    \node[dot] at (0,2){}; 
    \node[whitedot] at (0,3){}; 
    \node[dot] at (0,4){}; 
    \node[whitedot] at (0,5){}; 
    \node[dot] at (0,6){}; 

    \node[whitedot] at (1,0){}; 
    \node[whitedot] at (1,1){}; 
    \node[dot] at (1,2){}; 
    \node[dot] at (1,3){}; 
    \node[dot] at (1,4){}; 
    \node[dot] at (1,5){}; 
    \node[whitedot] at (1,6){}; 

    \node[whitedot] at (2,0){};   
    \node[whitedot] at (2,1){}; 
    \node[dot] at (2,2){};
    \node[dot] at (2,3){};   
    \node[dot] at (2,4){}; 
    \node[whitedot] at (2,5){}; 
    \node[whitedot] at (2,6){}; 

    \node[dot] at (3,0){}; 
    \node[dot] at (3,1){}; 
    \node[dot] at (3,2){}; 
    \node[dot] at (3,3){}; 
    \node[dot] at (3,4){}; 
    \node[dot] at (3,5){}; 
    \node[dot] at (3,6){}; 

\end{tikzpicture}
\end{adjustbox}
\caption{A rectangular toric like diagram for $E_8$}\label{e8dots}
\end{figure}

We begin by adding one more flavor brane to the $N_f=6$ brane configuration. Via successive applications of the Hanany-Witten transition, it is straightforward to see that it leads to a tuned $T_6$ diagram as in Figure \ref{webNf7andt6}.
\noindent The corresponding toric-like diagram is given in Figure \ref{T6e8toric}. 
\noindent 
As this toric-like diagram is a tuned $T_6$ diagram,\footnote{We note that although this toric-like diagram is a tune $T_6$, the number of white dots inside is different from the tuned $T_6$ in \cite{Benini:2009gi}. Depending on how one triangulates while keeping one Coulomb modulus, an interior black dot near the boundary can be turned to a white dot.}
 it has manifest symmetry of $SU(6)\times SU(3)\times SU(2)$ which is a maximal compact subgroup of $E_8$. %If we compute the corresponding SW curve, then the curve will be expressed such that manifest symmetry is $SU(6)\times SU(3)\times SU(2)$. 
 As explained earlier, by performing the Hanany-Witten transition, the manifest symmetry structure is changed to another subgroup of $E_8$. For instance, if we perform the Hanany-Witten transition on one of 7-branes combining three 5-branes on the bottom of the toric-like diagram, Figure \ref{T6e8toric}, we get the corresponding toric-like diagram of a rectangular shape, Figure \ref{e8dots}. 
The diagram shows manifest symmetry of $S[U(6)\times U(3)]$. 

\begin{figure}[H]
\centering
\begin{adjustbox}{width=0.17\textwidth}
\begin{tikzpicture}
  %\draw[help lines] (-2,0) grid (2,4);
 [inner sep=0.5mm,
 dot/.style={fill=black,draw,circle,minimum size=1pt},
 whitedot/.style={fill=white,draw,circle,minimum size=1pt}]
%    \draw[help lines] (0,0) grid (6,3);
    \draw (0,0)  -- (2,0)  -- (2,1) -- (0,7) -- (0,0);
    \draw (0,1)  -- (1,1)  -- (1,3) -- (0,6);
    \draw (2,0)  -- (0,2) ;
    \draw (2,0)  -- (0,4);
    \draw (2,1)  -- (0,3);
    \draw (2,1)  -- (0,5);
    \draw (1,2)  -- (0,5);
    \draw (1,2)  -- (0,2);
    \draw (0,0)  -- (0,-1)  -- (3,-1) -- (2,0) ;
    \draw (2,1)  -- (3,-1)  -- (0,8) -- (0,7) ;

    \node[dot] at (2,0) {};
    \node[dot] at (0,0) {};
    \node[dot] at (0,1) {};
    \node[dot] at (1,1) {};
    \node[dot] at (2,1) {};
    \node[dot] at (0,2) {};
    \node[dot] at (1,2) {};
    \node[dot] at (0,3) {};
    \node[dot] at (1,3) {};
    \node[dot] at (0,4) {};
    \node[dot] at (0,5) {};
    \node[dot] at (0,6) {};
    \node[dot] at (0,7) {};
    \node[dot] at (0,-1) {};
    \node[dot] at (3,-1) {};
    \node[dot] at (0,8) {};

    \node[whitedot] at (1,0) {};
    \node[whitedot] at (1,4) {};
    \node[whitedot] at (1,-1) {};
    \node[whitedot] at (2,-1) {};
    \node[whitedot] at (2,2) {};
    \node[whitedot] at (1,4) {};
    \node[whitedot] at (1,5) {};

    \draw[->] (-0.3,-1.3) -- (0.7,-1.3); \node[] at (0.6, -1.9) [label=above:$T$] {};
    \draw[->] (-0.3,-1.3) -- (-0.3,-0.3); \node[] at (-0.6,-0.5) [label=above:$w$] {};
\end{tikzpicture}
\end{adjustbox}
\caption{A toric-like diagram for $E_8$ with manifest $SU(9)$ symmetry}\label{e8tridots}
\end{figure}
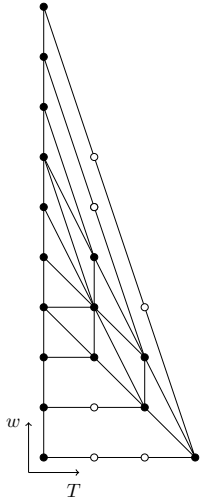

 We can further move all the $[1,0]$ 7-branes on the right hand sides to the left
by Hanany-Witten transition. 
Then, we obtain the toric-like diagram in Figure \ref{e8tridots}.
This toric-like diagram has manifest $SU(9)$ symmetry, which is the maximal subgroup of $E_8$.
Again, the SW curve can be computed from either of these three toric diagrams shown above.

\subsection{$N_f=7$ Seiberg-Witten curve from M-theory}\label{subsec:E8curve}

%We emphasize again the structure of the SW curve with the white dots. The curve does not depend on the detail of triangulation, but it does depend on white dots on the boundary edges.
%The white dots on the boundary yield degenerate polynomials of the curve and more importantly the white dots next to the edge also give rise to the degenerate polynomials of one less degree than those corresponding to white dots on the boundary, as explained in section \ref{generalprocedure}.

In the following, we compute the SW curve based on toric-like diagram of $N_f=7$ .
We start from the diagram in Figure \ref{T6e8toric} and obtain
the SW curve corresponding to the other diagrams by coordinate transformation.
Since the computation and the logic are quite parallel to section \ref{subsec:E7curve},
we summarize the computation briefly.

The SW curve is given by the special case of $T_6$ curve
\begin{align}
\sum_{p \ge 0, q\ge 0 \atop{p+q \le 6} } c_{pq} t^p w^q = 0.
\end{align}
We impose the following conditions
\begin{align}
w, t \to \infty  \quad :\quad &
\sum_{p=0}^6 c_{p,6-p} t^p w^{6-p} = c_{06} \prod_{i=1}^3 (w + L_i t)^2 ,
\cr
& \sum_{p=0}^5 c_{p,5-p} t^p w^{5-p} \propto \prod_{i=1}^3 (w + L_i t),
\\
t \to 0  \quad :\quad &
\sum_{q=0}^6 w^q = c_{06} \prod_{i=1}^6 (w-M_i),
\\
w \to 0  \quad :\quad &
\sum_{p=0}^6 c_{p0} t^p = c_{60} \prod_{i=1}^2 (t-N_i)^3, 
\qquad
\sum_{p=0}^6 c_{p1} t^p \propto \prod_{i=1}^2 (t-N_i)^2,
 \cr
& \sum_{p=0}^6 c_{p2} t^p \propto \prod_{i=1}^2 (t-N_i).
\end{align}
%
%We have 28 dots or 28 coefficients to be determined. Let us count conditions from the boundaries: The boundary condition for $w, t\to \infty$ give 9 conditions \eqref{tinftycone8}; the boundary condition for $t\to0$ yields a degree two and degree one degenerate polynomials \eqref{t0cone8} which gives 9 conditions; the boundary condition for $w\to \infty$ yields a degree three, two, one degenerate polynomials \eqref{winftycone8} which gives 6 conditions; likewise the boundary condition for $w\to0$ gives 6 conditions \eqref{w0cone8}. Taking into account the compatibility condition \eqref{WE-SNE8},
%in total, these give 26 conditions. Two undermined coefficients are an overall constant and one Coulomb modulus $U$, and thus one can completely determine the curve \eqref{e8precurve}.
%
This leads to the SW curve as%We can solve it and obtain
\begin{align}
0 = &\quad w^6  
- (S_1' t -2T_1') w^5 
+ \Big( S_2 t^2 - (S_1' T_1' + S_5' + R_1 T_2') t + (T_1'{}^2+ 2T_2')\Big)w^4
\cr
& - \Big(S_3' t^3 - U' t^2 + (S_1' T_2' + S_5' T_1' + 3R_1 + R_1 T_1' T_2' ) t - (2 T_1'T_2' +2) \Big) w^3
\cr
& + (t^2 - R_1t + 1)\Big(S_4'  t^2- (S_1' +  S_5' T_2' + R_1 T_1')t + (T_2'{}^2+ 2T_1')\Big) w^2 
\cr
& - (t^2 - R_1t + 1)^2(S_5' t- 2T_2') w
 + (t^2 - R_1t + 1)^3,
\end{align}
where we defined the characters of $SU(6)$, $SU(3)$, and $SU(2)$ as
\begin{align}
&S_n' =  \sum_{1 \le i_1 \le i_2 \le \cdots \le i_n} M_{i_1} M_{i_2} \cdots  M_{i_n},
\quad (n=1,\cdots, 5)
\cr
&T_n' =  \sum_{1 \le i_1 \le i_2 \le \cdots \le i_n} L_{i_1} L_{i_2} \cdots  L_{i_n},
\quad (n=1,\cdots, 2)
\cr
& R_1 = N_1 + N_2,
\end{align}
and we also have imposed
\begin{align}
\prod_{i=1}^3 L_i = \prod_{i=1}^6 M_i =  \prod_{i=1}^2 N_i = 1,
\end{align}
by using the rescaling of $t$ and $w$ together with the compatibility condition.

Let us consider the coordinate transformation
\begin{align}
w \to w (t-N_1)
\end{align}
which corresponds to the Hanany-Witten transition enabling us to obtain the toric-like diagram in Figure \ref{e8dots}.
After further scalings\footnote{$t \to N_1{}^{-1} t$ and $w \to N_1{}^{-4/3} w$.} of $w$ and $t$,
and introducing new parameter, we obtain
\begin{align}\label{e8poly1}
&(t-S_6)^3 w^6  -S_1(t-S_6)^2(t-2T_1S_6 S_1^{-1}) w^5 \cr
&+S_2(t-S_6)\Big(t^2 - S_2^{-1}(S_1S_6T_1 + S_5 + S_6T_2 + S_6^2 T_2)t + S_2^{-1}S_6^2(T_1^2+ 2T_2)\Big)w^4\cr
&+\Big(-S_3t^3 + U t^2 -\big[S_1S_6^2T_2+S_5S_6T_1 +3S_6+3 S_6^2 + S_6^2T_1T_2 + S_6^3 T_1T_2 \big]t+2S_6^3 T_1T_2+2S_6^2 \Big) w^3\cr
&+S_4(t-1)\Big(t^2- S_4^{-1}[S_1S_6 +  S_5S_6 T_2 +S_6 T_1+ S_6^2 T_1]t + S_4^{-1}S_6^2(S_6T_2^2+ 2T_1)\Big) w^2 \cr
& -S_5(t-1)^2(t- 2S_5^{-1}S_6^2T_2) w+S_6(t-1)^3=0,
\end{align}
where we introduced 
\begin{align}
&S_n = N_1{}^{\frac{n}{3}} S_n', 
\quad (n=1,\cdots, 5),
\qquad
S_6 = N_1{}^2,
\cr
&T_n = N_1{}^{-\frac{2n}{3}} T_n',
\quad (n=1,2),
\qquad
T_3 = N_1{}^{-2}.
\end{align}
We note that this curve is invariant under $S[U(6)\times U(3)]$.
This curve is again also directly obtained from the diagram in Figure \ref{e8dots}.
%This curve is again directly obtained from the diagram in Figure \ref{e8dots}. %as follows.
%We start from the form 
%\begin{align}\label{e8precurve}
%\sum^{3}_{i=0}\sum^{6}_{j=0} c_{ij}t^iw^j=0,
%\end{align}
%and impose the boundary condition
%\begin{align}
%t\to \infty  :& \quad \sum_{q=0}^6 c_{3q} w^q ~= ~ c_{36}\prod^{6}_{j=1}(w-\widetilde{m}_j),\label{tinftycone8}\\
%t\to 0  :& \quad  \sum_{q=0}^6 c_{0q} w^q   ~= ~ c_{06}\prod^{9}_{i=7}(w-\widetilde{m}_i)^2,
%%
%\qquad \sum_{q=0}^6 c_{1q} w^q ~\propto ~ \prod^{9}_{i=7}(w-\widetilde{m}_i),\label{t0cone8}
%\end{align}
%and
%\begin{align}
%w\to \infty \,\,:&\quad 
%\sum_{p=0}^3 c_{p6} t^p =  c_{36}(t-t_1)^3,
%%
%\quad \sum_{p=0}^3 c_{p5} t^p \propto(t-t_1)^2,
%%
%\quad \sum_{p=0}^3 c_{p4} t^p \propto (t-t_1),\label{winftycone8}\\
%%
%w\to 0 \,\, :&\quad 
%\sum_{p=0}^3 c_{p0} t^p =  c_{30}(t-t_2)^3,
%%
%\quad \sum_{p=0}^3 c_{p1} t^p \propto(t-t_2)^2,
%%
%\quad \sum_{p=0}^3 c_{p2} t^p \propto (t-t_2). \label{winftycone0}
%\end{align}
In terms of the fugacities associated with the mass parameters, $S_n$ and $T_n$ can be written as follows:
%Here, the parameters $\tilde{m}_i$ and $t_1$ are related with $S_n$ and $T_n$ as 
\begin{align*}%\label{STnotation}
&S_1= \sum^{6}_{i=1}\widetilde m_i, \quad S_2=\sum^{6}_{i,j=1,\, i<j}\widetilde m_i \widetilde m_j, \quad S_3=\sum^{6}_{i_1,i_2,i_3} \widetilde m_{i_1}\widetilde m_{i_2}\widetilde m_{i_3},~ \cdots,~ S_6=\prod^{6}_{i=1} \widetilde m_{i},\cr%=t_1
&T_1=\sum^{9}_{i=7}\widetilde m_i,\quad T_2=\sum^{9}_{i,j=7,\,
i<j}\widetilde m_i \widetilde m_j, \quad
T_3= \widetilde m_7\widetilde m_8 \widetilde m_9,\quad S_6T_3=1.
\end{align*}
% We also chose 
% \begin{align}
% t_2=1,\quad \qquad \prod^{9}_{i=1} \widetilde m_i=1,\quad\qquad c_{36}=1,
% \end{align}
% for convenience.
% The relation with the $SU(6) \times SU(3) \times SU(2)$ fugacities are
% \begin{align}
% \tilde{m}_i = N_1{}^{\frac{1}{3}} M_i \quad (i=1,\cdots,6),
% \qquad
% \tilde{m}_j = N_1{}^{-\frac{2}{3}} L_j \quad (j=7,\cdots,9),
% \qquad
% t_1 = N_1.
% \end{align}
% Then, we obtain the same SW curve \eqref{e8poly1}.
% compatibility condition
% \begin{align}\label{WE-SNE8}
% \frac{t_2^3}{t_1^3} = \frac{\prod^{9}_{i=7}\widetilde m_i^2}{\prod^{6}_{j=1}\widetilde m_j}.
% \end{align}
%%%%%%%%%%%%%%%%%%%%%%%%%%%%%%%%%%%%
Introducing the characters of the fundamental weights of $SU(9)$
\footnote{Dynkin diagram for $SU(9)$ is given by
\begin{flalign*}
 \underset{\mathclap{\mu_1}}{\circ} -\!\!\!-
 \underset{\mathclap{\mu_2}}{\circ}-\!\!\!-
  \underset{\mathclap{\mu_3}}{\circ}-\!\!\!-
 \underset{\mathclap{\mu_4}}{\circ}-\!\!\!-
\underset{\mathclap{\mu_5}}{\circ}-\!\!\!-
\underset{\mathclap{\mu_6}}{\circ} -\!\!\!-
\underset{\mathclap{\mu_7}}{\circ}-\!\!\!-
\underset{\mathclap{\mu_8}}{\circ}
\end{flalign*}
and $\mu_i$ are the fundamental weight associated with the Dynkin label.}
\begin{align}\label{su9chiitfsu63}
\chi_n= \sum_{i=0}^{n}S_{n-i}T_{i}\qquad \quad (S_0=1=T_0,\quad S_{n>6} =0 = T_{n>3}, \quad \chi_9=S_6T_3=1),
\end{align}
the curve \eqref{e8poly1} is expressed in a simple form as
\begin{align}
&\Big[\prod^{6}_{i=1}(w-\widetilde m_i)\Big] t^3 -S_6\Big[
3 w^6 -2 \chi_1 w^5 + (\chi_2 + \chi_8)w^4 + U w^3 + (\chi_1+\chi_7)w^2 -2\chi_8 w+3
\Big]t^2 \cr
&
+S_6^2\Big[\big( 3w^3-\chi_1 w^2 + \chi_8 w - 3\big)\prod^{9}_{j=7}(w-\widetilde m_j)\Big] t
-S_6^3\prod^{9}_{i=7}(w-\widetilde m_i)^2  =0. \label{E8Su6u3}
\end{align}
%%%%%%%%%%%%%%%%%%%%%%%%%%%%%%
We now take a further coordinate transformation
\begin{align*}
t\to -S_6\frac{\prod^{9}_{i=7}(w-\widetilde{m}_i)}{T}.
\end{align*}
%where $\widetilde{m}_{i}$ ($i=7,8,9$) are the fugacities of $U(3)$.
This corresponds to the Hanany-Witten transition that enables us 
to obtain the toric-like diagram in Figure \ref{e8tridots}.
% Introducing the characters of the fundamental weights of $SU(9)$
% \footnote{Dynkin diagram for $SU(9)$ is given by
% \begin{flalign*}
%  \underset{\mathclap{\mu_1}}{\circ} -\!\!\!-
%  \underset{\mathclap{\mu_2}}{\circ}-\!\!\!-
%   \underset{\mathclap{\mu_3}}{\circ}-\!\!\!-
%  \underset{\mathclap{\mu_4}}{\circ}-\!\!\!-
% \underset{\mathclap{\mu_5}}{\circ}-\!\!\!-
% \underset{\mathclap{\mu_6}}{\circ} -\!\!\!-
% \underset{\mathclap{\mu_7}}{\circ}-\!\!\!-
% \underset{\mathclap{\mu_8}}{\circ}
% \end{flalign*}
% and $\mu_i$ are the fundamental weight associated with the Dynkin label.}
% \begin{align}\label{su9chiitfsu63}
% \chi_n= \sum_{i=0}^{n}S_{n-i}T_{i}\qquad \quad (S_0=1=T_0,\quad S_{n>6} =0 = T_{n>3}, \quad \chi_9=S_6T_3=1),
% \end{align}
We then obtain an $SU(9)$ manifest curve
\begin{align}\label{su9curve1}
&T^3
+\Big(3w^3-\chi_1 w^2 + \chi_8 w - 3\Big)T^2 \cr
&+\Big(3 w^6 -2 \chi_1 w^5 + (\chi_2 + \chi_8)w^4 + U w^3 + (\chi_1+\chi_7)w^2 -2\chi_8 w+3\Big)
T \cr
&+ w^9 - \chi_1 w^8 + \chi_2 w^7 - \chi_3 w^6 + \chi_4 w^5 -\chi_5 w^4+ \chi_6 w^3 -\chi_7 w^2 +\chi_8 w -1=0.\quad
\end{align}
Again, this diagram can be also obtained directly from Figure \ref{e8tridots}.

\subsection{$E_8$ invariance}\label{sec:E8-inv}
We now compare the $SU(9)$ manifest SW curve for the $N_f=7$ case to the known $E_8$ manifest curve \cite{Eguchi:2002fc,Eguchi:2002nx} to check $E_8$ invariance of the curve \eqref{su9curve1}. To this end,
with 
\begin{align}
\widetilde T  = \frac{1}{w}\Big(T -w^3 +\frac13\chi_1 w^2-\frac13\chi_8w+1\Big),
\end{align}
we rewrite \eqref{su9curve1} as
\begin{align}\label{su9weir}
&{\widetilde T}^3+ \left[\Big(-\frac{\chi _1^2}{3}+\chi _2-\chi _8\Big) w^2+
\Big(U+\frac{2 \chi_1 \chi_8}{3}+6\Big) w
-\frac{\chi_8^2}{3}-\chi_1+\chi_7 \right] {\widetilde T}\cr
&
 +\Big(- U-\frac{2}{27} \chi _1^3+\frac13
   \chi _2 \chi _1-\frac13 \chi _8 \chi _1- \chi _3-3\Big) w^3 \cr
&
+\Big(\frac{U}{3}\chi _1+\frac{2}{9} \chi _8 \chi _1^2+\chi _1+\frac{\chi _8^2}{3}+\chi _4-\chi
   _7-\frac{\chi _2 \chi _8}{3}\Big)w^2 \cr
&   +\Big(-\frac13 U \chi _8-\frac13 \chi _1^2-\frac29 \chi _8^2
   \chi _1+\frac13 \chi _7 \chi _1+ \chi _2- \chi _5- \chi _8\Big)w\cr
&+U+\frac{2 \chi _8^3}{27}+\frac{\chi _1 \chi _8}{3}-\frac{\chi _7 \chi
   _8}{3}+\chi _6+3=0.
\end{align}
Observe that this curve is of mutually degree 3 polynomials in $\widetilde T$ and $w$. One can convert this into the standard Weierstrass form which makes it easier to compare to the known $E_8$ manifest curve (For an explicit coordinate transformation to the Weierstrass form, see, for example, \cite{Huang:2013yta}).

The $E_8$ manifest curve \cite{Eguchi:2002fc,Eguchi:2002nx} is given by
\begin{align}
y^2&= 4x^3+\Big[-u^2 + 4\chi^{E_8}_{\mu_1} - 100\chi^{E_8}_{\mu_8} + 9300\Big]x^2
 + \Big[(2\chi^{E_8}_{\mu_2} - 12\chi^{E_8}_{\mu_7} - 70\chi^{E_8}_{\mu_1} + 1840\chi^{E_8}_{\mu_8} - 115010)u \cr
 &+4\chi^{E_8}_{\mu_3} - 4\chi^{E_8}_{\mu_6} - 64\chi^{E_8}_{\mu_1}\chi^{E_8}_{\mu_8} + 824(\chi^{E_8}_{\mu_8})^2 -112\chi^{E_8}_{\mu_2}+ 680\chi^{E_8}_{\mu_7} + 8024\chi^{E_8}_{\mu_1} - 205744\chi^{E_8}_{\mu_8} + 9606776\Big] x \cr
 &+ 4 u^5 +\big(4\chi^{E_8}_{\mu_8} - 992\big)u^4 +\big(4\chi^{E_8}_{\mu_7} - 12\chi^{E_8}_{\mu_1} - 680\chi^{E_8}_{\mu_8} + 93620\big)u^3\cr
 & + \big( 4\chi^{E_8}_{\mu_6} - 8\chi^{E_8}_{\mu_1} \chi^{E_8}_{\mu_8} + 92 (\chi^{E_8}_{\mu_8})^2  - 28\chi^{E_8}_{\mu_2} - 540\chi^{E_8}_{\mu_7} + 2320\chi^{E_8}_{\mu_1} + 30608\chi^{E_8}_{\mu_8} -3823912 \big)u^2\cr
 &
+ \Big(4\chi^{E_8}_{\mu_5} - 4\chi^{E_8}_{\mu_1}\chi^{E_8}_{\mu_7} - 20\chi^{E_8}_{\mu_2}\chi^{E_8}_{\mu_8} + 116\chi^{E_8}_{\mu_7}\chi^{E_8}_{\mu_8} + 8 (\chi^{E_8}_{\mu_1})^2 -52\chi^{E_8}_{\mu_3} - 416\chi^{E_8}_{\mu_6} + 1436 \chi^{E_8}_{\mu_1}\chi^{E_8}_{\mu_8} \cr
&\quad
- 17776(\chi^{E_8}_{\mu_8})^2 + 4180\chi^{E_8}_{\mu_2} + 16580\chi^{E_8}_{\mu_7} -182832\chi^{E_8}_{\mu_1} + 1103956\chi^{E_8}_{\mu_8} + 18130536\Big)u\cr
&
 +4\chi^{E_8}_{\mu_4} - (\chi^{E_8}_{\mu_2})^2 + 4(\chi^{E_8}_{\mu_1})^2\chi^{E_8}_{\mu_8} - 40\chi^{E_8}_{\mu_3}\chi^{E_8}_{\mu_8} + 36\chi^{E_8}_{\mu_6}\chi^{E_8}_{\mu_8}
 + 248\chi^{E_8}_{\mu_1}(\chi^{E_8}_{\mu_8})^2 -2232(\chi^{E_8}_{\mu_8})^3\cr
 &+ 2\chi^{E_8}_{\mu_1}\chi^{E_8}_{\mu_2} - 232\chi^{E_8}_{\mu_5} + 224\chi^{E_8}_{\mu_1}\chi^{E_8}_{\mu_7} + 1124\chi^{E_8}_{\mu_2}\chi^{E_8}_{\mu_8} -6580\chi^{E_8}_{\mu_7}\chi^{E_8}_{\mu_8} - 457(\chi^{E_8}_{\mu_1})^2 + 4980\chi^{E_8}_{\mu_3} \cr
 &
 + 8708\chi^{E_8}_{\mu_6} - 88136\chi^{E_8}_{\mu_1}\chi^{E_8}_{\mu_8} +1129964(\chi^{E_8}_{\mu_8})^2 - 146282\chi^{E_8}_{\mu_2} + 66612\chi^{E_8}_{\mu_7} + 6123126\chi^{E_8}_{\mu_1}\cr
 &
 -104097420\chi^{E_8}_{\mu_8}+2630318907,
\end{align}
where the characters of $E_8$, $\chi^{E_8}_i$, are associated with the fundamental weights $\mu_i$ assigned to the $E_8$ Dynkin diagram as follows:
\begin{flalign*}
E_8 \qquad
 \underset{\mathclap{\mu_1}}{\circ} -\!\!\!-
 \underset{\mathclap{\mu_3}}{\circ}-\!\!\!-
\underset{\mathclap{\mu_4}}{\overset{\overset{\textstyle\circ_{\mathrlap{\mu_2}}}{\textstyle\vert}}{\circ}} -\!\!\!-
\underset{\mathclap{\mu_5}}{\circ}-\!\!\!-
\underset{\mathclap{\mu_6}}{\circ} -\!\!\!-
\underset{\mathclap{\mu_7}}{\circ}-\!\!\!-
\underset{\mathclap{\mu_8}}{\circ}&
\end{flalign*}

We first check the $j$-invariant for massless cases, where the character becomes the dimension of the representation. For the $E_8$ manifest curve,
it reads
\begin{align}
y^2=&~4 x^3-\left(u^2+14968\right) x^2+8 (41644 u+1022767) x\cr
&+ 4 u^5-76392 u^3+9420308 u^2-777374372 u+12225915472.
\end{align}
The $j$-invariant is then
\begin{align}\label{E8jin4massless}
\frac{ u^2}{1728 (u-432)}.
\end{align}
For the $SU(9)$ manifest curve, the corresponding curve is expressed as
\begin{align}
t^3 + (U+60)w t - (U+60) w^3 + 3(U+60)w^2 - 3(U+60)w + (U+60)=0,
\end{align}
and the corresponding $j$-invariant is
\begin{align}\label{jinv4ours}
\frac{(U+60 )^2}{1728 (U-372)}.
\end{align}
Note that two $j$-invariants, \eqref{E8jin4massless} and \eqref{jinv4ours}, look different, but a shift in the Coulomb moduli parameter can make them coincide with each other. In fact, if one checks the $j$-invariant for the case with generic masses,
in order to compare two curves, one needs to shift the Coulomb moduli parameters by a constant
\footnote{We can also compare \eqref{jinv4ours} to the $j$-invariant of $SO(16)$ manifest SW curve for $E_8$ \cite{Minahan:1997ch}, which reads in the massless case,
\begin{align*}
y^2 =x^3+u_N^2x^2- 2u_N^5 .
\end{align*}
The corresponding $j$-invariant is
\begin{align*}
\frac{u_N^2}{54 u_N - 729}.
\end{align*}
We then find that the two curves are related by a constant shift in the Coulomb modulus
\begin{align*}
U \to 4(8 u_N-15).
\end{align*}
}
\begin{align*}
U \to  u-60 .
\end{align*}

With this shift of the Coulomb moduli parameter, 
one can write \eqref{su9weir} in the standard Weierstrass form
\begin{align}\label{Weiersu9}
y^2= 4x^3- g_2^{SU(9)} x-g_3^{SU(9)},
\end{align}
where
\begin{align}\label{g2g3su9}
g_2^{SU(9)}=&~\frac1{12} u^4 -\frac{2}{3} \big(
2349 + \chi_1 \chi_2 - 27 \chi_3 + \chi_1 \chi_5 - 27 \chi_6 + \chi_2 \chi_7 -
 25 \chi_1 \chi_8 \cr
 &+ \chi_4 \chi_8 + \chi_7 \chi_8
\big)u^2+ \mathcal{O}(u),\cr
g_3^{SU(9)}=&~\frac1{216} u^6 -4 u^5-\frac{1}{18} \big(-15579 + \chi_1 \chi_2 + 45 \chi_3 + \chi_1 \chi_5 + 45 \chi_6 + \chi_2 \chi_7 \cr
&+   47 \chi_1 \chi_8 + \chi_4 \chi_8 + \chi_7 \chi_8\big)u^4+ \mathcal{O}(u^3).
\end{align}
The complete expressions of $g_2$ and $g_3$ are complicated and long, so we list them in Appendix \ref{g2gesu9}, from which one can compute the $j$-invariant. As the global symmetry for $N_f=7$ flavors is expected to be $E_8$, this $j$-invariant should coincide with that from the $E_8$ manifest curve ~\cite{Eguchi:2002nx,Huang:2013yta}
\begin{align}\label{Weiere8}
y^2 ~=& ~4 x^3 -g_2^{E_8}x -g_3^{E_8},
\end{align}
where
\begin{align}\label{g2g3e8}
g_2^{E_8}=&~\frac{1}{12}u^4 - \Big(\frac{2}{3}\chi_1^{E_8}-\frac{50}{3}\chi_8^{E_8}+1550\Big)u^2 +  \mathcal{O}(u),\cr
g_3^{E_8}=
     &~ \frac{1}{216} u^6 - 4u^5 - \Big(\frac{1}{18}\chi_1^{E_8}+\frac{47}{18}\chi_8^{E_8}-\frac{5177}{6} \Big)u^4  + \mathcal{O}(u^3).  
\end{align}
For the explicit form of $g_2^{E_8}$ and $g_3^{E_8}$, see Appendix \ref{App:E8}.
As done in the $N_f=6$ case, one can decompose the fundamental weights of $E_8$ into the fundamental weights of $SU(9)$. Such decomposition 
can be performed with help of computer programs, {\it e.g.} a Mathematica package like LieART \cite{Feger:2012bs} or a computer algebra package LiE \cite{LiE}. We here used a Mathematica package called `Susyno' \cite{Fonseca:2011sy} \footnote{The authors thank Renato Fonseca for email correspondence and also kindly sending us his Mathematica file confirming our result for $\chi_4^{E_8}$ and $\chi_5^{E_8}$, which we did not decompose but obtained indirectly in version 1. 
} to decompose the $E_8$ weights into the $SU(9)$ weights.  
The result of the decomposition is listed in Appendix \ref{che8tosu9}. Under this decomposition, the $g_2$ and $g_3$ for \eqref{g2g3su9} and \eqref{g2g3e8} 
are in agreement with each other, thus confirming that $E_8$ invariance of the SW curve \eqref{su9weir} obtained from the toric-like diagram for $N_f=7$ flavors.

%%%%%%%%%%%%%%%%%%%%%%%%%%%%%%%%%%%%%%%%%%%%%%%
\subsection{4d limit of 5d $E_8$ Seiberg-Witten curve}
We start with our SW curve which is of a manifest $S[U(3)\times U(6)]$ invariant form 
\eqref{E8Su6u3}. %\eqref{e8poly1}.
By redefining $-\frac{t}{S_6}$ by $t$, we get
\begin{align}\label{E8asSu3Su6-1}
&\Big[\prod^{6}_{i=1}(w-\widetilde m_i)\Big] t^3 +\Big[
3 w^6 -2 \chi_1 w^5 + (\chi_2 + \chi_8)w^4 + U w^3 + (\chi_1+\chi_7)w^2 -2\chi_8 w+3
\Big]t^2 \cr
&
+\Big[\big( 3w^3-\chi_1 w^2 + \chi_8 w - 3\big)\prod^{9}_{j=7}(w-\widetilde m_j)\Big] t
+\prod^{9}_{i=7}(w-\widetilde m_i)^2  =0,
\end{align}
where $w =e^{-\beta v}$ and $\widetilde m_i =e^{-\beta m_i}$.

Upon reduction to 4d by expanding $w$ and $\widetilde{m}_i$ in $\beta$ as well as the 5d Coulomb moduli parameter $U$ as a polynomial of $\beta$ as \eqref{Uexpand}, we obtain 4d SW curve at the order of $\beta^6$, expressed in terms of the symmetric product
$
D_n \equiv \sum^{9}_{i_1<\cdots <i_n}m_{i_1}\cdots m_{i_n},
$
as
\begin{align}\label{4dE8curve}
&t^3 \prod^{6}_{i=1}(v-m_i) +
 t^2\Big[
 3 v^6+2D_2 v^4 -2 D_3 v^3 + D_4\, v^2 -D_5 v +u^{4d}\Big]  \cr
&+ t \Big(3v^3 +D_2v -  D_3
\Big)\prod^{9}_{j=7}(v-m_j) + \prod^{9}_{j=7} (v-m_j)^2=0,
\end{align}
where
\begin{align}
U = 12 D_2 \beta^2-D_2^2 \beta^4 +
\left( u^{4d} -\frac{1}{60}\big(24 D_6 -4D_4D_2 +3 D_3^2 -2 D_2^3\big) \right) \beta^6
\end{align}

\begin{figure}[H]
\centering
\begin{adjustbox}{width=0.25\textwidth}
\begin{tikzpicture}
  [inner sep=0.5mm,
 dot/.style={fill=black,draw,circle,minimum size=1pt},
 whitedot/.style={fill=white,draw,circle,minimum size=1pt}]
\draw (0,0.3) circle [radius=3cm];

\draw (-2,-1) -- (-0.5,-1) -- (-0.5,0) -- (-2, 0) -- (-2,-1);
\draw (-1.5,-1) -- (-1.5,0);
\draw (-1,-1) -- (-1,0);
\draw (-2,-0.5) -- (-0.5,-0.5);
\node at (-1.25,- 1.25) [below] {$t=0$};

\draw (1,-1)--(2,-1)--(2,0.5)--(1,0.5)--(1,-1);
\draw (1.5,-1)--(1.5,0.5);
\draw (1,-0.5)--(2,-0.5);
\draw (1,0)--(2,0);
\node at (1.5,-1.2) [below] {$t=-1$};

\draw (-1.5,1.5) -- (1.5,1.5) -- (1.5,2) -- (-1.5,2) -- (-1.5,1.5);
\draw (-1,1.5) -- (-1,2);
\draw (-0.5,1.5) -- (-0.5,2);
\draw (0,1.5) -- (0,2);
\draw (0.5,1.5) -- (0.5,2);
\draw (1,1.5) -- (1,2);
\node at (0,1.3) [below] {$t=\infty$};
\end{tikzpicture}
\end{adjustbox}
\caption{Sphere with three punctures which corresponds to $E_8$ CFT.}
\label{surface E8}

\end{figure}
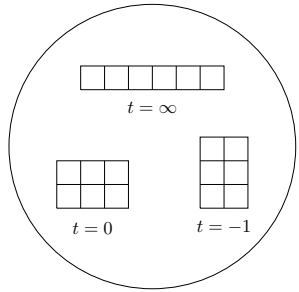

In the following, we check that the 4d SW curve (\ref{4dE8curve}) is exactly the SW curve for the 4d $E_8$ CFT found in \cite{Gaiotto:2009we,Benini:2009gi}, which is given by the sextuple cover of the sphere with three punctures of specific type. See Figure \ref{surface E8}.
 For convenience, we reparametrize the mass parameters as
\begin{align}
&m = \frac{1}{6}\sum_{i=1}^6 m_i = - \frac{1}{6}\sum_{i=5}^8 m_i, \cr
&\hat{m}_i = m_i - m \quad (i=1,2,3,4,5,6), \qquad
\hat{m}_i = m_i + 2 m \quad (i=7,8,9).
\end{align}
By changing the coordinate as
\begin{align}
v = xt - m \frac{t-2}{t+1},
\end{align}
we can write the curve in the way
\begin{align}
x^6 + \sum_{n=2}^6 \phi_n(t) x^{6-n} = 0,
\end{align}
with SW one-form $\lambda=x dt$.
Here, $\phi_n(t)$ has poles at $t=0,-1,\infty$ where the three punctures exist.
The residues at each pole are given by
\begin{align}
& \{ \hat{m}_7,\hat{m}_7,\hat{m}_8,\hat{m}_8,\hat{m}_9,\hat{m}_9 \} \quad \text{at} \quad t=0, \cr
& \{ -3m, -3m , -3m , 3m, 3m, 3m \} \quad \text{at} \quad t=-1, \cr
& \{ \hat{m}_1,\hat{m}_2,\hat{m}_3,\hat{m}_4,\hat{m}_5,\hat{m}_6 \} \quad \text{at} \quad t=\infty,
\end{align}
which we identify as mass parameters.
This is consistent with the type of each puncture.
The type of each puncture can be further checked by looking at the order of the pole
of $\phi_n$ at each puncture when we turn off the mass parameters associated with the corresponding puncture.
Denoting the order of the pole of $\phi_n$ as $p_n$, we can explicitly check
\begin{align}
&(p_2, p_3, p_4, p_5, p_6 ) = (1,2,2,3,4) \quad \text{at} \quad t=0 \quad \text{when} \quad \hat{m}_7 = \hat{m}_8 = \hat{m}_9 = 0, \cr
&(p_2, p_3, p_4, p_5, p_6 ) = (1,1,2,2,3) \quad \text{at} \quad t=-1 \quad \text{when} \quad m = 0, \cr
&(p_2, p_3, p_4, p_5, p_6 ) = (1,2,3,4,5) \quad \text{at} \quad t=\infty \quad \text{when} \quad \hat{m}_1 = \hat{m}_2 = \cdots = \hat{m}_6= 0.
\end{align}
This is again consistent with the type of punctures.
Thus, we have checked that our 4d curve (\ref{4dE8curve}) agree with
that of the 4d $E_8$ CFT.

%%%%%%%%%%%%%%%%%%%%%%%%%%%%%%%%%%%%%%%%%%%%%%%
\section{Mass decoupling limit}
In this section, we discuss ``mass decoupling'' limit of 5d theory with $E_{8}$ ($N_f=7$). Here masses are positions of semi-infinite $D7$ branes in $(p,q)$ web.  As in 4d, one can take large mass limit such that the flavor associated with large mass decouples which yields that rank of global symmetry group is reduced to lower one. As toric-like diagrams of $E_7$ theory can be naturally embedded into toric-like  diagrams of $E_8$ theory, one expects that mass decoupling limit of $E_8$ toric-diagram leads to a $E_7$ toric-like diagram.

%%%%%%%%%%%%%%%%%%%%%%%%%%%%%%%%%%%%
For mass decoupling limit from $E_8$ to $E_7$ SW curve, consider for the $SU(9)$ manifest curve for $N_f=7$ flavors, Figure \ref{e8tridots}.  We take the following scaling limit
\begin{align}
&\widetilde{m}_1 \to L^{-1}\,\widetilde{m}_1,\quad \widetilde{m}_9\to L\,,\quad
\widetilde{m}_i \to \widetilde{m}_i ~ (i = 2,\cdots, 8),
\end{align}
and 
\begin{align}
U\to L\,U,\qquad w\to w, \qquad t\to t.
\end{align}

\noindent This scaling leads that the fundamental weight of $SU(9)$ scales like
\begin{align}
 \chi_{i} ~\sim~L \,\chi^{U(7)}_{i-1} \qquad (i=2,\cdots, 8),
\end{align} 
where $\chi^{U(7)}_0\equiv1$ and $\chi^{U(7)}_{i}$ are the $U(7)$ fundamental characters. For instance, $\chi^{U(7)}_1=\sum^8_{i=2}\widetilde m_i$ and $\chi^{U(7)}_7=\prod^{8}_{i=2}\widetilde m_i$. It follows from the $SU(9)$ traceless condition 
%$\prod^9_{i=1} \widetilde m_i =1$ 
that in this scaling one obtains $SU(8)$ traceless condition $\prod^8_{i=1}\widetilde m_i =1$, and thus $\chi^{U(7)}_7=\widetilde m_1^{-1}$.

By taking large $L$ limit, we find that the $SU(9)$ manifest SW curve \eqref{su9curve1} becomes
\begin{align}
& 
\big(- w + \chi^{U(7)}_7  \big)T^2 +\Big(-2  w^4 + (\chi^{U(7)}_1 + \chi^{U(7)}_7)w^3 + U w^2 + (1+\chi^{U(7)}_6)w -2\chi^{U(7)}_7 \Big) 
T \cr
& -  w^7 + \chi^{U(7)}_1 w^6 - \chi^{U(7)}_2 w^5 + \chi^{U(7)}_3 w^4 -\chi^{U(7)}_4 w^3+ \chi^{U(7)}_5 w^2 -\chi^{U(7)}_6 w +\chi^{U(7)}_7  =0.\quad
\end{align}
This has $S[U(7)\times U(1)]$  which can be also read off from the corresponding toric-like diagram below. 

\begin{figure}[H]
\centering
\begin{tikzpicture}
 [inner sep=0.5mm,
 dot/.style={fill=black,draw,circle,minimum size=1pt},
 whitedot/.style={fill=white,draw,circle,minimum size=1pt}]
    \draw (0,0)  -- (2,0)  -- (2,1) -- (0,7) -- (0,0);
    \draw (0,1)  -- (1,1)  -- (1,3) -- (0,6);
    \draw (2,0)  -- (0,2) ;
    \draw (2,0)  -- (0,4);
    \draw (2,1)  -- (0,3);
    \draw (2,1)  -- (0,5);
    \draw (1,2)  -- (0,5);
    \draw (1,2)  -- (0,2);

    \node[dot] at (2,0) {};
    \node[dot] at (0,0) {};
    \node[dot] at (0,1) {};
    \node[dot] at (1,1) {};
    \node[dot] at (2,1) {};
    \node[dot] at (0,2) {};
    \node[dot] at (1,2) {};
    \node[dot] at (0,3) {};
    \node[dot] at (1,3) {};
    \node[dot] at (0,4) {};
    \node[dot] at (0,5) {};
    \node[dot] at (0,6) {};   
    \node[dot] at (0,7) {};

    \node[whitedot] at (1,0) {};
    \node[whitedot] at (1,4) {};

    \draw[->] (-0.3,-0.3) -- (-0.3,0.7); \node[] at (-0.7,0.4) [label=above:$w$] {};
    \draw[->] (-0.3,-0.3) -- (0.7,-0.3); \node[] at (0.6,-0.9) [label=above:$T$] {};  
\end{tikzpicture}
\caption{$E_7$ toric-like diagram of a manifest $S[U(7)\times U(1)]$ symmetry}\label{SU7U1}
\end{figure}
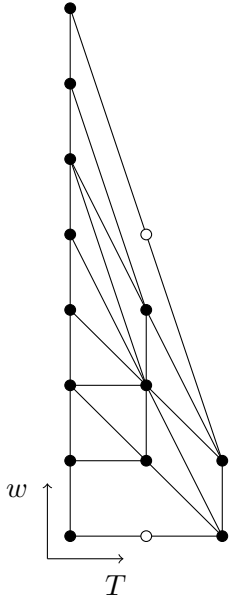
We now take the Hanany-Witten transition to move the 7-brane on the right to the left. It leads to the same $SU(8)$ manifest toric-like diagram as Figure \ref{dote7su8}. As Hanany-Witten effect is realized as a coordinate transformation, we have 
\begin{align}
T\to \big(-w + \chi^{U(7)}_7 \big)^{-1}\,t,
\end{align}
which gives 
\begin{align}
& 
t^2 +\Big(-2 w^4 + (\chi^{U(7)}_1 + \chi^{U(7)}_7)w^3 + U w^2 + (1+\chi^{U(7)}_6)w -2\chi^{U(7)}_7 \Big) 
t \\
& +\big(- w + \chi^{U(7)}_7  \big)\big(- w^7 + \chi^{U(7)}_1 w^6 - \chi^{U(7)}_2 w^5 + \chi^{U(7)}_3 w^4 -\chi^{U(7)}_4 w^3+ \chi^{U(7)}_5 w^2 -\chi^{U(7)}_6 w +\chi^{U(7)}_7\big)  =0.\nn
\end{align}
In order to compare to \eqref{su8manicurveTW}, we redefine the coordinate
\begin{align}
t\to \widetilde{m}_1^{-1}\,T, \qquad \quad
w\to \widetilde{m}_1^{-\frac14}\,w,
\end{align}
which leads to the the $SU(8)$ manifest curve for $N_f=6$ \eqref{su8manicurveTW} with the decomposition between the fundamental characters of $S[U(7)\times U(1)]$ and $SU(8)$ 
\begin{align}
\chi_n^{U(7)}\, \widetilde{m}_1^{\frac{n}{4}}+ \chi_{n-1}^{U(7)}\, \widetilde{m}_1^{\frac{n}{4}}~ =~ \chi_{n}^{SU(8)}\ ,
\end{align}
where $\chi_0^{U(7)}=1$, $\chi_7^{U(7)}=\widetilde m^{-1}$, and $n=1,\ldots,7$.

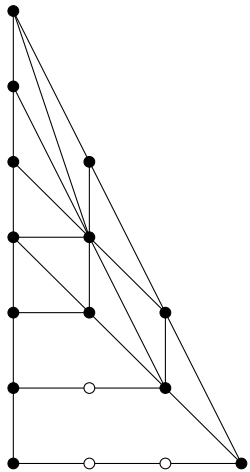
\begin{figure}[H]
\centering
\begin{tikzpicture}
 [inner sep=0.5mm,
 dot/.style={fill=black,draw,circle,minimum size=1pt},
 whitedot/.style={fill=white,draw,circle,minimum size=1pt}]
    \draw (0,0)  -- (3,0)  -- (0,6);
    \draw (0,1)  -- (2,1)  -- (2,2) -- (0,4);
    \draw (0,2)  -- (1,2)  -- (1,4);
    \draw (0,3)  -- (1,3)  -- (0,6);
    \draw (3,0)  -- (0,3);
    \draw (2,1)  -- (0,5);
    \draw (0,0)  -- (0,6);

    \node[whitedot] at (1,0) {};
    \node[whitedot] at (2,0) {};
    \node[whitedot] at (1,1) {};

    \node[dot] at (0,0) {};
    \node[dot] at (3,0) {};
    \node[dot] at (0,1) {};
    \node[dot] at (2,1) {};
    \node[dot] at (0,2) {};
    \node[dot] at (1,2) {};
    \node[dot] at (2,2) {};
    \node[dot] at (0,3) {};
    \node[dot] at (1,3) {};
    \node[dot] at (0,4) {};
    \node[dot] at (1,4) {};
    \node[dot] at (0,5) {};
    \node[dot] at (0,6) {};
\end{tikzpicture}
\caption{Toric-like diagram of $E_7$ theory with a manifest $SU(6)\times SU(3)$ symmetry}\label{SU6SU3}
\end{figure}

We note that there exits another scaling of mass parameters which is equivalent up to $E_8$ Weyl transformation
\begin{align}
\widetilde{m}_i \to  L^{\frac23}\,\widetilde{m}_i~~(i=1,2,3),\qquad
\widetilde{m}_j \to L^{-\frac13}\widetilde{m}_j ~~(j=4,\cdots,9),
\end{align}
as well as 
\begin{align}
U\to L U,\qquad w\to L^{-\frac13} w, \qquad T\to T.
\end{align}
As the shows two distinct scaling behaviors, it leads to the following decomposition of $SU(9)$ characters into $SU(3)\times SU(6)$ characters:
\begin{align}
\chi_n\to L^{\frac{2n}{3}}\chi^{SU(3)}_n~(n=1,2,3),\qquad \chi_n\to L^{\frac{9-n}{3}}\tilde\chi_{9-n}^{SU(6)}~(n=4,\cdots,8),
\end{align}
satisfying $\chi^{SU(3)}_3=1=\tilde\chi_6^{SU(6)}$.
Taking large $L$ limit, the $E_8$ SW curve \eqref{su9curve1} is expressed as
\begin{align}
&T^3
+\Big(-\chi^{SU(3)}_1 w^2 + \tilde\chi_1^{SU(6)} w - 3\Big)T^2 \cr
&+\Big( \chi^{SU(3)}_2 w^4 + U w^3 + (\chi^{SU(3)}_1+\tilde\chi_2^{SU(6)})w^2 -2\tilde\chi_1^{SU(6)} w+3\Big)
T \cr
&- w^6 + \tilde\chi_5^{SU(6)} w^5 -\tilde\chi_4^{SU(6)} w^4+ \tilde\chi_3^{SU(6)} w^3 -\tilde \chi_2^{SU(6)} w^2 +\tilde \chi_1^{SU(6)} w -1=0,\quad
\end{align}
which is of manifest $SU(6)\times SU(3)$ symmetry, a maximal compact subgroup of $E_7$. The corresponding toric-like diagram is given by Figure \ref{SU6SU3}.

\noindent One can show that this diagram is equivalent to toric-like diagram for $E_7$ global symmetry by applying the Hanany-Witten transition several times.

The mass decoupling limit from $E_7$ to $E_6$ and to lower $E_n$ can be done in a similar fashion, as lower $E_n$ toric (or toric-like) diagram is embedded into higher $E_n$ diagram. For example, 
 $E_6$ rectangular shape toric-like diagram in Figure \ref{webNf5-1} is embedded to $E_7$ toric-like diagram, Figure \ref{e7dot2}. 
 For $N_f\le 4$, mass decoupling limit is straightforward and given in Appendix \ref{SW4Nfle4}.
%%%%%%%%%%%%%%%%%%%%%%%%%%%%%%%%%%%%%%
%%%%%%%%%%%%%%%%%%%%%%%%%%%%%%%%%%%%%%
\section{Rank-$N$ $E_n$ curve}\label{sec:rankN}
Toric-like diagrams for higher rank $E_n$ theories are proposed
in \cite{Benini:2009gi} based on symmetry, dimension of the Higgs
branch as well as the Coulomb branch. As in the rank-1 case, they are embedded in $T_N$. More precisely, the rank-$N$ $E_6$, $E_7$, and $E_8$ theories are embedded in $T_{3N}$, $T_{4N}$, and $T_{6N}$, respectively, such that $N$ 5-branes are bound together with the same $7$-branes as in the rank-1 case, on each side of the multi-junction. This means that the number of $7$-branes does not change regardless of rank of the gauge group, $Sp(N)$, and hence global symmetry is still $E_n$.

In this section, we consider the SW curve for the higher rank $E_n$ theories based on toric-like diagram. Computing the curve for the corresponding toric-like diagram is straightforward, following the properties of the white dots in the previous sections, so here we do not give details of computation, rather we sketch how the computation can be done. As explained before, to find the SW curve, it is convenient to implement the Hanany-Witten transition on the (tuned) $T_N$ diagram so that the resultant diagram is of a rectangular shape.

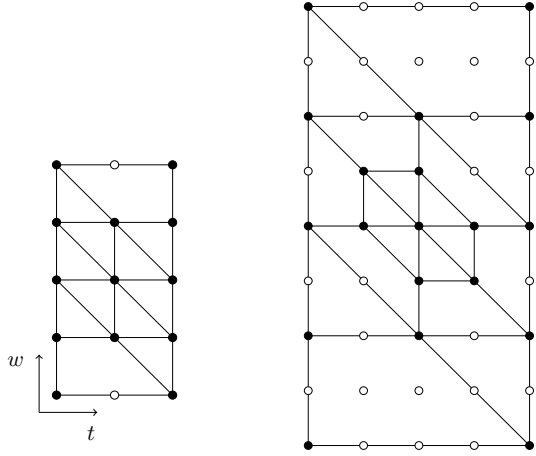
\begin{figure}[H]
\centering
\begin{adjustbox}{width=0.15\textwidth}
\begin{tikzpicture}
  [inner sep=0.5mm,
 dot/.style={fill=black,draw,circle,minimum size=1pt},
 whitedot/.style={fill=white,draw,circle,minimum size=1pt}]
    \draw (0,0) -- (2,0)  -- (2,4) -- (0,4) -- (0,0);
    \draw (0,1) -- (2,1);
    \draw (0,2) -- (2,2);
    \draw (0,3) -- (2,3);
    \draw (2,0) -- (0,2);
    \draw (2,1) -- (0,3);
    \draw (2,2) -- (0,4);
    \draw (1,1) -- (1,3);

    \node[dot] at (2,0) {};
    \node[dot] at (2,1) {};
    \node[dot] at (2,2) {};
    \node[dot] at (2,3) {};
    \node[dot] at (2,4) {};
    \node[whitedot] at (1,0) {};
    \node[dot] at (1,1) {};
    \node[dot] at (1,2) {};
    \node[dot] at (1,3) {};
    \node[whitedot] at (1,4){};
    \node[dot] at (0,0) {};
    \node[dot] at (0,1) {};
    \node[dot] at (0,2) {};
    \node[dot] at (0,3) {};
    \node[dot] at (0,4) {};

    \draw[->] (-0.3,-0.3) -- (-0.3,0.7); \node[] at (-0.7,0.4) [label=above:$w$] {};
    \draw[->] (-0.3,-0.3) -- (0.7,-0.3); \node[] at (0.6,-0.9) [label=above:$t$] {};

\end{tikzpicture}
\end{adjustbox}
\qquad \qquad
\begin{adjustbox}{width=0.2\textwidth}
\begin{tikzpicture}
  [inner sep=0.5mm,
 b/.style={fill=black,draw,circle,minimum size=1pt},
 w/.style={fill=white,draw,circle,minimum size=1pt}]
    \draw (0,0) -- (4,0)  -- (4,8) -- (0,8) -- (0,0);
    \draw (0,2) -- (4,2);
    \draw (0,4) -- (4,4);
    \draw (0,6) -- (4,6);
    \draw (4,0) -- (0,4);
    \draw (4,2) -- (0,6);
    \draw (4,4) -- (0,8);
    \draw (2,2) -- (2,6);
    \draw (2,3) -- (1,4) -- (1,5) -- (2,5) -- (3,4) -- (3,3) -- (2,3);

    \node[b] at (0,0) {};
    \node[w] at (1,0) {};
    \node[w] at (2,0) {};%[label=above:$c_{20}$] {};
    \node[w] at (3,0) {};
    \node[b] at (4,0) {};

    \node[w] at (0,1) {};
    \node[w] at (1,1) {};
    \node[w] at (2,1) {};%[label=above:$c_{20}$] {};
    \node[w] at (3,1) {};
    \node[w] at (4,1) {};

    \node[b] at (0,2) {};%[label=below:$c_{02}$] {};
    \node[w] at (1,2) {};%[label=right:$c_{12}$] {};
    \node[b] at (2,2) {};%[label=above:$c_{22}$] {};
    \node[w] at (3,2) {};
    \node[b] at (4,2) {};

    \node[w] at (0,3) {};
    \node[w] at (1,3) {};
    \node[b] at (2,3) {};
    \node[b] at (3,3) {};
    \node[w] at (4,3) {};

    \node[b] at (0,4) {};
    \node[b] at (1,4) {};
    \node[b] at (2,4) {};
    \node[b] at (3,4) {};
    \node[b] at (4,4) {};

    \node[w] at (0,5) {};
    \node[b] at (1,5) {};
    \node[b] at (2,5) {};
    \node[w] at (3,5) {};
    \node[w] at (4,5) {};

    \node[b] at (0,6) {};
    \node[w] at (1,6) {};
    \node[b] at (2,6) {};
    \node[w] at (3,6) {};
    \node[b] at (4,6) {};

    \node[w] at (0,7) {};
    \node[w] at (1,7) {};
    \node[w] at (2,7) {};
    \node[w] at (3,7) {};
    \node[w] at (4,7) {};

    \node[b] at (0,8) {};
    \node[w] at (1,8) {};
    \node[w] at (2,8) {};
    \node[w] at (3,8) {};
    \node[b] at (4,8) {};

    %\draw[->] (-0.3,-0.3) -- (-0.3,0.7); \node[] at (-0.7,0.4) [label=above:$t$] {};
    %\draw[->] (-0.3,-0.3) -- (0.7,-0.3); \node[] at (0.6,-0.9) [label=above:$w$] {};
\end{tikzpicture}
\end{adjustbox}
\caption{(Left) Rank-1 $E_7$ toric-like diagram, (Right) Rank-2 $E_7$ toric-like diagram}\label{e7rank2dia}
\end{figure}
Consider toric-like diagram for rank-1 and rank-$2$ $E_7$ curves above.
The toric-like diagram for rank-1 (on the left of Figure \ref{e7rank2dia}) is the same one as Figure \ref{e7dot2}. The black dot in the middle of the rank-1 diagram corresponds to the Coulomb modulus. The toric-like diagram for rank-2 (on the right of Figure \ref{e7rank2dia}) is obtained as follows: Given two dots which are next to each other along the outer edges of the rank-1 diagram, one inserts a white dot such that, from the point of view of $(p,q)$-web, the number of semi-infinite 5-branes is doubled while they are combined with the same 7-branes. The dots inside of the rank-2 diagram are introduced such that it does not break the s-rule, and dimension of the Coulomb moduli gets doubled to account for rank-2. This procedure of making multi-junction is also applicable to rank-$N$ diagrams with any $N_f (\le 7)$ flavors.

To compute the SW curve
\begin{align}
\sum^{4}_{i=0}\sum^{8}_{j=0} c_{ij}t^iw^j=0,
\end{align}
we write the boundary conditions
\begin{align}
t\to 0\quad &:\quad  \sum_{j=0}^8 c_{0j}w^j =c_{08}\prod^{4}_{j=1}(w-\widetilde{m}_j)^2;&&
\sum_{j=0}^8 c_{1j}t\,w^j \propto t\prod^{4}_{j=1}(w-\widetilde{m}_j),\label{leftbc4e7rk2}
\\
t\to \infty\quad &:\quad \sum_{j=0}^8 c_{4j}\,t^4w^j =c_{48}\,t^4\prod^{8}_{j=5}(w-\widetilde{m}_j)^2;&&\sum_{j=0}^8 c_{3j}\,t^3w^j \propto t^3\prod^{8}_{j=5}(w-\widetilde{m}_j),\label{rightbc4e7rk2}
\end{align}
and
\begin{align}
w\to 0 \quad &:\quad \sum_{i=0}^4c_{i0}t^i=c_{40}(t-t_2)^4;\quad
			 &&~\quad \sum_{i=0}^4c_{i1}t^iw\propto c_{41}w(t-t_2)^3;&\cr
 			 &~~\quad \sum_{i=0}^4c_{i2}t^iw^2\propto c_{42}w^2(t-t_2)^2;\quad
 			 &&~\quad \sum_{i=0}^4c_{i3}t^iw^3\propto c_{43}w^3(t-t_2),&\label{lowerbc4e7rk2}\\
w\to \infty\quad &:\quad  \sum_{i=0}^4c_{i8}t^iw^8 = c_{48}w^8(t-t_1)^4; \quad
			 &&~\quad \sum_{i=0}^4c_{i7}t^iw^7 \propto w^7 (t-t_1)^3; &\cr
			 	&~~\quad \sum_{i=0}^4c_{i6}t^iw^6 \propto w^6 (t-t_1)^2; \quad
			 	&&~\quad \sum_{i=0}^4c_{i1}t^iw^5 \propto w^5 (t-t_1).& \label{upperbc4e7rk2}
\end{align}
As explained in \eqref{relationtm}, not all parameters are independent
but they are constrained by the same compatibility condition as for the rank-1 case
\begin{align*}%\label{relationtmrank2}
\frac{t^2_2}{t^2_1}=
\frac{\prod^{8}_{i=5} \widetilde{m}_i}{\prod^{4}_{j=1} \widetilde{m}_j}.
\end{align*}
We note that unlike rank-1 case, it turns out that there is another set of coefficients which is related each other again by this compatibility condition.

Let us count the number of the coefficients and the conditions from the boundaries.
There are 45 dots (or non-vanishing coefficients) in the rank-2 $E_7$ toric-like diagram.
For the boundary conditions $t\to 0$ and $t\to \infty$, one finds $(8+4)\times 2=24$
conditions; for the boundary conditions $w\to 0$ and $w\to \infty$,
one finds $(4+3+2+1)\times 2=20$ conditions. Recall that the compatibility
condition indicates that not all boundary conditions are independent. For rank-2, there are two sets of conditions, as mentioned above. The compatibility condition for rank-2 tells us that the number of independent conditions is 42 instead of 44. Hence,
one is left with 3 undetermined coefficients;
One of them is an overall constant (or rescaling)
and the other two are two Coulomb moduli for the rank-2 theory.

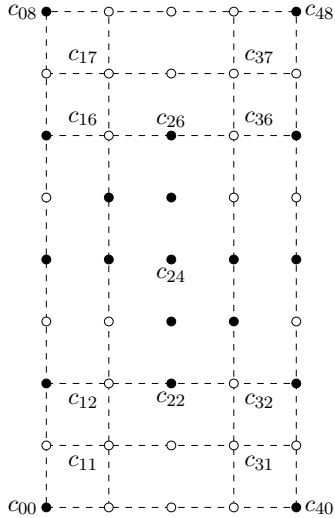
\begin{figure}[H]
\centering
\begin{adjustbox}{width=0.29\textwidth}
\begin{tikzpicture}
  [inner sep=0.5mm,
 b/.style={fill=black,draw,circle,minimum size=1pt},
 w/.style={fill=white,draw,circle,minimum size=1pt},
 g/.style={fill=gray,draw,circle,minimum size=1pt}]
    \draw[dashed] (0,0) -- (4,0)  -- (4,8) -- (0,8) -- (0,0);
    \draw[dashed] (0,1) -- (4,1) ;
    \draw[dashed] (3,0)  -- (3,8);
    \draw[dashed] (1,0)  -- (1,8);
    \draw[dashed] (0,7) -- (4,7) ;
    \draw[dashed] (0,2) -- (4,2) ;
    \draw[dashed] (0,6) -- (4,6) ;

    \node[b] at (0,0) [label=left:$c_{00}$] {};
    \node[w] at (1,0) {};
    \node[w] at (2,0) {};%[label=above:$c_{20}$] {};
    \node[w] at (3,0) {};
    \node[b] at (4,0) [label=right:$c_{40}$] {};

    \node[w] at (0,1) {};
    \node[w] at (1,1) [label={[label distance=0.08cm]225:$c_{11}$}] {};
    \node[w] at (2,1) {};%[label=above:$c_{20}$] {};
    \node[w] at (3,1) [label={[label distance=0.08cm]-45:$c_{31}$}] {};
    \node[w] at (4,1) {};

    \node[b] at (0,2) {};%[label=below:$c_{02}$] {};
    \node[w] at (1,2) [label={[label distance=0.08cm]225:$c_{12}$}] {};
    \node[b] at (2,2) [label=below:$c_{22}$] {};
    \node[w] at (3,2) [label={[label distance=0.08cm]-45:$c_{32}$}] {};
    \node[b] at (4,2) {};

    \node[w] at (0,3) {};
    \node[w] at (1,3) {};
    \node[b] at (2,3) {};
    \node[b] at (3,3) {};
    \node[w] at (4,3) {};

    \node[b] at (0,4) {};
    \node[b] at (1,4) {};
    \node[b] at (2,4) [label=below:$c_{24}$] {};
    \node[b] at (3,4) {};
    \node[b] at (4,4) {};

    \node[w] at (0,5) {};
    \node[b] at (1,5) {};
    \node[b] at (2,5) {};
    \node[w] at (3,5) {};
    \node[w] at (4,5) {};

    \node[b] at (0,6) {};
    \node[w] at (1,6) [label={[label distance=0.08cm]135:$c_{16}$}] {};
    \node[b] at (2,6) [label=above:$c_{26}$]  {};
    \node[w] at (3,6) [label={[label distance=0.08cm]45:$c_{36}$}]  {};
    \node[b] at (4,6) {};

    \node[w] at (0,7) {};
    \node[w] at (1,7) [label={[label distance=0.08cm]135:$c_{17}$}] {};
    \node[w] at (2,7) {};
    \node[w] at (3,7) [label={[label distance=0.08cm]45:$c_{37}$}]{};
    \node[w] at (4,7) {};

    \node[b] at (0,8) [label=left:$c_{08}$] {};
    \node[w] at (1,8) {};
    \node[w] at (2,8) {};
    \node[w] at (3,8) {};
    \node[b] at (4,8) [label=right:$c_{48}$] {};

    % \draw[->] (-0.8,-0.5) -- (-0.8,0.5); \node[] at (-1,0.4) [label=above:$w$] {};
    % \draw[->] (-0.8,-0.5) -- (0.2,-0.5); \node[] at (0.2,-0.9) [label=above:$t$] {};
\end{tikzpicture}
\end{adjustbox}
\caption{Rank-2 $E_7$ toric-like diagram}\label{e7rank2diagrey}
\end{figure}
Let us give a bit more explanation for the compatibility condition. Along the four boundary edges, there are 24 dots, and we have 24 conditions
from the boundary conditions above. These boundary conditions interrelate
the 24 coefficients.
It turns that these 24 conditions do not have solutions unless the compatibility condition is satisfied. Given this compatibility condition, 23 out of 24 conditions are independent.
The undetermined coefficient is associated with choice of overall rescaling.

Now consider the dots $c_{11}, c_{21}, c_{31}$ along the next-to-boundary edge on the bottom.
As there are three dots here and we have 3 conditions from the property of the white dots, the coefficients $c_{11}, c_{21}, c_{31}$ are all determined.
The same logic applied to the dots $c_{17},c_{27},c_{37}$
along the next-to-boundary edge on the top.
We then consider the vertical five dots
$c_{12},\cdots,c_{16}$, we know there are only 4 conditions from the boundary conditions
\eqref{leftbc4e7rk2}.
Similarly, we have the vertical five dots $c_{32},\cdots,c_{36}$ and 4 conditions
\eqref{rightbc4e7rk2}.
We also have three dots $c_{32},c_{22},c_{12}$ and 2 conditions from \eqref{lowerbc4e7rk2}.
Likewise, three dots $c_{36},c_{26},c_{16}$ and 2 conditions from \eqref{upperbc4e7rk2}.
In total 12 dots with 12 conditions. However,
here the coefficients along this rectangular
are in interrelated, just like how the compatibility condition \eqref{relationtm}
was obtained. It thus gives another undetermined coefficient,
which is related to the Coulomb moduli for the rank-2 case.
Recall that the dot in the middle of the diagram is not determined,
and this dot is another undetermined coefficient $c_{24}$.
Together, these two unknown coefficients account for two Coulomb moduli of
the rank-2 theory. With suitable identification of these Coulomb moduli parameters, for instance, with $U_1+U_2$ for the first undermined coefficient and $U_1U_2$ for $c_{24}$, one finds that the rank-2 SW curve is factorized as the product of the rank-1 SW curves of different Coulomb moduli parameters
\begin{align}\label{SWcurvrk2}
&\bigg(\prod^{4}_{i=1}(w-\widetilde{m}_i)\, t^2 +
\left(-2 w^4+\chi^{SU(8)}_{\mu_1}\, w^3 + U_1 \,w^2 +\chi^{SU(8)}_{\mu_7}\,w -2\right) t+ \prod^{8}_{i=5}(w-\widetilde{m}_i)\bigg)\\
&\times
\bigg(\prod^{4}_{i=1}(w-\widetilde{m}_i)\, t^2 +
\left(-2 w^4+\chi^{SU(8)}_{\mu_1}\, w^3 + U_2 \,w^2 +\chi^{SU(8)}_{\mu_7}\,w -2\right) t+ \prod^{8}_{i=5}(w-\widetilde{m}_i)\bigg)
=0.\nn
\end{align}

We note that even though rank-2 in this case means $Sp(2)$ and the $Sp(2)$ gauge theories contain an additional antisymmetric hypermultiplet compare with $Sp(1)$ theory, the rank-2 SW curve \eqref{SWcurvrk2} does not describe a generic $Sp(2)$ SW curve with antisymmetric hypermultiplet, rather it describes the $Sp(2)$ SW curve with the vanishing mass of antisymmetric hypermultiplet.

Generalization to rank-$N$ is straightforward. Toric-like diagram for rank-$N$ is a generalization of the rank-2 diagram. The compatibility condition is the exactly the same as that for rank-1. As in rank-2 case, $N$ boundary conditions are redundant for rank-$N$ case.

For instance, if the number of dots of the corresponding toric-like diagram for rank-$N$ is $n$, then there are $n-1$ conditions from the boundary conditions but taking into account the compatibility conditions, the number of independent conditions is $n-1-N$. Hence, among $n$ coefficients of the SW curve, $n-1-N$ coefficients are fully specified by the parameters of the theory, and the undermined $N+1$ coefficients are one overall rescaling and $N$ Coulomb moduli parameters.

The product of SW curve for rank-$1$ satisfies the boundary conditions for the toric-like diagram of rank-$N$ theory proposed by \cite{Benini:2009gi}. We thus claim that
the SW curve for rank-$N$ theory is also factorized as the product form of the rank-1 SW curves
\begin{align}
SW_N(U_1, U_2,\cdots,U_N) = \prod_{i=1}^{N}SW_i(U_i),
\end{align}
where we denoted $SW_n$ as the SW curve for rank-$n$ with $E_{N_f+1}$ symmetry, and $U_i$ as the corresponding Coulomb moduli parameters.
This describes the $Sp(N)$ SW curve of $E_{N_f+1}$ symmetry with the vanishing masses of antisymmetric hypermultiplet.

%%%%%%%%%%%%%%%%%%%%%%%%%%%%%%%%%%%%%%%%%%%%%%%%
%%%%%%%%%%%%%%%%%%%%%%%%%%%%%%%%%%%%%%%%%%%%%%%
\section{Summary and discussion}

In this paper, we have proposed a systematic procedure for computing the Seiberg-Witten (SW) curve from generic toric-like diagram base on $(p,q)$ 5-brane web diagram.
Using this method, we computed the SW curve for five-dimensional $\cN=1$ $Sp(1)$ gauge theories with $N_f=6,7$ flavors, which are expected to have the UV fixed point with $E_7, E_8$ enhanced global symmetry, respectively. Enhancement of the global symmetry has been seen from various ways, for example, superconformal index \cite{Kim:2012gu,Hwang:2014uwa}, topological vertex amplitudes \cite{Iqbal:2012xm,Bao:2013pwa,Hayashi:2013qwa}, as well as fiber-base duality invariant Nekrasov partition functions \cite{Mitev:2014jza}. In this paper, the enhancement also appears in the SW curve. At first sight, $E_n$ global symmetry does not look manifest in our expression, rather only subgroup of $E_n$ can be seen in the SW curve, and the curves with different subgroups are related by simple coordinate transformations which correspond to the Hanany-Witten transition.
Our SW curves are computed using a totally different way from the method that led to the previously
known $E_n$ manifest results \cite{Eguchi:2002fc,Eguchi:2002nx}, which are computed by using E-string effective action.
By comparing the both $j$-invariants, or
by performing coordinate transformation to express as the  Weierstrass standard form,
we find that our result agrees with the known results \cite{Minahan:1997ch,Eguchi:2002fc,Eguchi:2002nx}. 
%%%%%%%%%%%%%%%%%%%%%%%%
Mass decoupling limits of the curve are discussed to connect the SW curve for the theory of less flavor, and the 4d limit reproduces 4d SW curves \cite{Minahan:1996fg,Minahan:1996cj}, as expected. 

We have also computed the SW curve for five-dimensional $Sp(2)$ gauge theory with six fundamental flavors and one massless
antisymmetric tensor, which is identified as rank-2 theory of $E_7$ symmetry, and
have shown that it reduces to just the two copies of SW curve for $Sp(1)$ gauge theory with six fundamental flavors. 
Our result strongly implies that the SW curve for the five-dimensional $Sp(N)$ gauge theory with $N_f$ fundamental flavors
and one massless antisymmetric tensor, which is rank-$N$ $E_{N_f+1}$ CFT,
is also factorized into $N$ copies of the SW curve of the rank-1 $E_{N_f+1}$ CFT.

As our method of obtaining the SW curve is applicable to any toric-like diagram,
there will be more applications to various theories.
One of interesting directions is to consider the five-dimensional uplift of the class S theory.
The toric diagram for the five-dimensional uplift of the $T_N$ theory was given in \cite{Benini:2009gi}, and the corresponding SW curve was studied in \cite{Benini:2009gi, Bao:2013pwa}.
By replacing some of the full punctures of $T_4$ and $T_6$
with certain type of degenerate punctures, the 5d theories of $E_7$ and $E_8$ global symmetries are obtained, respectively.  Degenerate punctures are nothing but white dots in toric-like diagram. %For instance, see Appendix \ref{App:relation} for $E_7$ case.
 It is straightforward to write down toric-like diagram corresponding to
a sphere with three punctures of arbitrary type,
which is the counterpart of the ``pants'' in the context of pants decomposition,
 and thus obtaining the corresponding SW curve is also straightforward. 
%We have already shown in this paper that the SW curve for some of such example can be factorized. 
It would be interesting to consider classification of the 5d uplifts of the pants.
Moreover, it would be natural to expect that 5d uplift of the class S theories
can be obtained by gluing such pants as in the 4d class S theories. 
%analogously to the pants decomposition in the 4d class S theories.
The corresponding SW curve for 5d uplifts of class S theories can also be computed based on the method developed in this paper.
%%%%%%%%%%%

The 5d uplift of the degenerate punctures are studied in various recent papers 
in the context of, for example, the topological string partition function and the superconformal index.
Especially, the way of realizing the degenerate puncture by tuning the K\"{a}hler moduli parameters 
of the Calabi-Yau geometry in the refined topological string is first proposed for the $E_7$ case in \cite{Hayashi:2013qwa}
and then generalized and/or applied to various cases in \cite{Hayashi:2014wfa,Bergman:2014kza,Hayashi:2014hfa,Mitev:2014jza,Isachenkov:2014eya}.
When we consider the limit where the $\Omega$-deformation parameters vanish,
their method is roughly translated into tuning the Coulomb moduli parameters and some of the mass parameters to be same.
%as well as putting some of the mass parameters to be the same $m_I=m_J$ in the gauge theory language. %The latter tuning ($m_I=m_J$) is an obvious tuning. 
This tuning then reduces the genus of the SW curve as in this paper.\footnote{
The tuning of mass parameters $m_I=m_J$ is equivalent to a constraint coming from the white dot, for example, the first equation in the first line of \eqref{bcE6}, which is straightforward to understand. 
The tuning of the Coulomb moduli parameters $a_i=m_J$ corresponds to another constraint coming from the white dot in the SW curve, for instance, the second equation in the first line of \eqref{bcE6}, which may not be obvious, % because their Coulomb moduli parameters $a_i$ are obtained by performing $A$-cycle integral of the corresponding SW curve.
because the $a_i$ are obtained by performing $A$-cycle integral of the corresponding SW curve.
However, by taking into account that the mass parameter is obtained by picking up the residue of the singularity, we can see that
the integral over the sum of the cycle giving $a_i$ and the path giving $-m_I$ will vanish.
This implies that a certain non-trivial cycle vanishes and thus the genus of the corresponding SW curve reduces.}
%We also see in this paper that the genus of the SW curve is reduced compared to the one before tuning.
Thus, we claim that the way of tuning in this paper is consistent with the way developed in the other papers.
%%%%%%%%%%%%%

When we consider the 5d uplift of class S theory,
an important issue is how to uplift the ``$\cN=2$ dualities'' \cite{Gaiotto:2009we},
which is a generalization of the electromagnetic duality of the $SU(2)$ SW theory
or Argyres-Seiberg duality for $SU(3)$ theory with six flavors.
Related issue has been addressed in various papers including \cite{Bao:2013pwa, Bergman:2014kza, Hayashi:2014hfa}.
Especially, it was pointed out in \cite{Bao:2013pwa},
that the SW curve of $E_6$ CFT is obtained by tuning some of the parameters of
the SW curve for $SU(3)$ gauge theory with six flavors with proper identification of the mass parameters, which we expect to be related to Argyres-Seiberg duality \cite{Argyres:2007cn}.
Using the result of this paper, we can find several more examples analogous to this.

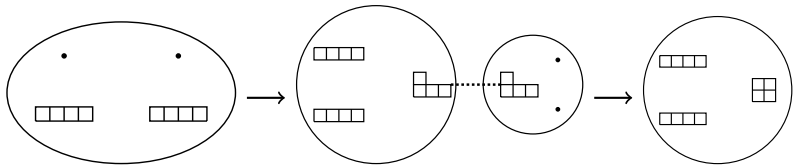
\begin{figure}[H]
\centering
\begin{adjustbox}{width=0.2\textwidth}
\begin{tikzpicture}
  [inner sep=0.5mm, line width=0.5mm,
 dot/.style={fill=black,draw,circle,minimum size=1pt},
 whitedot/.style={fill=white,draw,circle,minimum size=1pt}]
\draw (0,0) circle [x radius = 4cm, y radius=2.5cm];

\draw (-3,-1) -- (-1,-1) -- (-1,-0.5) -- (-3, -0.5) -- (-3,-1);
\draw (-2.5,-1) -- (-2.5,-0.5);
\draw (-2,-1) -- (-2,-0.5);
\draw (-1.5,-1) -- (-1.5,-0.5);

\draw (3,-1) -- (1,-1) -- (1,-0.5) -- (3, -0.5) -- (3,-1);
\draw (2.5,-1) -- (2.5,-0.5);
\draw (2,-1) -- (2,-0.5);
\draw (1.5,-1) -- (1.5,-0.5);

\node[dot] at (2,1.3) {};
\node[dot] at (-2,1.3) {};

\end{tikzpicture}
\end{adjustbox}
\begin{tikzpicture}
 \draw[->,thick] (-1.5,0) -- (-1, 0);
 \node[circle, fill=white] at (-1.2,-0.7) {};
%   \node at (0,0) [above] {S duality};
\end{tikzpicture}
\begin{adjustbox}{width=0.25\textwidth}
\begin{tikzpicture}
  [inner sep=0.5mm, line width=0.5mm,
 dot/.style={fill=black,draw,circle,minimum size=1pt},
 whitedot/.style={fill=white,draw,circle,minimum size=1pt}]
\draw (0,0) circle [radius = 3.2cm];

\draw (-2.5,-1.5) -- (-0.5,-1.5) -- (-0.5,-1) -- (-2.5, -1) -- (-2.5,-1.5);
\draw (-2,-1.5) -- (-2,-1);
\draw (-1.5,-1.5) -- (-1.5,-1);
\draw (-1,-1.5) -- (-1,-1);

\draw (-2.5,1.5) -- (-0.5,1.5) -- (-0.5,1) -- (-2.5, 1) -- (-2.5,1.5);
\draw (-2,1.5) -- (-2,1);
\draw (-1.5,1.5) -- (-1.5,1);
\draw (-1,1.5) -- (-1,1);

\draw (1.5, -0.5) -- (3,-0.5) -- (3,0) -- (1.5,0) -- (1.5, -0.5);
\draw (1.5, 0) -- (1.5,0.5) -- (2,0.5) -- (2,-0.5) ;
\draw (2.5,0) -- (2.5,-0.5);
%----------------------------------
\draw [dashed, line width=1mm] (3,0) -- (5,0);
%----------------------------------
\draw (6.3,0) circle [radius = 2cm];

\draw (5, -0.5) -- (6.5,-0.5) -- (6.5,0) -- (5,0) -- (5, -0.5);
\draw (5, 0) -- (5,0.5) -- (5.5,0.5) -- (5.5,-0.5) ;
\draw (6,0) -- (6,-0.5);
\node[dot] at (7.3,1) {};
\node[dot] at (7.3,-1) {};

\end{tikzpicture}
\end{adjustbox}
\begin{tikzpicture}
 \draw[->,thick] (-1.5,0) -- (-1, 0);
 \node[circle, fill=white] at (-1.2,-0.7) {};
\end{tikzpicture}
\begin{adjustbox}{width=0.13\textwidth}
\begin{tikzpicture}
  [inner sep=0.5mm, line width=0.5mm,
 dot/.style={fill=black,draw,circle,minimum size=1pt},
 whitedot/.style={fill=white,draw,circle,minimum size=1pt}]
\draw (0,0) circle [radius=3.2cm];

\draw (-2.5,-1.5) -- (-0.5,-1.5) -- (-0.5,-1) -- (-2.5, -1) -- (-2.5,-1.5);
\draw (-2,-1.5) -- (-2,-1);
\draw (-1.5,-1.5) -- (-1.5,-1);
\draw (-1,-1.5) -- (-1,-1);

\draw (-2.5,1.5) -- (-0.5,1.5) -- (-0.5,1) -- (-2.5, 1) -- (-2.5,1.5);
\draw (-2,1.5) -- (-2,1);
\draw (-1.5,1.5) -- (-1.5,1);
\draw (-1,1.5) -- (-1,1);

\draw (1.5, -0.5) -- (2.5,-0.5) -- (2.5,0.5) -- (1.5,0.5) -- (1.5, -0.5);
\draw (1.5,0) -- (2.5,0);
\draw (2,-0.5) -- (2,0.5);

\end{tikzpicture}
\end{adjustbox}
\caption{Construction of $E_7$ CFT. Start from $SU(4)$ with eight flavors, take the strong coupling limit, and Higgs one of the punctures.}
\label{construct E7}
\end{figure}

As an example, 4d $E_7$ CFT is constructed from $SU(4)$ gauge theory with eight flavors 
with $\cN=2$ dualities and Higgsing procedure to reduce the order of a puncture. See Figure \ref{construct E7}.
Especially, it means that the SW curve for 4d $E_7$ CFT is obtained from the SW curve for 4d $SU(4)$ gauge theory with eight flavors. This analogue can be seen in five dimensions.
From the toric-like diagram for $Sp(1)$ theory of $E_7$ symmetry in Figure \ref{e7dots}, 
one would find that it corresponds to the toric diagram for $SU(4)$ gauge theory with eight flavors
if all the dots were black.
Changing black dots to white dots corresponds to tuning the parameters
in a specific way that is discussed in this paper.
Therefore, we observe that the SW curve for 5d $E_7$ CFT is obtained 
from the SW curve for 5d $SU(4)$ gauge theory with eight flavors by tuning parameters. 
Especially, this tuning turns out to include the strong coupling limit in Figure \ref{construct E7}.
It is, therefore, natural to expect that our observation is related to the construction of 4d $E_7$ CFT, where the $\cN=2$ dualities play an essential role.

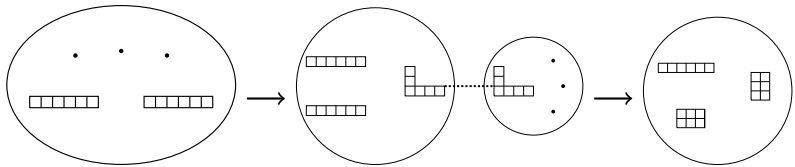
\begin{figure}[H]
\centering
\begin{adjustbox}{width=0.2\textwidth}
\begin{tikzpicture}
  [inner sep=0.5mm, line width=0.5mm,
 dot/.style={fill=black,draw,circle,minimum size=1pt},
 whitedot/.style={fill=white,draw,circle,minimum size=1pt}]
\draw (0,0) circle [x radius = 5cm, y radius=3.5cm];

\draw (-4,-1) -- (-1,-1) -- (-1,-0.5) -- (-4, -0.5) -- (-4,-1);
\draw (-3.5,-1) -- (-3.5,-0.5);
\draw (-3,-1) -- (-3,-0.5);
\draw (-2.5,-1) -- (-2.5,-0.5);
\draw (-2,-1) -- (-2,-0.5);
\draw (-1.5,-1) -- (-1.5,-0.5);

\draw (4,-1) -- (1,-1) -- (1,-0.5) -- (4, -0.5) -- (4,-1);
\draw (3.5,-1) -- (3.5,-0.5);
\draw (3,-1) -- (3,-0.5);
\draw (2.5,-1) -- (2.5,-0.5);
\draw (2,-1) -- (2,-0.5);
\draw (1.5,-1) -- (1.5,-0.5);

\node[dot] at (2,1.3) {};
\node[dot] at (-2,1.3) {};
\node[dot] at (0,1.5) {};

\end{tikzpicture}
\end{adjustbox}
\begin{tikzpicture}
 \draw[->,thick] (-1.5,0) -- (-1, 0);
 \node[circle, fill=white] at (-1.2,-0.7) {};
%   \node at (0,0) [above] {S duality};
\end{tikzpicture}
\begin{adjustbox}{width=0.25\textwidth}
\begin{tikzpicture}
  [inner sep=0.5mm, line width=0.5mm,
 dot/.style={fill=black,draw,circle,minimum size=1pt},
 whitedot/.style={fill=white,draw,circle,minimum size=1pt}]
\draw (0,0) circle [radius = 4cm];

\draw (-3.5,-1.5) -- (-0.5,-1.5) -- (-0.5,-1) -- (-3.5, -1) -- (-3.5,-1.5);
\draw (-3,-1.5) -- (-3,-1);
\draw (-2.5,-1.5) -- (-2.5,-1);
\draw (-2,-1.5) -- (-2,-1);
\draw (-1.5,-1.5) -- (-1.5,-1);
\draw (-1,-1.5) -- (-1,-1);

\draw (-3.5,1.5) -- (-0.5,1.5) -- (-0.5,1) -- (-3.5, 1) -- (-3.5,1.5);
\draw (-3,1.5) -- (-3,1);
\draw (-2.5,1.5) -- (-2.5,1);
\draw (-2,1.5) -- (-2,1);
\draw (-1.5,1.5) -- (-1.5,1);
\draw (-1,1.5) -- (-1,1);

\draw (1.5, -0.5) -- (3.5,-0.5) -- (3.5,0) -- (1.5,0) -- (1.5, -0.5);
\draw (1.5, 0) -- (1.5,1) -- (2,1) -- (2,-0.5) ;
\draw (2.5,0) -- (2.5,-0.5);
\draw (3,0) -- (3,-0.5);
\draw (1.5, 0.5) -- (2,0.5);
%----------------------------------
\draw [dashed, line width=1mm] (3.5,0) -- (6,0);
%----------------------------------
\draw (8,0) circle [radius = 2.5cm];

\draw (6, -0.5) -- (8,-0.5) -- (8,0) -- (6,0) -- (6, -0.5);
\draw (6, 0) -- (6,1) -- (6.5,1) -- (6.5,-0.5) ;
\draw (7,0) -- (7,-0.5);
\draw (7.5,0) -- (7.5,-0.5);
\draw (6,0.5) -- (6.5,0.5);
\node[dot] at (9,1.3) {};
\node[dot] at (9,-1.3) {};
\node[dot] at (9.5,0) {};

\end{tikzpicture}
\end{adjustbox}
\begin{tikzpicture}
 \draw[->,thick] (-1.5,0) -- (-1, 0);
 \node[circle, fill=white] at (-1.2,-0.7) {};
%   \node at (0,0) [above] {S duality};
\end{tikzpicture}
\begin{adjustbox}{width=0.13\textwidth}
\begin{tikzpicture}
  [inner sep=0.5mm, line width=0.5mm,
 dot/.style={fill=black,draw,circle,minimum size=1pt},
 whitedot/.style={fill=white,draw,circle,minimum size=1pt}]
\draw (-0.3,0) circle [radius = 4cm];

\draw (-2.5,-2) -- (-1,-2) -- (-1,-1) -- (-2.5, -1) -- (-2.5,-2);
\draw (-2,-2) -- (-2,-1);
\draw (-1.5,-2) -- (-1.5,-1);
\draw (-1,-2) -- (-1,-1);
\draw (-2.5,-1.5) -- (-1,-1.5);

\draw (-3.5,1.5) -- (-0.5,1.5) -- (-0.5,1) -- (-3.5, 1) -- (-3.5,1.5);
\draw (-3,1.5) -- (-3,1);
\draw (-2.5,1.5) -- (-2.5,1);
\draw (-2,1.5) -- (-2,1);
\draw (-1.5,1.5) -- (-1.5,1);
\draw (-1,1.5) -- (-1,1);

\draw (1.5, -0.5) -- (2.5,-0.5) -- (2.5,1) -- (1.5,1) -- (1.5, -0.5);
\draw (1.5, 0) -- (2.5,0);
\draw (1.5, 0.5) -- (2.5,0.5);
\draw (2,-0.5) -- (2,1.0);

\end{tikzpicture}
\end{adjustbox}
\caption{Construction of $E_8$ CFT. Start from $SU(6)^2$ theory with 6+6 flavors,
take the weak coupling limit in the S-dual frame, and Higgs two of the punctures.}
\label{construct E8}
\end{figure}

Another example is the construction of the 
$E_8$ CFT from $SU(6) \times SU(6)$ gauge theory with 6+6 flavors.
There exists such construction in 4d level. See Figure \ref{construct E8}.
From the toric-like diagram for 5d $E_8$ theory in Figure \ref{e8dots},
we observe that the SW curve for $E_8$ CFT is obtained from that for the 
 $SU(6) \times SU(6)$ gauge theory with 6+6 flavors in five dimensions.
This observation also implies that topological string amplitude for $E_8$ theory
computed in \cite{Hayashi:2014wfa}
can be obtained as a limit of that for $N_f=6~SU(6)\times N_f=6~SU(6)$ theory.

Although this observation is expected to be related to the $\cN=2$ duality, it is not clear how this observation is connected to what is discussed in \cite{Bergman:2014kza}.
We would like clarify this point in the future.

\section*{Acknowledgement}
We are grateful to Francesco Benini, Renato Fonseca, Axel Kleinschmidt, Vladimir Mitev, Elli Pomoni, Masato Taki, Piljin Yi, and Gabi Zafrir for discussion. We would like to thank Kimyeong Lee for helpful comments and discussion. We are also thankful to the workshops, ``Liouville, Integrability and Branes (10)'' at APCTP, ``Autumn Symposium on String/M Theory'' at KIAS, and
``International Workshop on Exceptional Symmetries and Emerging Spacetime'' at Nanyang Technological University.
%%%%%%%%%%%%%%%%%%%%%%%%%%%%%%%%%%%%%%%%%%%%%%%
%%%%%%%%%%%%%%%%%%%%%%%%%%%%%%%%%%%%%%%%%%%%%%%
%%%%%%%%%%%%%%%%%%%%%%%%%%%%%%%%%%%%%%%%%%%%%%%

\appendix

%%%%%%%%%%%%%%%%%%%%%%

\section{Seiberg-Witten curves for $N_f\le4$ flavors}\label{SW4Nfle4}
In this appendix, derivation of the Seiberg-Witten curve from toric-diagram is presented.
\subsection{$N_f=0$ with $E_1$ or $\widetilde E_1$ flavor symmetry}
\subsubsection{$E_1$ symmetry}
\begin{figure}[H]
\centering
\begin{tikzpicture}
  [inner sep=0.5mm, %line width=1.5pt,
 dot/.style={fill=black,draw,circle,minimum size=1pt},
 whitedot/.style={fill=white,draw,circle,minimum size=1pt},
 mark size=10pt, mark options={fill=white} ]
    \node[dot] at (0,0) [label={[label distance=0.03cm]45:$c_{11}$}] {};
    \node[dot] at (2,0) [label=right:$c_{21}$] {};
    \node[dot] at (0,2) [label=above:$c_{12}$] {};
    \node[dot] at (-2,0) [label=left:$c_{01}$]{};
    \node[dot] at (0,-2) [label=below:$c_{10}$] {};
    \draw (2,0) -- (0,2) -- (-2,0) -- (0,-2) -- (2,0);
	\draw (2,0) -- (-2,0);
	\draw (0,2) -- (0,-2);	
 \end{tikzpicture}
%\end{adjustbox}
\quad\quad
%\begin{adjustbox}{width=0.2\textwidth}
\begin{tikzpicture}
  [inner sep=0.5mm, line width=1.0pt,
 dot/.style={fill=black,draw,circle,minimum size=1pt},
 whitedot/.style={fill=white,draw,circle,minimum size=1pt},
 mark size=10pt, mark options={fill=white} ]
	% \draw[help lines] (-3,-3) grid (3,3);
 %    \foreach \x in {-3,-2,...,3} { \node [anchor=north] at (\x,-0.3) {\x}; }
 %    \foreach \y in {-3,-2,...,3} { \node [anchor=east] at (-0.3,\y) {\y}; }
 	\draw (1.0,-0.5)  -- (1.0,0.5) -- (-1.0, 0.5) -- (-1.0, -0.5) -- (1.0, -0.5);
	\draw (1.0,0.5) -- (2.5, 2);
    \draw (-1.0,0.5) -- (-2.5,2);
    \draw (-1.0,-0.5) -- (-2.5, -2);
    \draw (1.0,-0.5) -- (2.5,-2);
	\node[] at (2.5,2) [label=right:$\ell_1$]{};
   	\node[] at (-2.5,2) [label=left:$\ell_3$]{};
	\node[] at (-2.5,-2) [label=below:$\ell_4$]{};
	\node[] at (2.5,-2) [label=below:$\ell_2$]{};

\end{tikzpicture}
%\end{adjustbox}
\caption{Toric web diagram for $N_f=0$ of $E_1$ flavor symmetry}\label{Nf0E1}
\end{figure}
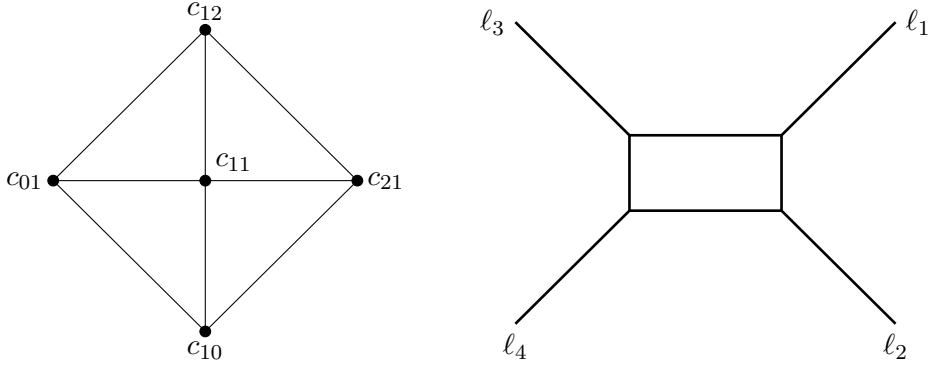

%\noindent{\bf $E_1$ symmetry}\\
The SW curve for the pure case ($N_f=0$) is of the form
\begin{align}
c_{01}\,w+c_{10}\,t+ c_{11}\,tw + c_{12}\,tw^2 + c_{21}\,t^2w=0.
\end{align}
Though we have thus five unknowns from the beginning, including three rescalings, one is left with only two parameters which are one Coulomb modulus and the dynamical scale $\Lambda_0$. We show how one can identify them below. We start with asymptotic boundary conditions:
\begin{align}
&|w|\sim |t|~{\rm large}:&&~ c_{12}tw^2+ c_{21}t^2w = c_{12}tw (w+\ell_1 t)\cr
&|w^{-1}|\sim |t|~{\rm large}:&&~ c_{10}t+ c_{21}t^2w = c_{21}t (\ell_2+ tw)\cr
&|w|\sim |t^{-1}|~{\rm large}:&&~ c_{12}tw^2+ c_{01}w = c_{12}w (tw+\ell_3)\cr
&|w^{-1}|\sim |t^{-1}|~{\rm large}:&&~ c_{01}w+ c_{10}t = c_{01} (w+\ell_4t).
\end{align}
The compatibility condition is then
\begin{align}\label{relationamongpara}
\ell_1\,\ell_2\,=\,\ell_3\,\ell_4.
\end{align}
As for three rescalings degrees of freedom, we take
\begin{align}\label{coeffNf0}
c_{01}=c_{21}=1, \quad c_{10}=c_{12}=\ell_2=\ell_1^{-1}=\ell_4,\quad c_{11}\propto U,
\end{align}
and
the SW curve for $N_f=0$ is expressed in terms of two parameters, the Coulomb modulus $U$, and $\ell_1$:
\begin{align}\label{SWforNf=0}
w+\ell_1^{-1}t\Big( w^2+ U w + 1\Big) + t^2 w=0.
\end{align}
The $\ell_1$ is in fact related to the dynamical scale $\Lambda_0$. We explain how to relate it to the dynamical scale.
By referring to the topological vertex results, the dynamical scale is determined as the geometric mean of differences $\Delta t$ evaluated in asymptotic values of $t$ for given $w$. This is consistent with the dependence of the dynamical scale over energy scales
\begin{align}
\sqrt{\Big(\frac{\ell_3 w_1^{-1}}{\ell_1^{-1}w_1}\Big)\Big(\frac{\ell_2 w_2^{-1}}{\ell_4^{-1}w_2}\Big)} = \Big(\frac{\ell_1\ell_3}{\ell_2\ell_4}\Big)^{\frac12}w_2w_1^{-1}\bigg|_{w_1=w_2=w_0} =\big(2\pi R\Lambda_0\big)^{2N_c-N_f}.
\end{align}
We then find
\begin{align}\label{t1valuenf0}
\ell_1 = \big(2\pi R \Lambda_0\big)^2,
\end{align}
and thus the SW curve for $N_f=0$ is written as
\begin{align}\label{SWforNf=0-1}
w+\frac{t}{(2\pi R \Lambda_0)^{2}}\Big( w^2+ U w + 1\Big) + t^2 w=0.
\end{align}
This has $E_1$ flavor symmetry. %

\subsubsection{$\widetilde E_1$ symmetry}
\begin{figure}[H]
\centering
\begin{tikzpicture}
  [inner sep=0.5mm, line width=0.5pt,
 dot/.style={fill=black,draw,circle,minimum size=1pt},
 whitedot/.style={fill=white,draw,circle,minimum size=1pt},
 mark size=10pt, mark options={fill=white} ]

    \node[dot] at (0,0) [label={[label distance=0.03cm]-45:$c_{11}$}] {};
    \node[dot] at (2,2) [label=right:$c_{22}$] {};
    \node[dot] at (0,2) [label=above:$c_{12}$] {};
    \node[dot] at (-2,0) [label=left:$c_{01}$]{};
    \node[dot] at (0,-2) [label=below:$c_{10}$] {};
    \draw (-2,0) -- (0,-2) -- (2,2) --(0,2) -- (-2,0);
    \draw (0,2) -- (0,-2);
    \draw (-2,0) -- (0,0) -- (2,2);
 \end{tikzpicture}
\quad\quad
\begin{tikzpicture}
  [inner sep=0.5mm, line width=1.0pt,
 dot/.style={fill=black,draw,circle,minimum size=1pt},
 whitedot/.style={fill=white,draw,circle,minimum size=1pt},
 mark size=10pt, mark options={fill=white} ]

 	\draw (1.0,-0.5)  -- (0,0.5) -- (-1.0, 0.5) -- (-1.0, -0.5) -- (1.0, -0.5);
	\draw (0.0,0.5) -- (0, 2);
    \draw (-1.0,0.5) -- (-2.5,2);
    \draw (-1.0,-0.5) -- (-2.5, -2);
    \draw (1.0,-0.5) -- (4,-2);
	\node[] at (0,2) [label=right:$t_1$]{};
   	\node[] at (-2.5,2) [label=left:$\ell_3$]{};
	\node[] at (-2.5,-2) [label=below:$\ell_4$]{};
	\node[] at (4,-2) [label=below:$\ell_2'$]{};

\end{tikzpicture}
%\end{adjustbox}
\caption{Toric web diagram for $N_f=0$ of $\widetilde E_1$ flavor symmetry}\label{Nf0E1t}
\end{figure}

 There exits an inequivalent toric diagram which has different asymptotic boundary conditions compared with the above $E_1$ case. It is known as $\widetilde{E}_1$ theory. The asymptotic boundary conditions are given by
\begin{align}
&|w^{-1}|\sim |t^{2}|~{\rm large}:&&~ c_{10}t+ c_{22}t^2w^2 = c_{22}t (tw^2+\ell'_2 )&&\Rightarrow&& c_{10}=\ell'_2 c_{22},\cr
&|w|~{\rm large}:&&~ c_{12}tw^2+ c_{22}t^2w^2 = c_{22}tw^2 (-t_1+ t)&& \Rightarrow&& c_{12}=-t_1c_{22},\cr
&|w|\sim |t^{-1}|~{\rm large}:&&~ c_{12}tw^2+ c_{01}w = c_{12}w (tw+\ell_3)&& \Rightarrow&& c_{01} = \ell_3 c_{12},\cr
&|w^{-1}|\sim |t^{-1}|~{\rm large}:&&~ c_{01}w+ c_{10}t = c_{01} (w+\ell_4t)&& \Rightarrow&& c_{10} = \ell_4 c_{01},
\end{align}
subject to the compatibility condition
\begin{align}
\ell_2' \,=\, -\,t_1\, \ell_3\,\ell_4.
\end{align}
We choose the three rescaling degrees of freedom to be
\begin{align}
c_{12}=c_{10},\qquad c_{22}=1=c_{01},
\end{align}
The dynamical scale $\widetilde\Lambda_0$ is
\begin{align}
\ell_4=(2\pi R\widetilde \Lambda_0)^{-2}.
\end{align}
One then obtains the SW curve for $\widetilde E_1$ as
\begin{align}\label{E1tildeSWcurve}
w +\frac{t}{(2\pi R\widetilde \Lambda_0)^2}\,(w^2+ \tilde U w + 1)+ t^2w^2 =0.
\end{align}

We note that the 4d limit of $E_1$ and $\tilde E_1$ gives the same SW curve at $\mathcal{O}(\beta^0)$,
\begin{align}
1+t^2 + \frac{t(u+v^2)}{\Lambda_0}=0.
\end{align}

\subsubsection{$E_0$ Seiberg-Witten curve}
It is interesting to take a limit that leads to the $E_0$ SW curve from the $\widetilde E_1$ curve. For this, we decouple the coefficient $c_{12}$ or the $tw^2$-term. Instead of \eqref{E1tildeSWcurve}, we start a generic form of the $\widetilde E_1$ curve
\begin{align}
c_{01}\Big(w + \ell_4 \,t + U \,t w + \frac{1}{\ell_3}\, tw^2+ \frac{\ell_4}{\ell_2'} t^2w^2 \Big)=0.
\end{align}
We then take $\ell_3\to \infty, t_1\to0$ while $-t_1\ell_3$ fixed, which gives
\begin{align}
w + \ell_4 \,t + U \,t w + \frac{\ell_4}{\ell_2'} t^2w^2 =0.
\end{align}
Using the remaining rescaling degrees of freedom associated with $t$ and $w$, we can fix $\ell_4=1, \ell_2'=1$, yielding the $E_0$ SW curve
\begin{align}
w +  \,t + U \,t w + t^2w^2 =0.
\end{align}

\subsection{$N_f=1$}
\begin{figure}[H]
\centering
%\begin{adjustbox}{width=0.2\textwidth}
\begin{tikzpicture}
  [inner sep=0.5mm, %line width=1.5pt,
 dot/.style={fill=black,draw,circle,minimum size=1pt},
 whitedot/.style={fill=white,draw,circle,minimum size=1pt},
 mark size=10pt, mark options={fill=white} ]
	% \draw[help lines] (-3,-3) grid (3,3);
 %    \foreach \x in {-3,-2,...,3} { \node [anchor=north] at (\x,-0.3) {\x}; }
 %    \foreach \y in {-3,-2,...,3} { \node [anchor=east] at (-0.3,\y) {\y}; }

    \node[dot] at (0,0) [label={[label distance=0.03cm]45:$c_{11}$}] {};
    \node[dot] at (2,0) [label=right:$c_{21}$] {};
    \node[dot] at (0,2) [label=above:$c_{12}$] {};
    \node[dot] at (-2,0) [label=left:$c_{01}$]{};
    \node[dot] at (0,-2) [label=below:$c_{10}$] {};
    \node[dot] at (2,2) [label=right:$c_{22}$] {};

    \draw (2,0) -- (0,2) -- (-2,0) -- (0,-2) -- (2,0);
	\draw (2,0) -- (-2,0);
	\draw (0,2) -- (0,-2);
	\draw (0,2) -- (2,2) -- (2,0);
 \end{tikzpicture}
%\end{adjustbox}
\quad\quad
%\begin{adjustbox}{width=0.2\textwidth}
\begin{tikzpicture}
  [inner sep=0.5mm, line width=1.0pt,
 dot/.style={fill=black,draw,circle,minimum size=1pt},
 whitedot/.style={fill=white,draw,circle,minimum size=1pt},
 mark size=10pt, mark options={fill=white} ]
	% \draw[help lines] (-3,-3) grid (3,3);
 %    \foreach \x in {-3,-2,...,3} { \node [anchor=north] at (\x,-0.3) {\x}; }
 %    \foreach \y in {-3,-2,...,3} { \node [anchor=east] at (-0.3,\y) {\y}; }
 	\draw (1.0,-0.5)  -- (1.0,0.5) -- (-1.0, 0.5) -- (-1.0, -0.5) -- (1.0, -0.5);
	\draw (1.0,0.5) -- (1.7, 1.2) -- (1.7,2); \draw (1.7,1.2) -- (2.5, 1.2);
    \draw (-1.0,0.5) -- (-2.5,2);
    \draw (-1.0,-0.5) -- (-2.5, -2);
    \draw (1.0,-0.5) -- (2.5,-2);
	\node[] at (2.5,1.2) [label=right:$\widetilde m_1$]{};
	\node[] at (1.7,2) [label=left:$t_1$]{};
   	\node[] at (-2.5,2) [label=left:$\ell_3$]{};
	\node[] at (-2.5,-2) [label=below:$\ell_4$]{};
	\node[] at (2.5,-2) [label=below:$\ell_2$]{};

   % \draw (4,8.2) arc (-90:90:0.3cm);
 % 	\foreach \plm in {otimes*} \draw plot[mark=\plm] coordinates {(0.25,-2.2)} ;
     % \draw[->] (-3,-2.5) -- (-3,-1.5); \node[] at (-3,-1.5) [label=above:$w$] {};
\end{tikzpicture}
%\end{adjustbox}
\caption{Toric web diagram for $N_f=1$ of $E_2$ flavor symmetry}\label{Nf=1}
\end{figure}
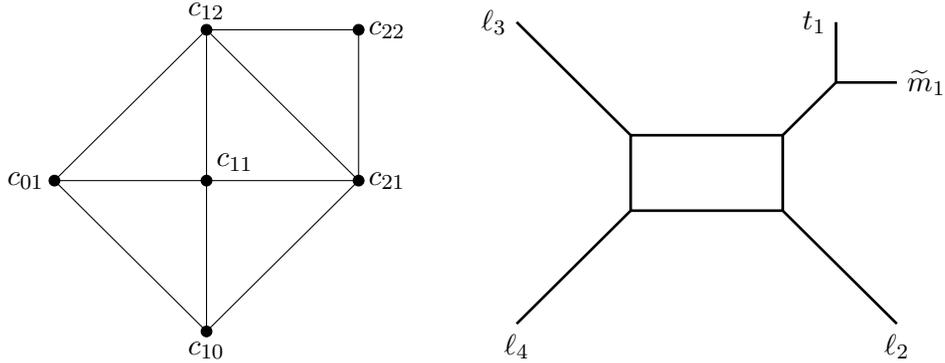

For $N_f=1$, we start with
\begin{align}
c_{01}\,w+c_{10}\,t+ c_{11}\,tw + c_{12}\,tw^2 + c_{21}\,t^2w+ c_{22}t^2w^2=0.
\end{align}
The boundary conditions are
\begin{align}
&|w^{-1}|\sim |t|~{\rm large}:&&~ c_{10}t+ c_{21}t^2w = c_{21}t (\ell_2+ tw) &&\Rightarrow\quad
c_{10} = \ell_2c_{21},\cr
&|w|\sim |t^{-1}|~{\rm large}:&&~ c_{12}tw^2+ c_{01}w = c_{12}w (tw+\ell_3)&&\Rightarrow\quad
c_{01} = \ell_3c_{12},
\cr
&|w^{-1}|\sim |t^{-1}|~{\rm large}:&&~ c_{01}w+ c_{10}t = c_{01} (w+\ell_4t)&&\Rightarrow\quad
c_{10} = \ell_4c_{01}.
\end{align}
In addition to this, there is an extra contribution from a flavor:
\begin{align}
&c_{21} \,t^2\,w+ c_{22}\, t^2\,w^2 = c_{22}\,t^2\, w\, (-\widetilde m_1+ w)&& \Rightarrow&& c_{21} = -\widetilde m_1\,c_{22},\cr
&c_{12} \,t\,w^2+ c_{22}\, t^2\,w^2 = c_{22}\,t \,w^2\, (-t_1+ t)&& \Rightarrow &&c_{12} = - t_1\,c_{22}.
\end{align}
The compatibility condition is
\begin{align}\label{relnf=1}
\frac{\widetilde m_1}{t_1}\,\ell_2= \ell_3\,\ell_4.
\end{align}
As in the  $N_f=0$ case, for three rescaling degrees of freedom, we take
\begin{align}
c_{01}=c_{21}=1\quad  {\rm and}\quad  c_{10}=c_{12},
\end{align}
or equivalently,
\begin{align}\label{condfornf1}
\frac{t_1}{\ell_2} = \widetilde m_1,\qquad  t_1\ell_3= {\widetilde m_1} ~~({\rm or}~~ \ell_2=\ell_4).
\end{align}
With dynamical scale is given by
\begin{align}\label{t1valuenf1}
t_1^2\,\widetilde m_1^{-\frac32}=\big(2\pi R_B \Lambda_1\big)^{-3} ,
\end{align}
the SW curve take the following form
\begin{align}\label{SWforNf=1}
w\,+\,t_1\,\widetilde m_1^{-1}\,t\,\Big( w^2\,+ U\, w \,+\, 1\Big) + t^2\, w -\widetilde m_1^{-1} \,t^2 \,w^2=0.
\end{align}

\subsubsection{$E_1$ limit}

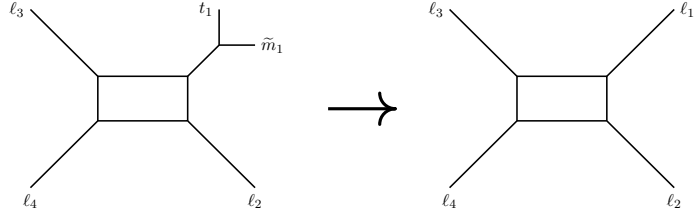
\begin{figure}[H]
\centering
\begin{adjustbox}{width=0.6\textwidth}
\begin{tikzpicture}
  [inner sep=0.5mm, line width=1.0pt,
 dot/.style={fill=black,draw,circle,minimum size=1pt},
 whitedot/.style={fill=white,draw,circle,minimum size=1pt},
 mark size=10pt, mark options={fill=white} ]
 	\draw (1.0,-0.5)  -- (1.0,0.5) -- (-1.0, 0.5) -- (-1.0, -0.5) -- (1.0, -0.5);
	\draw (1.0,0.5) -- (1.7, 1.2) -- (1.7,2); \draw (1.7,1.2) -- (2.5, 1.2);
    \draw (-1.0,0.5) -- (-2.5,2);
    \draw (-1.0,-0.5) -- (-2.5, -2);
    \draw (1.0,-0.5) -- (2.5,-2);
	\node[] at (2.5,1.2) [label=right:$\widetilde m_1$]{};
	\node[] at (1.7,2) [label=left:$t_1$]{};
   	\node[] at (-2.5,2) [label=left:$\ell_3$]{};
	\node[] at (-2.5,-2) [label=below:$\ell_4$]{};
	\node[] at (2.5,-2) [label=below:$\ell_2$]{};

\end{tikzpicture}
\quad \quad
\begin{tikzpicture}[line width=2pt]
 \draw (-1.5,0) -- (0.0, 0);
 \draw[join=round,cap=round]
        (-10pt,10pt) arc (180:270:10pt) arc (90:180:10pt);
 %\draw[->,thick] (-1.5,1) -- (0.0, 1);
   \node[circle, fill=white] at (0,-2.1) {};
\end{tikzpicture}
\quad
\begin{tikzpicture}
  [inner sep=0.5mm, line width=1.0pt,
 dot/.style={fill=black,draw,circle,minimum size=1pt},
 whitedot/.style={fill=white,draw,circle,minimum size=1pt},
 mark size=10pt, mark options={fill=white} ]
 	\draw (1.0,-0.5)  -- (1.0,0.5) -- (-1.0, 0.5) -- (-1.0, -0.5) -- (1.0, -0.5);
	\draw (1.0,0.5) -- (2.5, 2);
    \draw (-1.0,0.5) -- (-2.5,2);
    \draw (-1.0,-0.5) -- (-2.5, -2);
    \draw (1.0,-0.5) -- (2.5,-2);
	\node[] at (2.5,2) [label=right:$\ell_1$]{};
   	\node[] at (-2.5,2) [label=left:$\ell_3$]{};
	\node[] at (-2.5,-2) [label=below:$\ell_4$]{};
	\node[] at (2.5,-2) [label=below:$\ell_2$]{};
\end{tikzpicture}
\end{adjustbox}
\caption{One obtains $E_1$ curve from the $E_2$ curve by taking $\widetilde m_1\to\infty$ and $t_1\to \infty$, while $\widetilde m_1 t_1^{-1}=\ell_1$ fixed.}
\label{E2toE1}
\end{figure}

It is clear from Figure \ref{E2toE1} that by taking the mass decoupling limit
\begin{align}
\widetilde m_1\to \infty \quad ({\rm while~} {t^{-1}_1}{\widetilde m_1}= \ell_1 {\rm~ fixed}),
\end{align}
one reproduces the $N_f=0$,  $E_1$ curve, \eqref{SWforNf=0}. It follows that the condition \eqref{condfornf1} becomes the condition for $N_f=0$, \eqref{coeffNf0}. The dynamical scale for $N_f=1$ and $N_f=0$ are related by
\begin{align}
\widetilde m_1^\frac12 (2\pi R\Lambda_1)^3 = (2\pi R\Lambda_0)^4,
\end{align}
which relates \eqref{t1valuenf1} to \eqref{t1valuenf0}.

\subsubsection{$\widetilde E_1$ limit}
\begin{figure}[H]
\centering
\begin{adjustbox}{width=0.6\textwidth}
\begin{tikzpicture}
  [inner sep=0.5mm, line width=1.0pt,
 dot/.style={fill=black,draw,circle,minimum size=1pt},
 whitedot/.style={fill=white,draw,circle,minimum size=1pt},
 mark size=10pt, mark options={fill=white} ]
 	\draw (1.0,-0.5)  -- (1.0,0.5) -- (-1.0, 0.5) -- (-1.0, -0.5) -- (1.0, -0.5);
	\draw (1.0,0.5) -- (1.7, 1.2) -- (1.7,2); \draw (1.7,1.2) -- (2.5, 1.2);
    \draw (-1.0,0.5) -- (-2.5,2);
    \draw (-1.0,-0.5) -- (-2.5, -2);
    \draw (1.0,-0.5) -- (2.5,-2);
	\node[] at (2.5,1.2) [label=right:$\widetilde m_1$]{};
	\node[] at (1.7,2) [label=left:$t_1$]{};
   	\node[] at (-2.5,2) [label=left:$\ell_3$]{};
	\node[] at (-2.5,-2) [label=below:$\ell_4$]{};
	\node[] at (2.5,-2) [label=below:$\ell_2$]{};

\end{tikzpicture}
\quad \quad
\begin{tikzpicture}[line width=2pt]
 \draw (-1.5,0) -- (0.0, 0);
 \draw[join=round,cap=round]
        (-10pt,10pt) arc (180:270:10pt) arc (90:180:10pt);
 %\draw[->,thick] (-1.5,1) -- (0.0, 1);
   \node[circle, fill=white] at (0,-2.1) {};
\end{tikzpicture}
\quad
\begin{tikzpicture}
  [inner sep=0.5mm, line width=1.0pt,
 dot/.style={fill=black,draw,circle,minimum size=1pt},
 whitedot/.style={fill=white,draw,circle,minimum size=1pt},
 mark size=10pt, mark options={fill=white} ]
	 	\draw (1.0,-0.5)  -- (0,0.5) -- (-1.0, 0.5) -- (-1.0, -0.5) -- (1.0, -0.5);
	\draw (0.0,0.5) -- (0, 2);
    \draw (-1.0,0.5) -- (-2.5,2);
    \draw (-1.0,-0.5) -- (-2.5, -2);
    \draw (1.0,-0.5) -- (4,-2);
	\node[] at (0,2) [label=right:$t_1$]{};
   	\node[] at (-2.5,2) [label=left:$\ell_3$]{};
	\node[] at (-2.5,-2) [label=below:$\ell_4$]{};
	\node[] at (4,-2) [label=below:$\ell_2'$]{};

  \end{tikzpicture}
\end{adjustbox}
\caption{One obtains $\widetilde E_1$ curve from the $E_2$ curve by taking $\widetilde m_1\to 0$ and $\ell_2\to \infty$, while $-\widetilde m_1 \ell_2=\ell_2'$ fixed.}
\label{E2toE1tilde}
\end{figure}
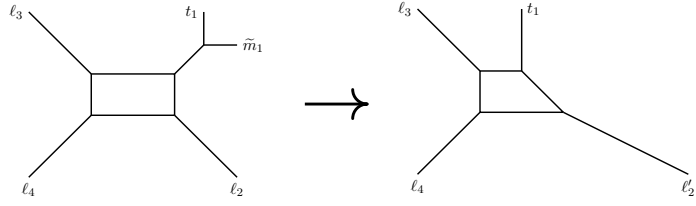

We can take a distinct mass decoupling limit such that $\widetilde m_1\to 0$, keeping $-\widetilde m_1 \ell_2~ (=\ell_2')$ fixed. In this case, the relation \eqref{relnf=1} is expressed as
\begin{align}
-t_1\ell_3\ell_4=-\widetilde m_1\ell_2 ~ (\equiv\ell_2').
\end{align}
In this limit, the $t^2w$-term drops out and the SW curve \eqref{SWforNf=1} becomes
\begin{align}
w\,-\,t_1\,\ell_2'^{-1}\, \ell_4\,t\,\Big( w^2\,+ \tilde U\, w \,-\, t_1^{-1}\ell_2'\Big)  -\ell_2'^{-1}\ell_4 \,t^2 \,w^2=0.
\end{align}
With the choice $-t_1^{-1}\ell_2'=1$ and $\ell_2'^{-1}\ell_4=1$, or equivalently $c_{12}=c_{10}$ and $c_{01}=c_{22}$, the
SW curve is written in term of two parameters, $\ell_4$, $\tilde U$,
\begin{align}
w + \ell_4 \,t \big(1+ \tilde U \, w +  w^2 \big)+ t^2w^2 =0,
\end{align}
which is nothing but the SW curve for $\widetilde E_1$, \eqref{E1tildeSWcurve}.

\subsection{$N_f=2$}
\begin{figure}[H]
\centering
%\begin{adjustbox}{width=0.2\textwidth}
\begin{tikzpicture}
  [inner sep=0.5mm, %line width=1.5pt,
 dot/.style={fill=black,draw,circle,minimum size=1pt},
 whitedot/.style={fill=white,draw,circle,minimum size=1pt},
 mark size=10pt, mark options={fill=white} ]
	
    \node[dot] at (0,0) [label={[label distance=0.03cm]45:$c_{11}$}] {};
    \node[dot] at (2,0) [label=right:$c_{21}$] {};
    \node[dot] at (0,2) [label=above:$c_{12}$] {};
    \node[dot] at (-2,0) [label=left:$c_{01}$]{};
    \node[dot] at (0,-2) [label=below:$c_{10}$] {};
    \node[dot] at (2,2) [label=right:$c_{22}$] {};
    \node[dot] at (2,-2) [label=right:$c_{20}$] {};

    \draw (2,0) -- (0,2) -- (-2,0) -- (0,-2) -- (2,0);
	\draw (2,0) -- (-2,0);
	\draw (0,2) -- (0,-2);
	\draw (0,2) -- (2,2) -- (2,0);
	\draw (2,0) -- (2,-2) -- (0, -2);
 \end{tikzpicture}
%\end{adjustbox}
\quad\quad
%\begin{adjustbox}{width=0.2\textwidth}
\begin{tikzpicture}
  [inner sep=0.5mm, line width=1.0pt,
 dot/.style={fill=black,draw,circle,minimum size=1pt},
 whitedot/.style={fill=white,draw,circle,minimum size=1pt},
 mark size=10pt, mark options={fill=white} ]
 	\draw (1.0,-0.5)  -- (1.0,0.5) -- (-1.0, 0.5) -- (-1.0, -0.5) -- (1.0, -0.5);
	\draw (1.0,0.5) -- (1.7, 1.2) -- (1.7,2); \draw (1.7,1.2) -- (2.5, 1.2);
    \draw (-1.0,0.5) -- (-2.5,2);
    \draw (-1.0,-0.5) -- (-2.5, -2);
    \draw (1.0,-0.5) -- (1.5,-1) -- (2.5,-1);
    \draw (1.5,-1) -- (1.5,-2);
	\node[] at (2.5,1.2) [label=right:$\widetilde m_1$]{};
	\node[] at (1.7,2) [label=left:$t_1$]{};
   	\node[] at (-2.5,2) [label=left:$\ell_3$]{};
	\node[] at (-2.5,-2) [label=below:$\ell_4$]{};
	\node[] at (2.5,-1) [label=right:$\widetilde m_2$]{};
	\node[] at (1.5,-2) [label=below:$t_2$]{};
\end{tikzpicture}
%\end{adjustbox}
\caption{Toric and web diagrams for $N_f=2$ of $E_3$ flavor symmetry}\label{nf2toric}
\end{figure}
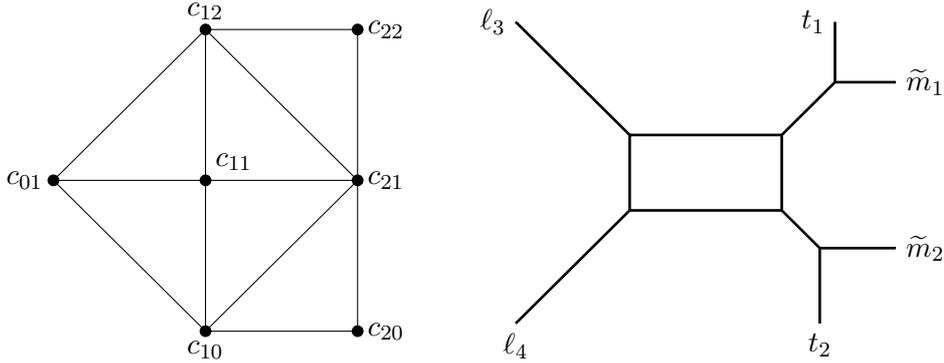
For $N_f=2$, we start with
\begin{align}
c_{01}\,w+c_{10}\,t+ c_{11}\,tw + c_{12}\,tw^2 + c_{21}\,t^2w+ c_{22}\,t^2w^2+ c_{20}\,t^2=0.
\end{align}
The boundary conditions are given by
\begin{align}
&|w|~{\rm large}:&&~ c_{12}tw^2+ c_{22}t^2w^2 = c_{22}tw^2 (-t_1+t)\cr
&|w|~{\rm small}:&&~ c_{10}t+ c_{20}t^2 = c_{20}t(-t_2+t)\cr
&|w|\sim |t^{-1}|~{\rm large}:&&~ c_{12}tw^2+ c_{01}w = c_{12}w (tw+\ell_3)\cr
&|w^{-1}|\sim |t^{-1}|~{\rm large}:&&~ c_{01}w+ c_{10}t = c_{01} (w+\ell_4t),
\end{align}
and the following extra boundary condition: When $|t|$ is very large,
\begin{align}
c_{20}t^2 + c_{21}t^2 w+ c_{22}t^2w^2 = c_{22}t^2(w-\widetilde m_1)(w-\widetilde m_2).
\end{align}
The compatibility condition is 
\begin{align}
\widetilde m_1\, t_1^{-1} \, \widetilde m_2\, t_2 = \ell_3\,\ell_4.
\end{align}
For three rescaling degrees of freedom, we choose
 \begin{align}
c_{01}=c_{21}=1,\quad c_{10}=c_{12},
\end{align}
or equivalently
\begin{align}\label{condfornf2}
\frac{t_1}{t_2} = \widetilde m_1\widetilde m_2,\qquad  t_1\ell_3= \widetilde m_1+ \widetilde m_2.
\end{align}
The dynamical scale is given by
\begin{align}
\Big(\frac{t_1t_2\ell_4}{\ell_3}\Big)^\frac12 =\big(2\pi R_B \Lambda_2\big)^{-2} ,
\end{align}
or equivalently,
\begin{align}\label{dynsclaNf2}
t_1^2\,\widetilde m_1^{-\frac12}\,\widetilde m_2^{-\frac12} (\widetilde m_1+\widetilde m_2)^{-1} =\big(2\pi R_B \Lambda_2\big)^{-2} .
\end{align}
%where $t_1$ is given by \eqref{t1value}.
The SW curve for $N_f=2$ is then
\begin{align}\label{SWforNf=2}\,
w+\,t\,\frac{t_1}{\widetilde m_1+\widetilde m_2}\Big(  w^2+ U w + 1\Big) -  t^2 \frac{1}{\widetilde m_1+\widetilde m_2}\, \big(w - \widetilde m_1 \big)\big(w- \widetilde m_2\big) =0.
\end{align}
In the mass decoupling limit $m_2\to \infty$, correspondingly $\widetilde m_2\to 0$, while keeping $\widetilde m_2t_2=\ell_2$ fixed, it is straightforward to see that the SW curve
 for $N_f=2$, \eqref{SWforNf=2}, becomes that for $N_f=1$, \eqref{SWforNf=1}. The condition that we used, \eqref{condfornf2} becomes the condition for $N_f=1$, \eqref{condfornf1}, and the dynamical scales for $N_f=2$ and $N_f=1$ are related in this decoupling limit as
\begin{align}
\widetilde m_2^{-\frac12} (2 \pi R \Lambda_2)^2 = (2\pi R \Lambda_1)^3
\end{align}
which relates \eqref{dynsclaNf2} to \eqref{t1valuenf1}.

\subsection{$N_f=3$}
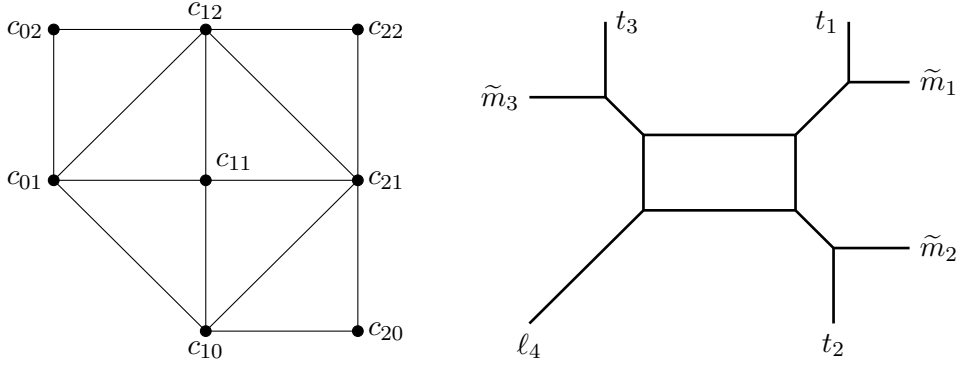
\begin{figure}[H]
\centering
%\begin{adjustbox}{width=0.2\textwidth}
\begin{tikzpicture}
  [inner sep=0.5mm, %line width=1.5pt,
 dot/.style={fill=black,draw,circle,minimum size=1pt},
 whitedot/.style={fill=white,draw,circle,minimum size=1pt},
 mark size=10pt, mark options={fill=white} ]
    \node[dot] at (0,0) [label={[label distance=0.03cm]45:$c_{11}$}] {};
    \node[dot] at (2,0) [label=right:$c_{21}$] {};
    \node[dot] at (0,2) [label=above:$c_{12}$] {};
    \node[dot] at (-2,0) [label=left:$c_{01}$]{};
    \node[dot] at (0,-2) [label=below:$c_{10}$] {};
    \node[dot] at (2,2) [label=right:$c_{22}$] {};
    \node[dot] at (2,-2) [label=right:$c_{20}$] {};
    \node[dot] at (-2,2) [label=left:$c_{02}$] {};

    \draw (2,0) -- (0,2) -- (-2,0) -- (0,-2) -- (2,0);
	\draw (2,0) -- (-2,0);
	\draw (0,2) -- (0,-2);
	\draw (0,2) -- (2,2) -- (2,0);
	\draw (2,0) -- (2,-2) -- (0, -2);	
	\draw (0,2) -- (-2,2) -- (-2,0);
 \end{tikzpicture}
%\end{adjustbox}
\quad\quad
%\begin{adjustbox}{width=0.2\textwidth}
\begin{tikzpicture}
  [inner sep=0.5mm, line width=1.0pt,
 dot/.style={fill=black,draw,circle,minimum size=1pt},
 whitedot/.style={fill=white,draw,circle,minimum size=1pt},
 mark size=10pt, mark options={fill=white} ]
 	\draw (1.0,-0.5)  -- (1.0,0.5) -- (-1.0, 0.5) -- (-1.0, -0.5) -- (1.0, -0.5);
	\draw (1.0,0.5) -- (1.7, 1.2) -- (1.7,2); \draw (1.7,1.2) -- (2.5, 1.2);
    \draw (-1.0,0.5) -- (-1.5,1) -- (-2.5,1);    \draw (-1.5,1) -- (-1.5,2);
    \draw (-1.0,-0.5) -- (-2.5, -2);
    \draw (1.0,-0.5) -- (1.5,-1) -- (2.5,-1);    \draw (1.5,-1) -- (1.5,-2);
	\node[] at (2.5,1.2) [label=right:$\widetilde m_1$]{};
	\node[] at (1.7,2) [label=left:$t_1$]{};
 	\node[] at (-2.5,1) [label=left:$\widetilde m_3$]{};
 	\node[] at (-1.5,2) [label=right:$t_3$]{};
	\node[] at (-2.5,-2) [label=below:$\ell_4$]{};
	\node[] at (2.5,-1) [label=right:$\widetilde m_2$]{};
	\node[] at (1.5,-2) [label=below:$t_2$]{};
\end{tikzpicture}
%\end{adjustbox}
\caption{Toric web diagram for $N_f=3$ of $E_4$ flavor symmetry}\label{nf3toric}
\end{figure}
For $N_f=3$, we start with
\begin{align}
c_{01}\,w+c_{10}\,t+ c_{11}\,tw + c_{12}\,tw^2 + c_{21}\,t^2w+ c_{22}\,t^2w^2+ c_{20}\,t^2+c_{02}w^2=0.
\end{align}
The boundary conditions are given by
\begin{align}
&|w|~{\rm small}:&&~ c_{10}t+ c_{20}t^2 = c_{20}t(-t_2+t),\cr
%&|w|\sim |t^{-1}|~{\rm large}:&&~ c_{12}tw^2+ c_{01}w = c_{12}w (tw+\ell_3)\cr
&|w^{-1}|\sim |t^{-1}|~{\rm large}:&&~ c_{01}w+ c_{10}t = c_{01} (w+\ell_4t),\cr
&|t|~{\rm large}: &&~ c_{20}t^2 + c_{21}t^2 w+ c_{22}t^2w^2 = c_{22}t^2(w-\widetilde m_1)(w-\widetilde m_2),
\end{align}
and when $|w|$ large, the boundary condition is given by
\begin{align}
c_{02}w^2+c_{12}tw^2+ c_{22}t^2w^2 = c_{22}w^2 (t-t_1)(t-t_3),
\end{align}
and the boundary condition for $|t|$ small ($|w|$ not small) is given by
\begin{align}
c_{01}w+ c_{02}w^2 = c_{02}w(w-\widetilde m_3).
\end{align}
The compatibility is given by
\begin{align}
\widetilde m_1\, t_1^{-1} \, \widetilde m_2\, t_2 = \widetilde m_3\,t_3\,\ell_4.
\end{align}
For three rescaling degrees of freedom, we choose
\begin{align}
c_{01}=c_{21}=1,\qquad c_{10}=c_{12},
\end{align}
or equivalently
\begin{align}\label{condforNf3}
\frac{t_1+t_3}{t_2}=  \widetilde m_1 \widetilde m_2,\qquad
t_1  t_3 = \frac{\widetilde m_1+ \widetilde m_2}{\widetilde m_3}.
\end{align}
The dynamical scale is given by
\begin{align}
\Big(\frac{t_1 t_2\ell_4}{t_3}\Big)^\frac12=\big(2\pi R_B \Lambda_3\big)^{-1},
\end{align}
or
\begin{align}\label{t1valuenf3}
t_1 = \Big(\frac{\widetilde m_1 + \widetilde m_2}{\widetilde m_3}\Big)^\frac12 \bigg( \frac{(\widetilde m_1\widetilde m_2\widetilde m_3)^\frac12}{2\pi R \Lambda_3}-1\bigg)^\frac12.
\end{align}
The SW curve for $N_f=3$ is then
\begin{align}\label{SWforNf=3}
w+\,t\,\Big(\frac{t_1}{\widetilde m_1+\widetilde m_2} +\frac{1}{t_1\widetilde m_3}\Big)\Big(  w^2+ U w + 1\Big) -  t^2 \frac{1}{\widetilde m_1+\widetilde m_2}\, \big(w - \widetilde m_1 \big)\big(w- \widetilde m_2\big)-\frac{1}{\widetilde m_3}w^2=0.
\end{align}

In the mass decoupling limit $\widetilde m_3\to \infty$ and $t_3\to 0$, while $\widetilde m_3\,t_3=\ell_3$ fixed,
the SW curve for $N_f=3$, \eqref{SWforNf=4}, becomes that for $N_f=2$, \eqref{SWforNf=3}, and the condition that we used, \eqref{condforNf3} becomes the condition for $N_f=3$, \eqref{condfornf2}, and the dynamical scales for $N_f=3$ and $N_f=2$ are related in this decoupling limit as
\begin{align}
\widetilde m_3^\frac12 (2 \pi R \Lambda_3) = (2\pi R \Lambda_2)^2
\end{align}
which relates \eqref{t1valuenf3} to \eqref{dynsclaNf2}.

\subsection{$N_f=4$}
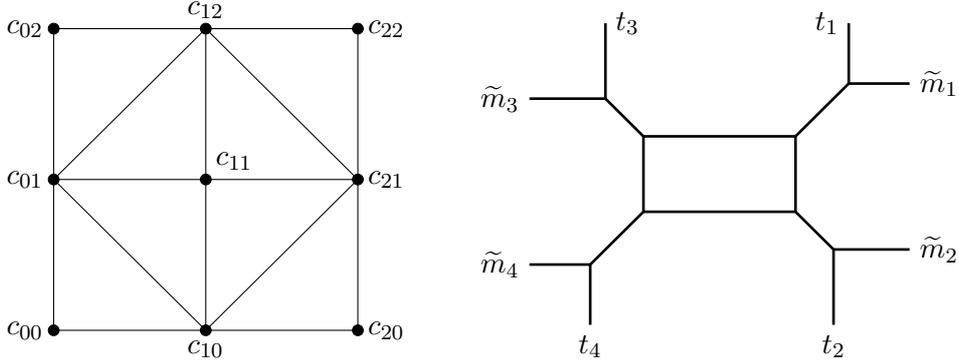
\begin{figure}[H]
\centering
%\begin{adjustbox}{width=0.2\textwidth}
\begin{tikzpicture}
  [inner sep=0.5mm, %line width=1.5pt,
 dot/.style={fill=black,draw,circle,minimum size=1pt},
 whitedot/.style={fill=white,draw,circle,minimum size=1pt},
 mark size=10pt, mark options={fill=white} ]
    \node[dot] at (0,0) [label={[label distance=0.03cm]45:$c_{11}$}] {};
    \node[dot] at (2,0) [label=right:$c_{21}$] {};
    \node[dot] at (0,2) [label=above:$c_{12}$] {};
    \node[dot] at (-2,0) [label=left:$c_{01}$]{};
    \node[dot] at (0,-2) [label=below:$c_{10}$] {};
    \node[dot] at (2,2) [label=right:$c_{22}$] {};
    \node[dot] at (2,-2) [label=right:$c_{20}$] {};
    \node[dot] at (-2,2) [label=left:$c_{02}$] {};
    \node[dot] at (-2,-2) [label=left:$c_{00}$] {};

    \draw (2,0) -- (0,2) -- (-2,0) -- (0,-2) -- (2,0);
	\draw (2,0) -- (-2,0);
	\draw (0,2) -- (0,-2);
	\draw (0,2) -- (2,2) -- (2,0);
	\draw (2,0) -- (2,-2) -- (0, -2);	
	\draw (0,2) -- (-2,2) -- (-2,0);
	\draw (-2,0) -- (-2,-2) -- (0, -2);	
 \end{tikzpicture}
%\end{adjustbox}
\quad\quad
%\begin{adjustbox}{width=0.2\textwidth}
\begin{tikzpicture}
  [inner sep=0.5mm, line width=1.0pt,
 dot/.style={fill=black,draw,circle,minimum size=1pt},
 whitedot/.style={fill=white,draw,circle,minimum size=1pt},
 mark size=10pt, mark options={fill=white} ]
 	\draw (1.0,-0.5)  -- (1.0,0.5) -- (-1.0, 0.5) -- (-1.0, -0.5) -- (1.0, -0.5);
	\draw (1.0,0.5) -- (1.7, 1.2) -- (1.7,2); \draw (1.7,1.2) -- (2.5, 1.2);
    \draw (-1.0,0.5) -- (-1.5,1) -- (-2.5,1);    \draw (-1.5,1) -- (-1.5,2);
    \draw (-1.0,-0.5) -- (-1.7, -1.2) -- (-1.7,-2); \draw (-1.7,-1.2) -- (-2.5, -1.2);
    \draw (1.0,-0.5) -- (1.5,-1) -- (2.5,-1);    \draw (1.5,-1) -- (1.5,-2);
	\node[] at (2.5,1.2) [label=right:$\widetilde m_1$]{};
	\node[] at (1.7,2) [label=left:$t_1$]{};
 	\node[] at (-2.5,1) [label=left:$\widetilde m_3$]{};
 	\node[] at (-1.5,2) [label=right:$t_3$]{};
	\node[] at (1.5,-2) [label=below:$t_2$]{};
	\node[] at (2.5,-1) [label=right:$\widetilde m_2$]{};
	\node[] at (-2.5,-1.2) [label=left:$\widetilde m_4$]{};
	\node[] at (-1.7,-2) [label=below:$t_4$]{};
\end{tikzpicture}
%\end{adjustbox}
\caption{Toric web diagram for $N_f=4$ of $E_5$ flavor symmetry}\label{nf4toric}
\end{figure}

For $N_f=4$, we start with
\begin{align}
c_{01}\,w+c_{10}\,t+ c_{11}\,tw + c_{12}\,tw^2 + c_{21}\,t^2w+ c_{22}\,t^2w^2+ c_{20}\,t^2+c_{02}w^2 + c_{00}=0.
\end{align}
The boundary conditions are 
\begin{align}
&|w|~{\rm large}:&&~c_{02}w^2+c_{12}tw^2+ c_{22}t^2w^2 = c_{22}w^2 (t-t_1)(t-t_3),\cr
&|w|~{\rm small}: &&~ c_{20}t^2 + c_{10}t + c_{00} = c_{20}(t-t_2)(t-t_4),\cr
&|t|~{\rm large}: &&~ c_{20}t^2 + c_{21}t^2 w+ c_{22}t^2w^2 = c_{22}t^2(w-\widetilde m_1)(w-\widetilde m_2),\cr
&|t|~{\rm small}: &&~ c_{02}w^2 + c_{01}w + c_{00} = c_{02}(w-\widetilde m_3)(w-\widetilde m_4).
\end{align}
The compatibility condition is given by
\begin{align}\label{conditionsamongcoef}
\widetilde m_1\, t_1^{-1} \,\, \widetilde m_2\, t_2 = \widetilde m_3\,t_3\,\,\widetilde m_4\, t_4^{-1}.
\end{align}
For three rescaling degrees of freedom, we choose
\begin{align}
c_{01}=c_{21}=1,\qquad c_{10}=c_{12},
\end{align}
or equivalently,
\begin{align}\label{condforNf4}
\frac{t_1+t_3}{t_2+t_4}=  \widetilde m_1 \widetilde m_2,\qquad
t_1 t_3 = \frac{\widetilde m_1+ \widetilde m_2}{\widetilde m_3+ \widetilde m_4}.
\end{align}
The SW curve for $N_f=4$ is then
\begin{align}\label{SWforNf=4}
 &-\frac{1}{\widetilde m_3+\widetilde m_4}\big(w-\widetilde m_3\big)\big(w-\widetilde m_4\big)+\,t\,\Big(\frac{t_1}{\widetilde m_1+\widetilde m_2} +\frac{1}{t_1(\widetilde m_3+\widetilde m_4)}\Big)\big(  w^2+ U w + 1\big) \cr
 &-  t^2 \frac{1}{\widetilde m_1+\widetilde m_2}\, \big(w - \widetilde m_1 \big)\big(w- \widetilde m_2\big)=0.
\end{align}
In this case, there is no dynamical scale but one can define the gauge coupling as geometric average of $t_i$
\begin{align}\label{t3value}
\Big(\frac{t_1t_2}{t_3t_4}\Big)^{\frac12}=q^{-1}.
\end{align}
It follows from \eqref{conditionsamongcoef}, \eqref{condforNf4}, and \eqref{t3value} that $t_1$ are expressed in terms of masses and the gauge coupling as
\begin{align}\label{t1S}
t_1=
 \Big(\frac{\widetilde m_1+\widetilde m_2}{\widetilde m_3+\widetilde m_4} \Big)^{\frac12}\cdot\Big(\frac{q^{-1}S^\frac12-1}{1-qS^\frac12 }\Big)^\frac12
, \quad
S\equiv \widetilde m_1\widetilde m_2\widetilde m_3\widetilde m_4.
\end{align}
The SW curve is also written in a symmetric way as
\begin{align}\label{SWt1t3-1}
 t^2 \, \big(w - \widetilde m_1 \big)\big(w- \widetilde m_2\big)-\,t\,(t_1+t_3)\big(  w^2+ U w + 1\big)+t_1t_3\big(w-\widetilde m_3\big)\big(w-\widetilde m_4\big)=0,
 \end{align}
 with
\begin{align}\label{SWt1t3-2}
t_1+t_3 = \Big(\frac{\widetilde m_1+\widetilde m_2}{\widetilde m_3+\widetilde m_4} \Big)^{\frac12}
q^{-\frac12} S^\frac14 \bigg(\frac{S^\frac12 - q S}{S^\frac12 -q} \bigg)^{\frac12} \bigg(1+ q\frac{S^\frac12 -q}{1- q S^\frac12}\bigg).
\end{align}

In the mass decoupling limit where $\widetilde m_4\to 0$ and $t_4\to 0$, while $\widetilde m_4\,t_4^{-1}=\ell_4$ fixed,
the SW curve for $N_f=4$, \eqref{SWforNf=4}, becomes naturally the SW curve for $N_f=3$, \eqref{SWforNf=3}, and the condition that we used, \eqref{condforNf4}, becomes the condition for $N_f=3$, \eqref{condforNf3}. We note that in this mass decoupling limit, dynamical scales for $N_f=4$ and $N_f=3$ are related as
\begin{align}
q^{-1}\widetilde m_4^\frac12 =(2\pi R \Lambda_3)^{-1},
\end{align}
which is consistent with that \eqref{t1S} for $N_f=4$ turns into that of $N_f=3$, \eqref{t1valuenf3}. We note that all the parameters including dynamical scales except for $q$ are not physical. This is due to our choice of three rescaling degrees of freedom. The choice,
$c_{01}=c_{21}=1$ and $c_{10}=c_{12}$, we have chosen, makes mass decoupling limit easier.
We comment that the SW curve with unphysical parameters can be related to the SW curve with physical ones by coordinate transformation below.
\\

\noindent\underline{\bf SW curve with physical masses}\\
Different choices for three rescaling degrees of freedom lead to a differently looking curve which is related by coordinate transformation. We consider a choice that the position of the exponential of masses $\widetilde m_i$ is measured from the center of the Coulomb branch. With $c_{22}=1=t_4$ in Figure \ref{nf4toric}, the SW curve is given by

\begin{align}\label{SWphysnf4}
 &{\sf t}^2 \, \big({\sf w} - \widetilde m'_1 \big)\big({\sf w}- \widetilde m'_2\big)-\,{\sf t}\,\Big( \widetilde m'_1\widetilde m'_2(1+qS'{}^{-\frac12}) {\sf w}^2+ U' {\sf w} + (1+qS'{}^{\frac12})\widetilde m'_1\widetilde m'_2\Big)\cr
 &+qS'{}^{-\frac12}\widetilde m'_1{}^2\widetilde m'_2{}^2\big({\sf w}-\widetilde m'_3\big)\big({\sf w}-\widetilde m'_4\big)=0,
 \end{align}
 where $S'\equiv\widetilde m'_1\widetilde m'_2\widetilde m'_3\widetilde m'_4$.
In the 4d limit, by expanding ${\sf w}\equiv e^{-\beta \sf v}$ and $\widetilde m'_i\equiv e^{-\beta \sf m_i}$ with respect to $\beta$ while keeping ${\sf t}$ and $q$ as they are, one finds \cite{Eguchi:2009gf,Bao:2013wqa}
 \begin{align}\label{4dlimitsu2nf4standard}
 {\sf t}^2({\sf v}-{\sf m_1})({\sf v}-{\sf m_2})+{\sf t}\Big(-(1+q){\sf v}^2 +q \,{\sf v} \sum^4_{{\sf i}=1}{\sf m_i} + u\Big) + q({\sf v}-{\sf m_3})({\sf v}-{\sf m_4})=0,
  \end{align}
where the masses $\sf m_i$ are physical masses. 
The coordinate transformation that connects \eqref{SWphysnf4} to \eqref{SWt1t3-1} is as follows:
\begin{align}
{\sf w} &\to  T w,\qquad
{\widetilde m_i'} \to T \widetilde m_i,\qquad T=\bigg(\frac{q- S^\frac12}{qS - S^{\frac12}}\bigg)^\frac12,
\cr
{\sf t} &\to \sqrt{ q  S^{-\frac12}T^2 \widetilde m_1^2\widetilde m_2^2\Big(\frac{\widetilde m_3+\widetilde m_4}{\widetilde m_1+\widetilde m_2}\Big)}\, t  ,
\end{align}
where $S$ is the product of masses given in \eqref{t1S}.

\subsection{Higher rank curve}
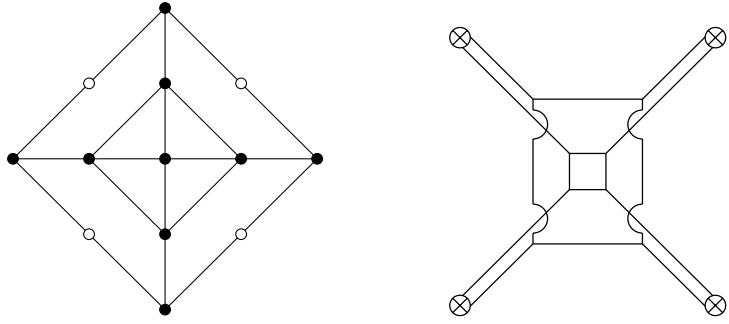
\begin{figure}[H]
\centering
\begin{tikzpicture}
  [inner sep=0.5mm,
 b/.style={fill=black,draw,circle,minimum size=1pt},
 w/.style={fill=white,draw,circle,minimum size=1pt},
 g/.style={fill=gray,draw,circle,minimum size=1pt},
 mark size=8pt, mark options={fill=white}]
%    \draw[help lines] (0,0) grid (6,3);
    \draw (2,0) -- (0,2) -- (-2,0) -- (0,-2) -- (2,0);
    \draw (2,0) -- (-2,0);
    \draw (0,2) -- (0,-2);
    \draw (1,0) -- (0,1) -- (-1,0) -- (0,-1) -- (1,0);

    \node[b] at (0,0) {};
	\node[b] at (0,1) {};
	\node[b] at (0,2) {};
	\node[b] at (0,-1) {};
	\node[b] at (0,-2 ) {};	
	\node[b] at (1,0) {};
	\node[b] at (2,0) {};
	\node[b] at (-1,0) {};
	\node[b] at (-2,0 ) {};
	
	\node[w] at (1,1 ) {};
	\node[w] at (-1,1 ) {};
	\node[w] at (-1,-1 ) {};
	\node[w] at (1,-1 ) {};

   \end{tikzpicture}
   \qquad \qquad
\begin{adjustbox}{width=0.24\textwidth}
\begin{tikzpicture}
  [inner sep=0.5mm,line width=1.0pt,
 b/.style={fill=black,draw,circle,minimum size=1pt},
 w/.style={fill=white,draw,circle,minimum size=1pt},
 g/.style={fill=gray,draw,circle,minimum size=1pt},
 mark size=8pt, mark options={fill=white}]
%    \draw[help lines] (0,0) grid (6,3);
    \draw (0.5,0.5) -- (-0.5,0.5)  -- (-0.5,-0.5) -- (0.5, -0.5) -- (0.5, 0.5);
    \draw (0.5,0.5) -- (3.5,3.5) ;
    \draw (-0.5,0.5) -- (-3.5,3.5) ;
    \draw (-0.5,-0.5) -- (-3.5,-3.5) ;
    \draw (0.5,-0.5) -- (3.5,-3.5) ;

    \draw (1.5,1.7) -- (1.5,2) -- (-1.5,2) ;
    \draw (1.5,1.7) arc (90:270:0.4cm);
    \draw (1.5,0.9) -- (1.5,-0.9);
    \draw (1.5,-1.7) -- (1.5,-2) -- (-1.5,-2) ;
    \draw (1.5,-0.9) arc (90:270:0.4cm);

    \draw (-1.5,1.7) -- (-1.5,2) -- (-1.5,2) ;
    \draw (-1.5,1.7) arc (90:-90:0.4cm);
    \draw (-1.5,0.9) -- (-1.5,-0.9);
    \draw (-1.5,-1.7) -- (-1.5,-2) -- (-1.5,-2) ;
    \draw (-1.5,-0.9) arc (90:-90:0.4cm);

	\draw (1.5,2) -- (3.25,3.75) ;
	\draw (-1.5,2) -- (-3.25,3.75) ;
	\draw (-1.5,-2) -- (-3.25,-3.75) ;
	\draw (1.5,-2) -- (3.25,-3.75) ;

	\foreach \plm in {otimes*} \draw plot[mark=\plm] coordinates {(3.5,3.7)} ;
	\foreach \plm in {otimes*} \draw plot[mark=\plm] coordinates {(-3.5,3.7)} ;
	\foreach \plm in {otimes*} \draw plot[mark=\plm] coordinates {(-3.5,-3.7)} ;
	\foreach \plm in {otimes*} \draw plot[mark=\plm] coordinates {(3.5,-3.7)} ;

   \end{tikzpicture}
\end{adjustbox}
\caption{
Toric-like diagram for rank-2 $E_1$ theory and corresponding web diagram.
}\label{fig:purerank2}
\end{figure}

The toric-like diagram for higher rank-$N$ with $N_f$ flavors can be obtained in the same way as explained section \ref{sec:rankN}. 
The corresponding SW curve is again factorized as the product of that of rank-1. As an example, we show the toric-like diagram for rank-2 $E_1$ theory ($N_f=0$) and the corresponding $(p,q)$ web in Figure \ref{fig:purerank2}.

%%%%%%%%%%%%%%%%%%%%%%
\section{The Seiberg-Witten curves from $E_8$ to $E_0$}\label{App:E8}
We here list the result of decomposition of the characters of $E_n$ into $E_{n-1}\times U(1)$ and then factoring out the $U(1)$ part, introduced in \cite{Eguchi:2002nx}:  \\
(i) Rescale the all variables
\begin{align}\label{scalingL}
&(u, x, y) \to (Lu,L^2x, L^3y),\qquad \\
&(\chi_{\mu_1}, \chi_{\mu_2}, \chi_{\mu_3}, \chi_{\mu_4}, \chi_{\mu_5}, \chi_{\mu_6}, \chi_{\mu_7}, \chi_{\mu_8})\to (L^2\chi_{\mu_1}, L^3\chi_{\mu_2}, L^4\chi_{\mu_3}, L^6\chi_{\mu_4}, L^5\chi_{\mu_5}, L^4\chi_{\mu_6}, L^3\chi_{\mu_7}, L^2\chi_{\mu_8}),\nn
\end{align}
where the power of $L$ in the character is the {\it marks} of $E_8$.\\
(ii) Set $\chi_{\mu_8}$ to 1 when reducing from $E_8$ to $E_7$, and take $L\to \infty$ limit, and \\
(iii) When reducing from $E_7$ to $E_6$, likewise the corresponding character to be unity while keeping the scaling.

The $E_n$ manifest curve is of the standard Weierstrass form~\cite{Minahan:1997ch, Eguchi:2002nx,Huang:2013yta}
\begin{align}
y^2 ~=& ~4 x^3 -g_2^{E_n} x -g_3^{E_n},
\end{align}
where
$g_2$ and $g_3$ are given according to $E_n$ as follows.\\
\ni \underline{For $E_8$:}
{\small{
\begin{align}\label{Huange8}
g_2^{E_8}=&~\frac{1}{12}u^4 - \Big(\frac{2}{3}\chi_1^{E_8}-\frac{50}{3}\chi_8^{E_8}+1550\Big)u^2
- \Big(-70 \chi_1^{E_8}+2 \chi_2^{E_8}-12 \chi_7^{E_8} +1840 {\chi_8^{E_8}}-115010\Big)u\cr
  & +\frac{4} {3}{\chi_1^{E_8}}{\chi_1^{E_8}}-\frac{8 }{3}{\chi_1^{E_8}}
   {\chi_8^{E_8}}-1824 {\chi_1^{E_8}}+112
   {\chi_2^{E_8}}-4 {\chi_3^{E_8}}+4 {\chi_6^{E_8}}\cr
    &-680 \chi_7^{E_8}+\frac{28}{3}\chi_8^{E_8} \chi_8^{E_8} +50744 \chi_8^{E_8}-2399276, \cr
g_3^{E_8}=
    &~ \frac{1}{216} u^6 - 4u^5 - \Big(\frac{1}{18}\chi_1^{E_8}+\frac{47}{18}\chi_8^{E_8}-\frac{5177}{6} \Big)u^4 \cr
&
    - \Big(-\frac{107}{6}\chi_1^{E_8}+\frac{1}{6}\chi_2^{E_8}+3\chi_7^{E_8} -\frac{1580}{3}\chi_8^{E_8}+\frac{504215}{6}\Big)u^3 \cr
&
   - \Big(-\frac{2}{9}\chi_1^{E_8}\chi_1^{E_8}-\frac{20}{9}\chi_1^{E_8} \chi_8^{E_8}+\frac{5866}{3}\chi_1^{E_8}-\frac{112}{3}\chi_2^{E_8}+\frac{1}{3} \chi_3^{E_8} \cr
   &+\frac
   {11}{3}\chi_6^{E_8}-\frac{1450}{3}\chi_7^{E_8}+\frac{196}{9}\chi_8^{E_8}\chi_8^{E_8}+39296\chi_8^{E_8}-\frac{12673792}{3} \Big)u^2\cr
&
    - \Big(\frac{94}{3}\chi_1^{E_8}\chi_1^{E_8}-\frac{2 }{3}\chi_1^{E_8}\chi_2^{E_8}+\frac{718}{3}\chi_1^{E_8}\chi_8^{E_8}-\frac{270736}{3}\chi_1^{E_8}-\frac{10 }{3}\chi_2^{E_8}\chi_8^{E_8}+2630 \chi_2^{E_8}-52\chi_3^{E_8}+4 \chi_5^{E_8} \cr
   & -416 \chi_6^{E_8}+16
   \chi_7^{E_8} \chi_8^{E_8}+25880
   \chi_7^{E_8}-\frac{7328
   }{3}\chi_8^{E_8} \chi_8^{E_8} -\frac{3841382
   }{3}\chi_8^{E_8}+107263286 \Big)u \cr
   &- \frac{8}{27}\chi_1^{E_8}\chi_1^{E_8}\chi_1^{E_8}-\frac{28
   }{9}\chi_1^{E_8}\chi_1^{E_8} \chi_8^{E_8}+1065
   \chi_1^{E_8}\chi_1^{E_8}-\frac{118 }{3}\chi_1^{E_8}
   \chi_2^{E_8}+\frac{4 }{3}\chi_1^{E_8}
   \chi_3^{E_8}-\frac{4 }{3}\chi_1^{E_8}\chi_6^{E_8} \cr
   &
   +\frac{8 }{3}\chi_1^{E_8}
   \chi_7^{E_8}+\frac{40}{9}\chi_1^{E_8}
   \chi_8^{E_8}\chi_8^{E_8}+\frac{19264 }{3}\chi_1^{E_8}
   \chi_8^{E_8}-\frac{4521802
   }{3}\chi_1^{E_8}+\chi_2^{E_8}\chi_2^{E_8}-\frac{572}{3}\chi_2^{E_8} \chi_8^{E_8} \cr
   &  +59482
   \chi_2^{E_8}+\frac{20 }{3}\chi_3^{E_8}
   \chi_8^{E_8}-1880 \chi_3^{E_8}-4
   \chi_4^{E_8}+232 \chi_5^{E_8}-\frac{8
   }{3}\chi_6^{E_8} \chi_8^{E_8}-11808
   \chi_6^{E_8}+\frac{2740 }{3}\chi_7^{E_8}
   \chi_8^{E_8}\cr
   &  +460388
   \chi_7^{E_8}-\frac{136}{27}\chi_8^{E_8}\chi_8^{E_8}\chi_8^{E_8}-\frac{205492
   }{3}\chi_8^{E_8}\chi_8^{E_8} -\frac{45856940}{3}\chi_8^{E_8}+1091057493  .
\end{align}
}}
\ni \underline{For $E_7$:}
{\small{
\begin{align}\label{Huange7}
g_2^{E_7}=&~\frac{1}{12}u^4 - \big(\frac{2}{3}\chi_1^{E_7}-\frac{50}{3}\big)u^2
- \big(2 \chi_2^{E_7}-12 \chi_7^{E_7} \big)u
+\frac{4} {3}\chi_1^{E_7}\chi_1^{E_7}-\frac83\chi_1^{E_7} -4 {\chi_3^{E_7}}+4 {\chi_6^{E_7}}+ \frac{28}{3}, \cr
g_3^{E_7}=
    &~ \frac{1}{216} u^6  -\big(\frac{1}{18}\chi_1^{E_7}+\frac{47}{18}\big)u^4
    - \big(\frac{1}{6}\chi_2^{E_7}+3\chi_7^{E_7} \big)u^3
    + \big(\frac{2}{9}\chi_1^{E_7}\chi_1^{E_7} +\frac{20}{9}\chi_1^{E_7}  -\frac{1}{3} \chi_3^{E_7}
       -\frac{11}{3}\chi_6^{E_7}-\frac{196}{9}\big)u^2\cr
&
    + \big(\frac{2}{3}\chi_1^{E_7}\chi_2^{E_7} +\frac{10}{3}\chi_2^{E_7}-4 \chi_5^{E_7} - 16\chi_7^{E_7}\big)u
 - \frac{8}{27}\chi_1^{E_7}\chi_1^{E_7}\chi_1^{E_7} -\frac{28}{9}\chi_1^{E_7}\chi_1^{E_7}
   +\frac{4}{3}\chi_1^{E_7}\chi_3^{E_7}\cr
&
   -\frac{4 }{3}\chi_1^{E_7}\chi_6^{E_7}
   +\frac{40}{9}\chi_1^{E_7}+\chi_2^{E_7}\chi_2^{E_7}+\frac{20}{3}\chi_3^{E_7}
   -4\chi_4^{E_7} -\frac83\chi_6^{E_7} -\frac{136}{27}
   .
\end{align}
}}
\ni \underline{For $E_6$:}
{\small{
\begin{align}\label{Huange6}
g_2^{E_6}=&~\frac{1}{12}u^4 - \frac{2}{3}\chi_1^{E_6}u^2
- \big(2 \chi_2^{E_6}-12  \big)u
+\frac{4} {3}\chi_1^{E_6}\chi_1^{E_6}-4 \chi_3^{E_6}+4 {\chi_6^{E_6}}, \cr
g_3^{E_6}=
    &~ \frac{1}{216} u^6  -\frac{1}{18}\chi_1^{E_6}u^4
    - \big(\frac{1}{6}\chi_2^{E_6}+3 \big)u^3
    + \big(\frac{2}{9}\chi_1^{E_6}\chi_1^{E_6}  -\frac{1}{3} \chi_3^{E_6}
       -\frac{11}{3}\chi_6^{E_6}\big)u^2 + \big(\frac{2}{3}\chi_1^{E_6}\chi_2^{E_6} -4 \chi_5^{E_6} \big)u\cr
&
 - \frac{8}{27}\chi_1^{E_6}\chi_1^{E_6}\chi_1^{E_6}
   +\frac{4}{3}\chi_1^{E_6}\chi_3^{E_6}
   -\frac{4 }{3}\chi_1^{E_6}\chi_6^{E_6}
 +\chi_2^{E_6}\chi_2^{E_6}
   -4\chi_4^{E_6}
   .
\end{align}
}}
\ni \underline{For $E_5=SO(10)$:}
{\small{
\begin{align}\label{Huange5}
g_2^{E_5}=&~\frac{1}{12}u^4 - \frac{2}{3}\chi_1^{E_5}u^2
- 2\chi_2^{E_5}u
+\frac{4} {3}\chi_1^{E_5}\chi_1^{E_5}-4 \chi_3^{E_5}+4 {}, \cr
g_3^{E_5}=
    &~ \frac{1}{216} u^6  -\frac{1}{18}\chi_1^{E_5}u^4
    - \frac{1}{6}\chi_2^{E_5}u^3
    + \big(\frac{2}{9}\chi_1^{E_5}\chi_1^{E_5}  -\frac{1}{3} \chi_3^{E_5}
       -\frac{11}{3}\big)u^2   + \big(\frac{2}{3}\chi_1^{E_5}\chi_2^{E_5} -4 \chi_5^{E_5} \big)u \cr
&
 - \frac{8}{27}\chi_1^{E_5}\chi_1^{E_5}\chi_1^{E_5}
   +\frac{4}{3}\chi_1^{E_5}\chi_3^{E_5}
   -\frac{4 }{3}\chi_1^{E_5}
 +\chi_2^{E_5}\chi_2^{E_5}
   -4\chi_4^{E_5}
   .
\end{align}
}}
\ni \underline{For $E_4=SU(5)$:}
{\small{
\begin{align}\label{Huange4}
g_2^{E_4}=&~\frac{1}{12}u^4 - \frac{2}{3}\chi_1^{E_4}u^2
- 2\chi_2^{E_4}u
+\frac{4} {3}\chi_1^{E_4}\chi_1^{E_4}-4 \chi_3^{E_4}, \cr
g_3^{E_4}=
    &~ \frac{1}{216} u^6  -\frac{1}{18}\chi_1^{E_4}u^4
    - \frac{1}{6}\chi_2^{E_4}u^3
    + \big(\frac{2}{9}\chi_1^{E_4}\chi_1^{E_4}  -\frac{1}{3} \chi_3^{E_4}
       \big)u^2   + \big(\frac{2}{3}\chi_1^{E_4}\chi_2^{E_4} -4  \big)u \cr
&
 - \frac{8}{27}\chi_1^{E_4}\chi_1^{E_4}\chi_1^{E_4}
   +\frac{4}{3}\chi_1^{E_4}\chi_3^{E_4}
 +\chi_2^{E_4}\chi_2^{E_4}
   -4\chi_4^{E_4}
   .
\end{align}
}}
\ni \underline{For $E_3=SU(3)\times SU(2)$:}
{\small{
\begin{align}\label{Huange3}
g_2^{E_3}=&~\frac{1}{12}u^4 - \frac{2}{3}\chi_1^{E_3}u^2
- 2\chi_2^{E_3}u
+\frac{4} {3}\chi_1^{E_3}\chi_1^{E_3}-4 \chi_3^{E_3}, \cr
g_3^{E_3}=
    &~ \frac{1}{216} u^6  -\frac{1}{18}\chi_1^{E_3}u^4
    - \frac{1}{6}\chi_2^{E_3}u^3
    + \big(\frac{2}{9}\chi_1^{E_3}\chi_1^{E_3}  -\frac{1}{3} \chi_3^{E_3}
       \big)u^2   + \frac{2}{3}\chi_1^{E_3}\chi_2^{E_3}u \cr
&
 - \frac{8}{27}\chi_1^{E_3}\chi_1^{E_3}\chi_1^{E_3}
   +\frac{4}{3}\chi_1^{E_3}\chi_3^{E_3}
 +\chi_2^{E_3}\chi_2^{E_3}
   -4.
\end{align}
}}
\ni \underline{For $E_2=SU(2)\times U(1)$:}
{\small{
\begin{align}\label{Huange2}
g_2^{E_2}=&~\frac{1}{12}u^4 - \frac{2}{3}\chi_1^{E_2}u^2
- 2\chi_2^{E_2}u
+\frac{4} {3}\chi_1^{E_2}\chi_1^{E_2}-4 , \\
g_3^{E_2}=
    &~ \frac{1}{216} u^6  -\frac{1}{18}\chi_1^{E_2}u^4
    - \frac{1}{6}\chi_2^{E_2}u^3
    + \big(\frac{2}{9}\chi_1^{E_2}\chi_1^{E_2}  -\frac{1}{3}
       \big)u^2   + \frac{2}{3}\chi_1^{E_2}\chi_2^{E_2}u
  - \frac{8}{27}\chi_1^{E_2}\chi_1^{E_2}\chi_1^{E_2}
   +\frac{4}{3}\chi_1^{E_2}
 +\chi_2^{E_2}\chi_2^{E_2}.\nn
\end{align}
}}
\ni \underline{For $\widetilde{E}_1=U(1)$: %(by following the decoupling procedure (i), (ii), and (iii) from $E_2$)
}
{\small{
\begin{align}\label{Huange1t}
g_2^{\widetilde{E}_1}=&~\frac{1}{12}u^4 - \frac{2}{3}u^2
- 2\chi_2^{\widetilde{E}_1}u
+\frac{4} {3} , \cr
g_3^{\widetilde{E}_1}=
    &~ \frac{1}{216} u^6  -\frac{1}{18}u^4
    - \frac{1}{6}\chi_2^{\widetilde{E}_1}u^3
    +\frac{2}{9} u^2   + \frac{2}{3}\chi_2^{\widetilde{E}_1}u
   -\frac{8}{27}
 +\chi_2^{\widetilde{E}_1}\chi_2^{\widetilde{E}_1}.
\end{align}
}}
\ni \underline{For $E_1=SU(2)$: %(only by setting $\chi_2\to 0$ from $E_2$)
}
{\small{
\begin{align}\label{Huange1}
g_2^{E_1}=&~\frac{1}{12}u^4 - \frac{2}{3}\chi_1^{E_1}u^2
+\frac{4} {3}\chi_1^{E_1}\chi_1^{E_1} -4, \\
g_3^{E_1}=
    &~ \frac{1}{216} u^6  -\frac{1}{18}\chi_1^{E_1}u^4
    + \big(\frac{2}{9}\chi_1^{E_1}\chi_1^{E_1}  -\frac{1}{3}
       \big)u^2   
  - \frac{8}{27}\chi_1^{E_1}\chi_1^{E_1}\chi_1^{E_1}
   +\frac{4}{3}\chi_1^{E_1}.\nn
\end{align}
}}
\noindent Here we note that the $E_1$ curve is obtained only by setting $\chi_2\to 0$ from the $E_2$ curve, which does not involve the scaling.\\

\ni \underline{For $E_0$: (from $\widetilde{E}_1$)}
{\small{
\begin{align}\label{Huange0}
g_2^{E_0}=&~\frac{1}{12}u^4
- 2u
, \qquad
g_3^{E_0}=
    ~ \frac{1}{216} u^6
    - \frac{1}{6}u^3
    %+ \frac{2}{3}u
 +1.
\end{align}
}}
In summary, one stars from $E_8$ curve and obtains lower $E_n$ curves:
\begin{align}
E_8\to E_7 \to E_6 \to E_5 \to E_4 \to E_3\to E_2&\to \widetilde E_1 \to E_0\cr
&\to E_1.
\end{align}
All the holomorphic SW one form is of the standard form:
\begin{align}
\omega_{SW} = \frac{dx}{y}.
\end{align}

%%%%%%%%%%%%%%%%%%%%%%%%%%%%%%%%%%%%
\section{The $j$-invariant}\label{app:jinv}
The Weierstrass form for elliptic curve is given
\begin{align}\label{Weierstrassform}
y^2= 4 z^3 -g_2 z -g_3,
\end{align}
then the $j$-invariant which is $SL(2,\mathbb{Z})$ invariant is define by
\begin{align}
j(\tau) =\frac{g_2^3}{g_2^3 -27g_3^2},
\end{align}
where the denominator is proportional to the discriminant of the Weierstrass form.
For \eqref{Weierstrassform}, $\Delta=16(g^3_2 - 27 g^2_3)$, and thus the denominator of the $j$-invariant is $\frac{\Delta}{16}$.

For an elliptic curve given by
\begin{align}\label{ellipcubic}
y^2=Ax^3+Bx^2+Cx+D,
\end{align}
let us find how the $j$-invariant is given. It is straightforward to rewrite \eqref{ellipcubic} into the standard Weierstrass form
\begin{align}
{\tilde y}^2= 4 {\tilde x}^3 -\frac{4}{3A^2} \big({B^2}-3AC\big){\tilde x} -\frac{4}{27 A^3}\big(
9ABC-2B^3-27A^2D\big),
\end{align}
where $\tilde{y}=\frac{y}{\sqrt{A}/2}$, $\tilde{x}= x+\frac{B}{3A}$, yielding
\begin{align}
g_2&=\frac{4}{3A^2} \big({B^2}-3AC\big),\quad
g_3= \frac{4}{27 A^3}\big( 9ABC-2B^3-27A^2D\big).
\end{align}
As the discriminant $\Delta$ for a cubic equation $Ax^3+Bx^2+Cx+D=0$ is given by
\begin{align}
\Delta = B^2C^2-4AC^3-4B^3D -27A^2D^2+18ABCD,
\end{align}
one finds that
\begin{align}
g_2^3-27g_3^2=\frac{16}{A^4} \Delta.
\end{align}

An elliptic curve may also be expressed as a quartic polynomial
\begin{align}
y^2=ax^4+bx^3+cx^2+dx +e.
\end{align}
The forms of $g_2$ and $g_3$ for the curve are given by
\begin{align}
g_2&= \frac{4}{a^3}(c^2-3bd+12ae),\qquad
g_2^3- 27 g^2_3=\frac{16}{a^6}\Delta,
\end{align}
where the discriminant for the quartic equation is given by
\begin{align*}
\Delta =& 256a^3e^3-192a^2bde^2-128a^2c^2e^2+144a^2cd^2e-27a^2d^4+144ab^2ce^2-6ab^2d^2e
-80abc^2de\cr
&+18abcd^3+16ac^4e-4ac^3d^2-27b^4e^2+18b^3cde-4b^3d^3-4b^2c^3e+b^2c^2d^2.
\end{align*}

\section{The Weierstrass from for $SU(8)$ manifest curve of $E_7$ theory}\label{g2gesu8}
Following \cite{Huang:2013yta}, we rewrite \eqref{su8manicurve2} into the standard Weierstrass form
\begin{align}
y^2= 4x^3- g_2 x-g_3,
\end{align}
where
\begin{align}
g_2=&-\frac43 \Big(-64 - 48 \chi_1 \chi_3 + 64 \chi_4 - 16 \chi_4^2 + 48 \chi_3 \chi_5 +
   48 \chi_1^2 \chi_6 - 192 \chi_2 \chi_6 + 16 \chi_1 \chi_7
   \cr
   &+
   16 \chi_1 \chi_4 \chi_7 - 48 \chi_5 \chi_7 - 16 \chi_1^2 \chi_7^2 +
   48 \chi_2 \chi_7^2 + 24 \chi_1^2 u - 192 \chi_2 u + 24 \chi_1 \chi_5 u
   \cr
   &
   -    192 \chi_6 u + 24 \chi_3 \chi_7 u + 24 \chi_7^2 u - 208 u^2 + 8 \chi_4 u^2 +
   8 \chi_1 \chi_7 u^2 - u^4\Big),
\end{align}
and
\begin{align}
g_3=&\frac8{27} \Big(-512 - 216 \chi_1^4 + 864 \chi_1^2 \chi_2 - 576 \chi_1 \chi_3 +
   768 \chi_4 + 288 \chi_1 \chi_3 \chi_4 - 384 \chi_4^2 
 \\
   &+ 64 \chi_4^3
   +
   432 \chi_1^3 \chi_5 - 1728 \chi_1 \chi_2 \chi_5 + 576 \chi_3 \chi_5 -
   288 \chi_3 \chi_4 \chi_5 - 216 \chi_1^2 \chi_5^2 + 864 \chi_2 \chi_5^2
    \cr
   & -
   1152 \chi_1^2 \chi_6 + 4608 \chi_2 \chi_6 + 864 \chi_3^2 \chi_6 +
   576 \chi_1^2 \chi_4 \chi_6 - 2304 \chi_2 \chi_4 \chi_6 + 192 \chi_1 \chi_7
    \cr
   & -
   144 \chi_1^2 \chi_3 \chi_7 + 96 \chi_1 \chi_4 \chi_7 - 96 \chi_1 \chi_4^2 \chi_7 -
   576 \chi_5 \chi_7 + 144 \chi_1 \chi_3 \chi_5 \chi_7 + 288 \chi_4 \chi_5 \chi_7
    \cr
   & -
   288 \chi_1^3 \chi_6 \chi_7 + 1152 \chi_1 \chi_2 \chi_6 \chi_7 -
   1728 \chi_3 \chi_6 \chi_7 + 336 \chi_1^2 \chi_7^2 - 1152 \chi_2 \chi_7^2 -
   216 \chi_3^2 \chi_7^2
 \cr
   &
   - 96 \chi_1^2 \chi_4 \chi_7^2 + 576 \chi_2 \chi_4 \chi_7^2 -
   144 \chi_1 \chi_5 \chi_7^2 + 864 \chi_6 \chi_7^2 + 64 \chi_1^3 \chi_7^3 -
   288 \chi_1 \chi_2 \chi_7^3
 \cr
   &
   + 432 \chi_3 \chi_7^3 - 216 \chi_7^4 -
   576 \chi_1^2 u + 4608 \chi_2 u + 864 \chi_3^2 u + 720 \chi_1^2 \chi_4 u -
   2304 \chi_2 \chi_4 u
 \cr
   &
   - 1440 \chi_1 \chi_5 u - 144 \chi_1 \chi_4 \chi_5 u +
   864 \chi_5^2 u + 4608 \chi_6 u + 864 \chi_1 \chi_3 \chi_6 u -
   2304 \chi_4 \chi_6 u
 \cr
   &
   - 144 \chi_1^3 \chi_7 u + 288 \chi_1 \chi_2 \chi_7 u -
   1440 \chi_3 \chi_7 u - 144 \chi_3 \chi_4 \chi_7 u - 144 \chi_1^2 \chi_5 \chi_7 u
 \cr
   &
   +
   864 \chi_2 \chi_5 \chi_7 u + 288 \chi_1 \chi_6 \chi_7 u - 576 \chi_7^2 u -
   144 \chi_1 \chi_3 \chi_7^2 u + 720 \chi_4 \chi_7^2 u - 144 \chi_1 \chi_7^3 u
 \cr
   &
   +
   4416 u^2 + 792 \chi_1 \chi_3 u^2 - 2112 \chi_4 u^2 - 48 \chi_4^2 u^2 +
   72 \chi_3 \chi_5 u^2 + 72 \chi_1^2 \chi_6 u^2 + 576 \chi_2 \chi_6 u^2
 \cr
   &
   -
   456 \chi_1 \chi_7 u^2 - 24 \chi_1 \chi_4 \chi_7 u^2 + 792 \chi_5 \chi_7 u^2 -
   48 \chi_1^2 \chi_7^2 u^2 + 72 \chi_2 \chi_7^2 u^2 + 36 \chi_1^2 u^3
 \cr
   &
   +
   576 \chi_2 u^3 + 36 \chi_1 \chi_5 u^3 + 576 \chi_6 u^3 +
   36 \chi_3 \chi_7 u^3 + 36 \chi_7^2 u^3 + 552 u^4 + 12 \chi_4 u^4    +
   12 \chi_1 \chi_7 u^4 - u^6\Big).\nn
\end{align}

\subsection{Decomposition of the $E_7$ characters into $SU(8)$}\label{che7tosu8}
We list decomposition of the characters $\chi^{E_7}_i\equiv\chi^{E_7}_{\mu_i}$ of $E_7$ fundamental weights into the characters $\chi_i$ of $SU(8)$ fundamental weights
\begin{flalign*}
E_7 ~{\rm Dynkin~ diagram}\qquad & \underset{\mathclap{\mu_1}}{\circ} -\!\!\!-
 \underset{\mathclap{\mu_3}}{\circ}-\!\!\!-
\underset{\mathclap{\mu_4}}{\overset{\overset{\textstyle\circ_{\mathrlap{\mu_2}}}{\textstyle\vert}}{\circ}} -\!\!\!-
\underset{\mathclap{\mu_5}}{\circ}-\!\!\!-
\underset{\mathclap{\mu_6}}{\circ} -\!\!\!-
\underset{\mathclap{\mu_7}}{\circ}
\end{flalign*}

{\small
\begin{align}
\chi^{E_7}_1 =~& -1 + \chi_1 \chi_7 +  \chi_4 \cr
\chi^{E_7}_2 =~& \chi_1^2+\chi_7^2+   \chi_3\chi_7+\chi_1\chi_5 -2\chi_2 -2\chi_6\cr
\chi^{E_7}_3 =~& 1 - 2 \chi_4 + \chi_3 \chi_5 + \chi_1^2 \chi_6 - 3 \chi_2 \chi_6 - \chi_1 \chi_7 +
 \chi_1 \chi_4 \chi_7 + \chi_2 \chi_7^2\cr
 \chi^{E_7}_4 =~&-2 + \chi_2^2 - \chi_1 \chi_3 + 2 \chi_4 - \chi_4^2 + \chi_1^3 \chi_5 -
 3 \chi_1 \chi_2 \chi_5 + 2 \chi_3 \chi_5 + \chi_2 \chi_5^2 - \chi_1^2 \chi_6 \cr
 &+
 3 \chi_2 \chi_6 + \chi_3^2 \chi_6 + \chi_1^2 \chi_4 \chi_6 - 4 \chi_2 \chi_4 \chi_6 -
 \chi_1 \chi_5 \chi_6 + \chi_6^2 + 2 \chi_1 \chi_7 - \chi_2 \chi_3 \chi_7 \cr
&- \chi_5 \chi_7 +
 \chi_1 \chi_3 \chi_5 \chi_7 - 3 \chi_3 \chi_6 \chi_7 - \chi_2 \chi_7^2 +
 \chi_2 \chi_4 \chi_7^2 + \chi_3 \chi_7^3 \cr
\chi^{E_7}_5 =~&\chi_3^2 + \chi_1^2 \chi_4 - 3 \chi_2 \chi_4 - \chi_1 \chi_5 + \chi_5^2 +
 \chi_1 \chi_3 \chi_6 - 3 \chi_4 \chi_6 - \chi_3 \chi_7 + \chi_2 \chi_5 \chi_7 +
 \chi_4 \chi_7^2\cr
 \chi^{E_7}_6 =~&-1 + \chi_1 \chi_3 - 2 \chi_4 + \chi_2 \chi_6 + \chi_5 \chi_7\cr
\chi^{E_7}_7 =~& \chi_2 +\chi_6.
\end{align}
}
%\newpage
\section{The Weierstrass from for $SU(9)$ manifest curve of $E_8$ theory}\label{g2gesu9}
Following \cite{Huang:2013yta}, we rewrite \eqref{su9weir} with $U\to u-60$ into the standard Weierstrass form
\begin{align}
y^2= 4x^3- g_2 x-g_3,
\end{align}
where
{\footnotesize{
\begin{align}
g_2=~&\frac{1}{972}
\bigg(16 (\chi_1^2 - 3 \chi_2 + 3 \chi_8)^2 (3 \chi_1 - 3 \chi_7 + \chi_8^2)^2 \cr
&
+   16 (3 \chi_1 - 3 \chi_7 + \chi_8^2) \Big[ 2 \chi_1^2 \chi_8
 +3 (3 \chi_4 - 3 \chi_7 - \chi_2 \chi_8 + \chi_8^2) +
      3 \chi_1 (-57 + u)\Big]^2\cr
      & -8 (\chi_1^2 - 3 \chi_2 + 3 \chi_8)
       (3 \chi_1 - 3 \chi_7 + \chi_8^2) \Big(2 \chi_1 \chi_8 + 3 (-54 + u)\Big)^2
      + (-162 + 2 \chi_1 \chi_8 + 3 u)^4
\cr
&
      -
   16 (\chi_1^2 - 3 \chi_2 + 3 \chi_8) \Big[ 2 \chi_1^2 \chi_8 +
      3 (3 \chi_4 - 3 \chi_7 - \chi_2 \chi_8 + \chi_8^2)+ 3 \chi_1 (-57 + u)\Big]\cr
&
\times(-1539 + 27 \chi_6 + 9 \chi_1 \chi_8 -
      9 \chi_7 \chi_8 + 2 \chi_8^3 + 27 u) \cr
&
-
   8 \Big[2 \chi_1^2 \chi_8 + 3 (3 \chi_4 - 3 \chi_7 - \chi_2 \chi_8 + \chi_8^2) +
      3 \chi_1 (-57 + u)) (2 \chi_1 \chi_8 + 3 (-54 + u)\Big]\cr
&\times (3 \chi_1^2 -9 \chi_2 + 9 \chi_5 - 3 \chi_1 \chi_7 - 171 \chi_8 + 2 \chi_1 \chi_8^2 +
      3 \chi_8 u)\cr
&
 +
   16 (\chi_1^2 - 3 \chi_2 + 3 \chi_8) (3 \chi_1^2 - 9 \chi_2 + 9 \chi_5 -
      3 \chi_1 \chi_7 - 171 \chi_8 + 2 \chi_1 \chi_8^2 + 3 \chi_8 u)^2 \cr
&
+
   8 \Big(2 \chi_1 \chi_8 + 3 (-54 + u)\Big) \Big[-1539 + 27 \chi_6 + 9 \chi_1 \chi_8 -
      9 \chi_7 \chi_8 + 2 \chi_8^3 + 27 u) \cr
&
      \times(2 \chi_1^3 -  9 \chi_1 (\chi_2 - \chi_8) + 27 (-57 + \chi_3 + u)\Big] -
   16 (3 \chi_1 - 3 \chi_7 + \chi_8^2) (3 \chi_1^2 - 9 \chi_2 + 9 \chi_5 \cr
&
-
      3 \chi_1 \chi_7 - 171 \chi_8 + 2 \chi_1 \chi_8^2 + 3 \chi_8 u) \Big[2 \chi_1^3 -
      9 \chi_1 (\chi_2 - \chi_8) + 27 (-57 + \chi_3 + u)\Big]\bigg),
      \end{align}
}}
and
{\footnotesize{
\begin{align}
g_3=&~\frac1{157464} \bigg(-64 (\chi_1^2 - 3 \chi_2 + 3 \chi_8)^3 (3 \chi_1 - 3 \chi_7 + \chi_8^2)^3 +
    192 (\chi_1^2 - 3 \chi_2 + 3 \chi_8)
\\&
    \times (3 \chi_1 - 3 \chi_7 +   \chi_8^2)^2 \Big[2 \chi_1^2 \chi_8 + 3 (3 \chi_4 - 3 \chi_7 - \chi_2 \chi_8 + \chi_8^2) +3 \chi_1 (-57 + u)\Big]^2
\cr&
     + 48 (\chi_1^2 - 3 \chi_2 + 3 \chi_8)^2 (3 \chi_1 - 3 \chi_7 + \chi_8^2)^2 (2 \chi_1 \chi_8 + 3 (-54 + u))^2
\cr&
     +24 (3 \chi_1 - 3 \chi_7 + \chi_8^2) \Big[2 \chi_1^2 \chi_8 +  3 (3 \chi_4 - 3 \chi_7 - \chi_2 \chi_8 + \chi_8^2) + 3 \chi_1 (-57 + u)\Big]^2
\cr&
    \times (2 \chi_1 \chi_8 + 3 (-54 + u))^2 -
12 (\chi_1^2 - 3 \chi_2 + 3 \chi_8) (3 \chi_1 - 3 \chi_7 + \chi_8^2)(-162 + 2 \chi_1 \chi_8 + 3 u)^4
\cr&
  + (-162 + 2 \chi_1 \chi_8 + 3 u)^6 -
  96 (\chi_1^2 - 3 \chi_2 + 3 \chi_8)^2
\cr&
  \times (3 \chi_1 - 3 \chi_7 + \chi_8^2) \Big[2 \chi_1^2 \chi_8 + 3 (3 \chi_4 - 3 \chi_7 - \chi_2 \chi_8 +
  \chi_8^2) + 3 \chi_1 (-57 + u)\Big]
\cr&
  \times(-1539 + 27 \chi_6 + 9 \chi_1 \chi_8 -  9 \chi_7 \chi_8 + 2 \chi_8^3 + 27 u)
\cr&
- 32 \Big[2 \chi_1^2 \chi_8  + 3 (3 \chi_4 - 3 \chi_7 - \chi_2 \chi_8 + \chi_8^2) + 3 \chi_1 (-57 + u)\Big]^3
\cr&
 \times (-1539 + 27 \chi_6 + 9 \chi_1 \chi_8 -  9 \chi_7 \chi_8 + 2 \chi_8^3 + 27 u) +
72 (\chi_1^2 - 3 \chi_2 + 3 \chi_8)
\cr&
\times\Big[ 2 \chi_1^2 \chi_8 + 3 (3 \chi_4 - 3 \chi_7 - \chi_2 \chi_8 + \chi_8^2) +
    3 \chi_1 (-57 + u)\Big] (2 \chi_1 \chi_8 + 3 (-54 + u))^2
\cr&
  \times (-1539 + 27 \chi_6 + 9 \chi_1 \chi_8 - 9 \chi_7 \chi_8 + 2 \chi_8^3 + 27 u) +
   32 (\chi_1^2 - 3 \chi_2 + 3 \chi_8)^3
\cr&
  \times(-1539 + 27 \chi_6 + 9 \chi_1 \chi_8 -  9 \chi_7 \chi_8 + 2 \chi_8^3 + 27 u)^2  -   240 (\chi_1^2 - 3 \chi_2 + 3 \chi_8)(3 \chi_1 - 3 \chi_7 +  \chi_8^2)\times \cr %\nn
% \end{align}
% }}
% {\footnotesize{
% \begin{align}
& \times \Big[2 \chi_1^2 \chi_8 +
 3 (3 \chi_4 - 3 \chi_7 - \chi_2 \chi_8 + \chi_8^2) +
  3 \chi_1 (-57 + u)\Big]
\cr&
  \times
  (2 \chi_1 \chi_8 + 3 (-54 + u))
   (3 \chi_1^2 - 9 \chi_2 + 9 \chi_5 - 3 \chi_1 \chi_7 - 171 \chi_8 + 2 \chi_1 \chi_8^2 + 3 \chi_8 u)
   \cr
&
   - 12 \Big[ 2 \chi_1^2 \chi_8    + 3 (3 \chi_4 - 3 \chi_7 - \chi_2 \chi_8 + \chi_8^2) +
      3 \chi_1 (-57 + u)\Big] (2 \chi_1 \chi_8 + 3 (-54 + u))^3
\cr&
  \times
      (3 \chi_1^2 -
      9 \chi_2 + 9 \chi_5 - 3 \chi_1 \chi_7 - 171 \chi_8 + 2 \chi_1 \chi_8^2 +
      3 \chi_8 u) -
   96 (\chi_1^2 - 3 \chi_2 + 3 \chi_8)^2
\cr&
  \times
  (2 \chi_1 \chi_8 +
      3 (-54 + u)) (-1539 + 27 \chi_6 + 9 \chi_1 \chi_8 - 9 \chi_7 \chi_8 +
      2 \chi_8^3 + 27 u)
\cr&
  \times
       (3 \chi_1^2 - 9 \chi_2 + 9 \chi_5 - 3 \chi_1 \chi_7 -
      171 \chi_8 + 2 \chi_1 \chi_8^2 + 3 \chi_8 u) +
   192 (\chi_1^2 - 3 \chi_2 + 3 \chi_8)^2 (3 \chi_1 - 3 \chi_7 +
      \chi_8^2)
\cr&
  \times (3 \chi_1^2 - 9 \chi_2 + 9 \chi_5 - 3 \chi_1 \chi_7 - 171 \chi_8 +
      2 \chi_1 \chi_8^2 + 3 \chi_8 u)^2 +
   24 (2 \chi_1^2 \chi_8 + 3 (3 \chi_4 - 3 \chi_7 - \chi_2 \chi_8 + \chi_8^2)
\cr&
   + 3 \chi_1 (-57 + u))^2 (3 \chi_1^2 - 9 \chi_2 + 9 \chi_5 - 3 \chi_1 \chi_7 -
      171 \chi_8 + 2 \chi_1 \chi_8^2 + 3 \chi_8 u)^2 +
   24 (\chi_1^2 - 3 \chi_2 + 3 \chi_8)
\cr&
  \times
    (2 \chi_1 \chi_8 +
      3 (-54 + u))^2 (3 \chi_1^2 - 9 \chi_2 + 9 \chi_5 - 3 \chi_1 \chi_7 -
      171 \chi_8 + 2 \chi_1 \chi_8^2 + 3 \chi_8 u)^2
\cr&
       -
   96 (3 \chi_1 - 3 \chi_7 + \chi_8^2)^2 \Big[2 \chi_1^2 \chi_8 +
      3 (3 \chi_4 - 3 \chi_7 - \chi_2 \chi_8 + \chi_8^2) +
      3 \chi_1 (-57 + u)\Big] (2 \chi_1 \chi_8 + 3 (-54 + u))
\cr&
  \times
  (2 \chi_1^3 -
      9 \chi_1 (\chi_2 - \chi_8) + 27 (-57 + \chi_3 + u)) +
   48 (\chi_1^2 - 3 \chi_2 + 3 \chi_8) (3 \chi_1 - 3 \chi_7 +
      \chi_8^2)
\cr&
  \times
  (2 \chi_1 \chi_8 + 3 (-54 + u)) (-1539 + 27 \chi_6 +
      9 \chi_1 \chi_8 - 9 \chi_7 \chi_8 + 2 \chi_8^3 + 27 u)
\cr&
  \times
  \Big[2 \chi_1^3 -
      9 \chi_1 (\chi_2 - \chi_8) + 27 (-57 + \chi_3 + u)\Big] -
   20 (2 \chi_1 \chi_8 + 3 (-54 + u))^3
\cr&
  \times
   (-1539 + 27 \chi_6 + 9 \chi_1 \chi_8 -
      9 \chi_7 \chi_8 + 2 \chi_8^3 + 27 u) \Big[2 \chi_1^3 -
      9 \chi_1 (\chi_2 - \chi_8) + 27 (-57 + \chi_3 + u)\Big]
\cr&
       -
   96 (\chi_1^2 - 3 \chi_2 + 3 \chi_8) (3 \chi_1 - 3 \chi_7 +
      \chi_8^2)^2 (3 \chi_1^2 - 9 \chi_2 + 9 \chi_5 - 3 \chi_1 \chi_7 -
      171 \chi_8 + 2 \chi_1 \chi_8^2 + 3 \chi_8 u)
\cr&
  \times
       (2 \chi_1^3 -
      9 \chi_1 (\chi_2 - \chi_8) + 27 (-57 + \chi_3 + u)) +
   72 (3 \chi_1 - 3 \chi_7 + \chi_8^2) (2 \chi_1 \chi_8 +
      3 (-54 + u))^2
\cr&
  \times
      (3 \chi_1^2 - 9 \chi_2 + 9 \chi_5 - 3 \chi_1 \chi_7 -
      171 \chi_8 + 2 \chi_1 \chi_8^2 + 3 \chi_8 u) (2 \chi_1^3 -
      9 \chi_1 (\chi_2 - \chi_8) + 27 (-57 + \chi_3 + u))
\cr&
   + 48 \Big[2 \chi_1^2 \chi_8 + 3 (3 \chi_4 - 3 \chi_7 - \chi_2 \chi_8 + \chi_8^2) +
      3 \chi_1 (-57 + u)\Big] (-1539 + 27 \chi_6 + 9 \chi_1 \chi_8 -
      9 \chi_7 \chi_8 + 2 \chi_8^3 + 27 u)
\cr&
  \times
      (3 \chi_1^2 - 9 \chi_2 + 9 \chi_5 -
      3 \chi_1 \chi_7 - 171 \chi_8 + 2 \chi_1 \chi_8^2 + 3 \chi_8 u) \Big[2 \chi_1^3 -
      9 \chi_1 (\chi_2 - \chi_8) + 27 (-57 + \chi_3 + u)\Big]
\cr&
      -
   32 (3 \chi_1^2 - 9 \chi_2 + 9 \chi_5 - 3 \chi_1 \chi_7 - 171 \chi_8 +
      2 \chi_1 \chi_8^2 + 3 \chi_8 u)^3 \Big[2 \chi_1^3 - 9 \chi_1 (\chi_2 - \chi_8) +
      27 (-57 + \chi_3 + u)\Big]
\cr&
+
   32 (3 \chi_1 - 3 \chi_7 + \chi_8^2)^3 (2 \chi_1^3 - 9 \chi_1 (\chi_2 - \chi_8) +
      27 (-57 + \chi_3 + u))^2
\cr&
       -
   8 (-1539 + 27 \chi_6 + 9 \chi_1 \chi_8 - 9 \chi_7 \chi_8 + 2 \chi_8^3 +
      27 u)^2 \Big[2 \chi_1^3 - 9 \chi_1 (\chi_2 - \chi_8) +
      27 (-57 + \chi_3 + u)\Big]^2\bigg).\nn\label{g3forSU(9)}
\end{align}
}}

%%%%%%%%%%%%%%%%
\subsection{Decomposition of the $E_8$ characters into $SU(9)$}\label{che8tosu9}
We list decomposition of the characters $\chi^{E_8}_i\equiv\chi^{E_8}_{\mu_i}$ of $E_8$ fundamental weights into the characters $\chi_i$ of $SU(9)$ fundamental weights. As mentioned in section \ref{sec:E8-inv}, $\chi_4$ and $\chi_5$ are determined such that \eqref{g3forSU(9)} agrees with \eqref{Huange8}.

\begin{flalign*}
E_8 ~{\rm Dynkin~ diagram}\qquad & \underset{\mathclap{\mu_1}}{\circ} -\!\!\!-
 \underset{\mathclap{\mu_3}}{\circ}-\!\!\!-
\underset{\mathclap{\mu_4}}{\overset{\overset{\textstyle\circ_{\mathrlap{\mu_2}}}{\textstyle\vert}}{\circ}} -\!\!\!-
\underset{\mathclap{\mu_5}}{\circ}-\!\!\!-
\underset{\mathclap{\mu_6}}{\circ} -\!\!\!-
\underset{\mathclap{\mu_7}}{\circ}-\!\!\!-
\underset{\mathclap{\mu_8}}{\circ}
\end{flalign*}
{\footnotesize{
\begin{align}
\chi^{E_8}_1 =~& -1 + \chi_1 \chi_2 - 2 \chi_3 + \chi_1 \chi_5 - 2 \chi_6 + \chi_2 \chi_7 + \chi_4 \chi_8 + \chi_7 \chi_8\\
%%%%%%%%%%%
\chi^{E_8}_2 =~& \chi_1{}^3 + \chi_1{}^2 \chi_4 - 4 (\chi_1 \chi_2 - \chi_3) + \chi_1 \chi_2 \chi_6 - 2 (\chi_2 \chi_7 - \chi_1 \chi_8) + \chi_1 \chi_2 \cr
&- \chi_1 \chi_3 \chi_8 + \chi_1 \chi_4 \chi_7 - 4 (\chi_1 \chi_5 - \chi_6) + 3 \chi_1 \chi_5 - \chi_1 \chi_6 \chi_8 + \chi_1 (\chi_7)^2 \cr
&+ \chi_1 \chi_8 - 4 (\chi_1 \chi_8 - 1) + \chi_2{}^2 \chi_8 - 2 \chi_2 \chi_4 + \chi_2 \chi_5 \chi_8 +
    2 \chi_2 \chi_7 \cr
 & - 4 (\chi_4 \chi_8 - \chi_3) - 3 \chi_3 \chi_6 +
    \chi_3 \chi_7 \chi_8 - 5 \chi_3 + \chi_4 \chi_5 + \chi_4 \chi_8 + \chi_5 (\chi_8)^2 \cr
&
 - 4 (\chi_7 \chi_8 - \chi_6) - 5 \chi_6 + \chi_7 \chi_8 + \chi_8{}^3 - 3 + \chi_2 \chi_4 - \chi_1 \chi_5 - \chi_5 \chi_7 + \chi_4 \chi_8\\
%%%%%%%%%%%
\chi^{E_8}_3 =~& \chi_2{}^3 + \chi_1{}^3 \chi_3 - 5 \chi_1 \chi_2 \chi_3 + 5 \chi_3{}^2 -
   \chi_2 \chi_4 + \chi_1 \chi_5 + \chi_1{}^2 \chi_2 \chi_5 - \chi_2{}^2 \chi_5 \cr
 &- \chi_1 \chi_3 \chi_5 - \chi_4 \chi_5 + \chi_2 \chi_5{}^2 - 2 \chi_1 \chi_2 \chi_6 +
5 \chi_3 \chi_6 + \chi_1{}^2 \chi_4 \chi_6 - 2 \chi_2 \chi_4 \chi_6\cr
 &- 3 \chi_1 \chi_5 \chi_6 + 5 \chi_6{}^2 + \chi_2 \chi_7 + \chi_1 \chi_2{}^2 \chi_7 -
\chi_1{}^2 \chi_3 \chi_7 - \chi_2 \chi_3 \chi_7 + \chi_1 \chi_4 \chi_7\cr
& + \chi_4{}^2 \chi_7
-\chi_5 \chi_7 + \chi_1 \chi_2 \chi_5 \chi_7 - 2 \chi_3 \chi_5 \chi_7 - \chi_2 \chi_6 \chi_7 - \chi_1 \chi_7{}^2  + \chi_1 \chi_3 \chi_7{}^2 \cr
&
 - \chi_4 \chi_7{}^2+\chi_7{}^3 - \chi_2{}^2 \chi_8 + \chi_1 \chi_3 \chi_8 + \chi_4 \chi_8 + \chi_1 \chi_2 \chi_4 \chi_8 - 3 \chi_3 \chi_4 \chi_8 \cr
&
 - (\chi_1)^2 \chi_5 \chi_8 + \chi_2 \chi_5 \chi_8 + \chi_1 \chi_4 \chi_5 \chi_8 + \chi_1 \chi_6 \chi_8 +
(\chi_2)^2 \chi_6 \chi_8 \cr
&- 2 \chi_1 \chi_3 \chi_6 \chi_8 - \chi_4 \chi_6 \chi_8 -
   2 \chi_3 \chi_7 \chi_8 + \chi_2 \chi_4 \chi_7 \chi_8+ \chi_1 \chi_5 \chi_7 \chi_8 - 5 \chi_6 \chi_7 \chi_8 \cr
&
   + \chi_2 (\chi_7)^2 \chi_8 - \chi_1 \chi_4 (\chi_8)^2  +
   \chi_3 \chi_5 (\chi_8)^2 - \chi_2 \chi_6 (\chi_8)^2
   + \chi_4 \chi_7 (\chi_8)^2 + \chi_6 (\chi_8)^3\\
%%%%%%%%%%%
\chi^{E_8}_5 =~& 3 + 2 \chi_2 \chi_4 + \chi_4^3 + \chi_1^4 \chi_5 + \chi_2^2 \chi_5 -
  2 \chi_4 \chi_5 + \chi_2 \chi_5^2 + \chi_5^3 - 3 \chi_6 + \chi_2^3 \chi_6 -
  3 \chi_2 \chi_4 \chi_6 \cr
&
  + \chi_2^2 \chi_5 \chi_6 - 4 \chi_4 \chi_5 \chi_6 -
  4 \chi_2 \chi_7 - 2 \chi_2^2 \chi_4 \chi_7 + \chi_4^2 \chi_7 + 2 \chi_5 \chi_7 +
  \chi_2 \chi_4 \chi_5 \chi_7 \cr
  &
  +   2 \chi_2 \chi_6 \chi_7 + \chi_2^2 \chi_7^2 +
  \chi_4 \chi_7^2 - 2 \chi_2 \chi_5 \chi_7^2 +
  \chi_1^3 (-1 + (-1 + \chi_3) \chi_6 + \chi_6^2 - \chi_5 \chi_7) - \chi_2^2 \chi_8
  \cr
  &
  -
  \chi_2 \chi_4^2 \chi_8 - 2 \chi_2 \chi_5 \chi_8 - \chi_5^2 \chi_8 -
  2 \chi_2^2 \chi_6 \chi_8 + 6 \chi_4 \chi_6 \chi_8 + 2 \chi_7 \chi_8 +
  2 \chi_2 \chi_4 \chi_7 \chi_8
\cr
  &
  + \chi_4 \chi_5 \chi_7 \chi_8 - \chi_2 \chi_7^2 \chi_8 +
  2 \chi_2 \chi_8^2 + \chi_2^2 \chi_4 \chi_8^2 - \chi_4^2 \chi_8^2 +
  \chi_2 \chi_6 \chi_8^2 - 4 \chi_4 \chi_7 \chi_8^2 \cr
  &
  + \chi_2 \chi_5 \chi_7 \chi_8^2 -
  \chi_8^3 - \chi_2 \chi_4 \chi_8^3 + \chi_4 \chi_8^4 +
  \chi_3^2 (8 \chi_6 - 4 \chi_7 \chi_8 + \chi_8^3) +
  \chi_1^2 (-\chi_5^2 + \chi_4 \chi_6 + 2 \chi_7
\cr
  &
  + \chi_3 \chi_7 + \chi_3 \chi_6 \chi_7 +
     \chi_4^2 \chi_8 - \chi_4 \chi_7 \chi_8 - \chi_3 \chi_8^2 - \chi_6 \chi_8^2 +
     \chi_2 (-4 \chi_5 + \chi_4 \chi_7 + \chi_8)
\cr
  &
     +
     \chi_5 (\chi_7^2 - (-2 + 2 \chi_3 + \chi_6) \chi_8)) +
  \chi_3 (-3 + 8 \chi_6^2 + 2 \chi_2 \chi_7 - 3 \chi_5 \chi_7 + \chi_7^3 +
     \chi_5^2 \chi_8 + 4 \chi_7 \chi_8 + \chi_5 \chi_8^2
\cr
  &
      - \chi_8^3 +
     \chi_4 (-4 \chi_5 + \chi_7^2 + 3 \chi_8 - 2 \chi_6 \chi_8) +
     \chi_6 (-3 - 2 \chi_2 \chi_7 - 6 \chi_7 \chi_8 + \chi_2 \chi_8^2 + \chi_8^3))
\cr
  &
     +
  \chi_1 (6 \chi_3 \chi_5 + \chi_4^2 (-1 + \chi_6) + 3 \chi_5 \chi_6 -
     2 \chi_3 \chi_5 \chi_6 - \chi_5^2 \chi_7 - \chi_7^2 - 2 \chi_3 \chi_7^2 -
     3 \chi_8 + \chi_3 \chi_8 + \chi_6 \chi_8
\cr
  &
     + \chi_3 \chi_6 \chi_8 -
     2 \chi_5 \chi_7 \chi_8 + \chi_3 \chi_5 \chi_7 \chi_8 + \chi_5^2 \chi_8^2 +
     \chi_7 \chi_8^2 + \chi_2^2 (-\chi_7 + \chi_5 \chi_8)
\cr
  &
     + \chi_4 (-2 \chi_7 + \chi_7^2 \chi_8 - \chi_8 (\chi_5 + (-2 + \chi_3 + 2 \chi_6) \chi_8)) +\chi_2 (2 - 4 \chi_6^2 + 2 \chi_5 \chi_7 - \chi_7 \chi_8
\cr
  &
+ \chi_3 \chi_7 \chi_8 - \chi_5 \chi_8^2 + \chi_4 (\chi_5 + (-2 + \chi_6) \chi_8) + \chi_6 (4 - 6 \chi_3 + \chi_7 \chi_8)))\\
%%%%%%%%%%%%%
\chi^{E_8}_6 =~& -1 + 3 \chi_3 - \chi_2 \chi_4 + \chi_1 \chi_4^2 + \chi_1 \chi_5 +
   \chi_2^2 \chi_5 - 2 \chi_1 \chi_3 \chi_5 - 2 \chi_4 \chi_5 + 3 \chi_6 +
   \chi_1^3 \chi_6 - 4 \chi_1 \chi_2 \chi_6 \cr
  &
   + 6 \chi_3 \chi_6 + \chi_2 \chi_4 \chi_6 - \chi_1 \chi_5 \chi_6 + \chi_1^2 \chi_3 \chi_7 - 2 \chi_2 \chi_3 \chi_7 - \chi_5 \chi_7 +
   \chi_3 \chi_5 \chi_7 + \chi_1^2 \chi_6 \chi_7 - 2 \chi_2 \chi_6 \chi_7\cr
&
    +
   \chi_4 \chi_7^2 + 2 \chi_1 \chi_8 - 2 \chi_1 \chi_3 \chi_8 + \chi_4 \chi_8 +
   \chi_1 \chi_2 \chi_4 \chi_8 - \chi_3 \chi_4 \chi_8 - \chi_1^2 \chi_5 \chi_8 +
   \chi_5^2 \chi_8 - 2 \chi_1  \chi_6 \chi_8\cr
&    + \chi_1 \chi_3 \chi_6 \chi_8 -
   2 \chi_4 \chi_6 \chi_8 + \chi_1 \chi_2 \chi_7 \chi_8 - 4 \chi_3 \chi_7 \chi_8 +
   \chi_1 \chi_5 \chi_7 \chi_8 - \chi_1^2 \chi_8^2 + \chi_2 \chi_3 \chi_8^2 \cr
&
- \chi_1 \chi_4 \chi_8^2 + \chi_2 \chi_6 \chi_8^2 + \chi_3 \chi_8^3
%%%%%%%%%%%
\end{align}
}}
{\footnotesize{
\begin{align}
\chi^{E_8}_7 =~& \chi_1{}^2 \chi_7 + \chi_1 \chi_3 \chi_8 - 2 \chi_1 \chi_5 + \chi_1 \chi_6 \chi_8 + 2 \chi_1 \chi_8 - 4 (\chi_1 \chi_8 - 1) + 2 \chi_2 \chi_4\cr
&
 -2 \chi_2 \chi_7 + \chi_2 \chi_8{}^2 + \chi_3 \chi_6 - \chi_3 - \chi_6 - 2 -\chi_2 \chi_4 + \chi_1 \chi_5 + \chi_5 \chi_7 - \chi_4 \chi_8\\
 %%%%%%%%%%%%%
\chi^{E_8}_8 =~& -1 + \chi_3 + \chi_6 + \chi_1 \chi_8\\
 %%%%%%%%%%%
\chi^{E_8}_4 =~&
 -3 + 6 \chi_3 - \chi_3^3 + \chi_1^5 \chi_4 + \chi_4^3 + 5 \chi_4 \chi_5 - 3 \chi_3 \chi_4 \chi_5 + \chi_5^3 + \chi_3 \chi_5^3 + 6 \chi_6 + 9 \chi_3 \chi_6 - 7 \chi_3^2 \chi_6 \cr
&
+ \chi_4^3 \chi_6 - 3 \chi_4 \chi_5 \chi_6 - 6 \chi_3 \chi_4 \chi_5 \chi_6 - 7 \chi_3 \chi_6^2 + 9 \chi_3^2 \chi_6^2 - \chi_6^3 + \chi_2^4 \chi_7 + \chi_3 \chi_4^2 \chi_7 \cr
&
+ \chi_5 \chi_7 + 2 \chi_3 \chi_5 \chi_7 - 2 \chi_3^2 \chi_5 \chi_7 - 2 \chi_4^2 \chi_6 \chi_7 - 4 \chi_5 \chi_6 \chi_7 + 4 \chi_3 \chi_5 \chi_6 \chi_7 + 3 \chi_3 \chi_4 \chi_7^2 \cr
&
+ 2 \chi_4 \chi_6 \chi_7^2 - \chi_7^3 - \chi_3 \chi_7^3 + \chi_3^2 \chi_7^3 - 4 \chi_4 \chi_8 - \chi_3 \chi_4 \chi_8 + 2 \chi_3^2 \chi_4 \chi_8 - \chi_4^2 \chi_5 \chi_8 \cr
&
- 3 \chi_5^2 \chi_8 + \chi_3 \chi_5^2 \chi_8 - 7 \chi_4 \chi_6 \chi_8 + 5 \chi_3 \chi_4 \chi_6 \chi_8 + 4 \chi_4 \chi_6^2 \chi_8 - 3 \chi_7 \chi_8 - 6 \chi_3 \chi_7 \chi_8 \cr
&
+ \chi_3^2 \chi_7 \chi_8 - \chi_4 \chi_5 \chi_7 \chi_8 + \chi_3 \chi_4 \chi_5 \chi_7 \chi_8 + \chi_6 \chi_7 \chi_8 + 6 \chi_3 \chi_6 \chi_7 \chi_8 - 6 \chi_3^2 \chi_6 \chi_7 \chi_8 \cr
&
+ 5 \chi_5 \chi_7^2 \chi_8 - 2 \chi_3 \chi_5 \chi_7^2 \chi_8 + \chi_4^2 \chi_8^2 - \chi_3 \chi_4^2 \chi_8^2 + 2 \chi_5 \chi_8^2 - \chi_3 \chi_5 \chi_8^2 + \chi_3^2 \chi_5 \chi_8^2 \cr
&
+ \chi_4^2 \chi_6 \chi_8^2 + 5 \chi_5 \chi_6 \chi_8^2 - 2 \chi_3 \chi_5 \chi_6 \chi_8^2 + 4 \chi_4 \chi_7 \chi_8^2 - 3 \chi_3 \chi_4 \chi_7 \chi_8^2 - 4 \chi_4 \chi_6 \chi_7 \chi_8^2 \cr
&
+ \chi_7^2 \chi_8^2 + \chi_8^3 + \chi_3 \chi_8^3 + \chi_3^2 \chi_8^3 - 2 \chi_6 \chi_8^3 - \chi_3 \chi_6 \chi_8^3 + \chi_3^2 \chi_6 \chi_8^3 - 5 \chi_5 \chi_7 \chi_8^3 + \chi_3 \chi_5 \chi_7 \chi_8^3 \cr
&
-\chi_4 \chi_8^4 + \chi_4 \chi_6 \chi_8^4 + \chi_5 \chi_8^5 + \chi_1^4 ((-1 + \chi_3) \chi_5 + \chi_7^2 - (1 + \chi_6) \chi_8) + \chi_2^2 (3 \chi_5 \chi_6 - 3 \chi_4 \chi_7 \cr
&
+ \chi_5^2 \chi_7 - 2 \chi_4 \chi_6 \chi_7 + 3 \chi_7^2 + \chi_6 \chi_7^2 + 2 \chi_8 - \chi_4 \chi_5 \chi_8 + \chi_6 \chi_8 - 2 \chi_6^2 \chi_8 + \chi_4 \chi_6 \chi_8^2 - 4 \chi_7 \chi_8^2 \cr
&
+ \chi_8^4 + \chi_3 (2 \chi_5 + \chi_7^2 + \chi_8 - 2 \chi_6 \chi_8)) + \chi_2^3 (-1 - \chi_6 + \chi_6^2 + \chi_5 (-2 \chi_7 + \chi_8^2)) + \chi_1^3 (1 + \chi_6 + \chi_6^2 \cr
&
+ \chi_4^2 \chi_7 - \chi_5 \chi_7 - \chi_4 \chi_8 - \chi_5^2 \chi_8 - \chi_7 \chi_8 + \chi_5 \chi_8^2 + \chi_3 (-2 - \chi_6 + \chi_6^2 - 2 \chi_5 \chi_7 + \chi_7 \chi_8) \cr
&
+ \chi_2 (\chi_4 (-5 + \chi_6) - \chi_6 \chi_7 + \chi_8 (\chi_5 + \chi_8))) + \chi_1^2 (\chi_5^2 + \chi_3 \chi_5^2 - \chi_5^2 \chi_6 + 2 \chi_3 \chi_7 + 2 \chi_3^2 \chi_7 \cr
&
+ 3 \chi_3 \chi_6 \chi_7 + \chi_3 \chi_5 \chi_7^2 + 2 \chi_5 \chi_8 + 3 \chi_3 \chi_5 \chi_8 + \chi_4^2 (-2 + \chi_6) \chi_8 + 4 \chi_5 \chi_6 \chi_8 - 2 \chi_3 \chi_5 \chi_6 \chi_8 \cr
&
- \chi_7^2 \chi_8 - \chi_3 \chi_7^2 \chi_8 - \chi_8^2 - 2 \chi_3 \chi_8^2 - 2 \chi_6 \chi_8^2 - \chi_5 \chi_7 \chi_8^2 + \chi_7 \chi_8^3 + \chi_2^2 (1 + \chi_5 \chi_7 - \chi_7 \chi_8) \cr
&
+ \chi_4 (2 + \chi_6^2 - 2 \chi_5 \chi_7 + \chi_7^3 - 2 \chi_7 \chi_8 + \chi_5 \chi_8^2 + \chi_8^3 - \chi_3 (-5 + 2 \chi_6 + \chi_7 \chi_8) - \chi_6 (1 + 2 \chi_7 \chi_8)) \cr
&
+ \chi_2 (-4 \chi_7^2 + \chi_8 + 3 \chi_3 \chi_8 + 5 \chi_6 \chi_8 - 2 \chi_3 \chi_6 \chi_8 + \chi_7 \chi_8^2 + \chi_5 (4 - 4 \chi_3 - 3 \chi_6 + \chi_4 \chi_8 - \chi_7 \chi_8) \cr
&
+ \chi_4 ((2 + \chi_6) \chi_7 - \chi_8^2))) + \chi_2 (\chi_5^2 \chi_6 + 4 \chi_3^2 \chi_7 - 6 \chi_6 \chi_7 + 4 \chi_6^2 \chi_7 - 3 \chi_5 \chi_7^2 + \chi_7^4 - \chi_5 \chi_8 \cr
&
- 3 \chi_5 \chi_6 \chi_8 - 2 \chi_5^2 \chi_7 \chi_8 + \chi_7^2 \chi_8 - 4 \chi_6 \chi_7^2 \chi_8 + 2 \chi_6 \chi_8^2 + 2 \chi_6^2 \chi_8^2 + 2 \chi_5 \chi_7 \chi_8^2 + \chi_5^2 \chi_8^3 \cr
&
+ \chi_4^2 (\chi_7^2 + \chi_8 - 2 \chi_6 \chi_8) + \chi_4 (1 - 2 \chi_6^2 + 3 \chi_5 \chi_7 - 2 \chi_7^3 + \chi_5^2 \chi_8 + 4 \chi_7 \chi_8 - 2 \chi_5 \chi_8^2 + \chi_7^2 \chi_8^2 \cr
&
- \chi_8^3 + \chi_3 (-4 + 4 \chi_6 - 2 \chi_7 \chi_8) + \chi_6 (2 + 4 \chi_7 \chi_8 - 2 \chi_8^3)) + \chi_3 (-2 \chi_5^2 + 2 \chi_7^2 \chi_8 + 3 \chi_6 \chi_8^2 \cr
&
+ \chi_5 (-2 \chi_7^2 + 3 \chi_8 + \chi_7 \chi_8^2) - \chi_7 (6 + 5 \chi_6 + \chi_8^3))) + \chi_1 (-4 \chi_5 - 7 \chi_3 \chi_5 + 4 \chi_3^2 \chi_5 - \chi_5 \chi_6 + 5 \chi_3 \chi_5 \chi_6 \cr
&
+ 2 \chi_5 \chi_6^2 + \chi_5^2 \chi_7 - 2 \chi_3 \chi_5^2 \chi_7 + 2 \chi_7^2 + \chi_3 \chi_7^2 - 2 \chi_3^2 \chi_7^2 + \chi_6 \chi_7^2 - 2 \chi_3 \chi_6 \chi_7^2 + 4 \chi_8 \cr
&
- 2 \chi_3 \chi_8 - 5 \chi_3^2 \chi_8 + \chi_2^3 (-1 + \chi_6) \chi_8 - 2 \chi_6 \chi_8 - 10 \chi_3 \chi_6 \chi_8 + 4 \chi_3^2 \chi_6 \chi_8 - 5 \chi_6^2 \chi_8 \cr
&
+ 4 \chi_3 \chi_6^2 \chi_8 + \chi_5 \chi_7 \chi_8 - \chi_7^3 \chi_8 + \chi_3 \chi_7^3 \chi_8 - 2 \chi_5^2 \chi_8^2 + \chi_3 \chi_5^2 \chi_8^2 + \chi_7 \chi_8^2 + 5 \chi_3 \chi_7 \chi_8^2 \cr
&
+ 3 \chi_6 \chi_7 \chi_8^2 - 2 \chi_3 \chi_6 \chi_7 \chi_8^2 - \chi_5 \chi_8^3 - \chi_8^4 - \chi_3 \chi_8^4 + \chi_4^2 (-3 + \chi_6 + \chi_5 \chi_7 + \chi_7 \chi_8 - \chi_8^3) \cr
&
+ \chi_2^2 ((1 - 4 \chi_3 + 2 \chi_6) \chi_7 - \chi_8 (5 \chi_5 - \chi_5 \chi_6 + \chi_8 + \chi_6 \chi_8) + \chi_4 (5 - 2 \chi_6 + \chi_7 \chi_8)) - \chi_4 (\chi_5^2 - (-5 + \chi_3) \chi_7^2 \chi_8 \cr
&
+ (-2 + 2 \chi_3 (-2 + \chi_6) - 3 \chi_6) \chi_8^2 + \chi_5 (\chi_7^2 + 2 (-2 + \chi_3 + \chi_6) \chi_8 - \chi_7 \chi_8^2) + \chi_7 (1 + 3 \chi_3 - 3 \chi_6 - \chi_8^3)) \cr
&
+ \chi_2 (-3 - 6 \chi_6 + \chi_6^2 - 2 \chi_4^2 \chi_7 + 4 \chi_5 \chi_7 - 2 \chi_5 \chi_6 \chi_7 + \chi_5^2 \chi_8 + 4 \chi_7 \chi_8 + \chi_6 \chi_7 \chi_8 + \chi_5 \chi_7^2 \chi_8 \cr
&
- 2 \chi_5 \chi_8^2 - \chi_5 \chi_6 \chi_8^2 - \chi_7^2 \chi_8^2 - \chi_8^3 + \chi_6 \chi_8^3 + \chi_3 (1 - 6 \chi_6^2 + 4 \chi_5 \chi_7 + \chi_7 \chi_8 - 2 \chi_5 \chi_8^2 \cr
&
+\chi_6 (6 + \chi_7 \chi_8)) + \chi_4 (\chi_8 - \chi_7 \chi_8^2 + \chi_5 (-1 + \chi_6 + \chi_7 \chi_8)))).
\end{align}
}}

%%%%%%%%%%%%%%%%

\section{Relation among the parameters for $E_7$ SW curve}\label{App:relation}

In this section, we clarify the the relation among various parameters appearing in the $E_7$ SW curve.
They are basically related by change of the basis of the simple roots.
First, we would like to write $E_7$ parameters in terms of 6 masses $m'_i$, 
$(i=1,\cdots 6)$ and 1 gauge coupling $m'_0 =1/(2g^2)$.
Here we put prime so that we distinguish them with the parameters $m_i$ related to $SU(8)$.
Since the mass parameters are related to $SO(12)$ subgroup of $E_7$,
we should decompose $E_7$ into $SO(12) \times SU(2)$.
In this case, it is convenient to choose the simple roots as 
\begin{eqnarray}
&&\alpha_1 = \mathbf{e}_1 - \mathbf{e}_2, \qquad
\alpha_2 = \mathbf{e}_2 - \mathbf{e}_3, \qquad
\alpha_3 = \mathbf{e}_3 - \mathbf{e}_4, 
\nonumber \\
&&\alpha_4 = \mathbf{e}_4 - \mathbf{e}_5, \qquad
\alpha_5 = \mathbf{e}_5 + \mathbf{e}_6, \qquad
\alpha_6 = \mathbf{e}_5 - \mathbf{e}_6, \qquad
\nonumber \\
&&
\alpha_7 = - \frac{1}{2} \left(
\mathbf{e}_1 + \mathbf{e}_2 + \mathbf{e}_3 
+ \mathbf{e}_4 + \mathbf{e}_5 - \mathbf{e}_6
+ \sqrt{2} \mathbf{e}_7 \right), \qquad
\nonumber \\
&&
\alpha_{-\gamma} = \sqrt{2} \mathbf{e}_7.
\end{eqnarray}
where $\mathbf{e}_i$ are the orthonormal basis of the root lattice.
The root $\alpha_{-\gamma}$ corresponds to the extra node
of the extended Dynkin diagram.
The $SO(12) \times SU(2)$ subgroup is given by removing the root $\alpha_7$.
In this convention, the weights of the fundamental representation of $SO(12)$
are given by $\pm \mathbf{e}_i$ ($i=1,\cdots, 6$).
When we compute the character, it is possible to parametrize the 
element of the Cartan subalgebra in such a way that 
the state with the weight $\pm \mathbf{e}_i$ has the eigenvalue $\pm m'_i$.
In this sense, we can identify the vector $\mathbf{e}_i$ with the mass parameter $m'_i$
($i=1,\cdots, 6$).
Analogously, we identify the vector $\mathbf{e}_7$ with the gauge coupling $m'_0$.

On the other hand, when we decompose $E_7$ into its subgroup $SU(8)$,
it is natural to choose the simple roots as
\begin{eqnarray}
&&\alpha_1 = \mathbf{e}'_1 - \mathbf{e}'_2, \qquad
\alpha_2 = \mathbf{e}'_2 - \mathbf{e}'_3, \qquad
\alpha_3 = \mathbf{e}'_3 - \mathbf{e}'_4,  \qquad
\alpha_4 = \mathbf{e}'_4 - \mathbf{e}'_5, \qquad
\nonumber \\
&& 
\alpha_5 = - \frac{1}{2} \left( 
\mathbf{e}'_1 + \mathbf{e}'_2 + \mathbf{e}'_3 + \mathbf{e}'_4
- \mathbf{e}'_5 - \mathbf{e}'_6 - \mathbf{e}'_7 - \mathbf{e}'_8
\right) ,
\nonumber \\
&&
\alpha_6 = \mathbf{e}'_5 - \mathbf{e}'_6, \qquad
\alpha_7 = \mathbf{e}'_6 - \mathbf{e}'_7, \qquad
\alpha_{-\gamma} = \mathbf{e}'_7 - \mathbf{e}'_8.
\end{eqnarray}
where we used the different orthonormal basis $\mathbf{e}'_i$.
The $SU(8)$ subgroup is obtained by removing the root $\alpha_5$.
We identify the vector $\mathbf{e}'_i$ with the parameters $m_i$.

Still another choice is the case when we decompose $E_7$ into its subgroup $SU(4) \times SU(4) \times SU(2)$;
\begin{eqnarray}
&&\alpha_1 = \mathbf{e}''_1 - \mathbf{e}''_2, \qquad
\alpha_2 = \mathbf{e}''_2 - \mathbf{e}''_3, \qquad
\alpha_3 = \mathbf{e}''_3 - \mathbf{e}''_4,  \qquad
\nonumber \\
&& 
\alpha_4 = - \frac{1}{4} \left( 
\mathbf{e}''_1 + \mathbf{e}''_2 + \mathbf{e}''_3 - 3 \mathbf{e}''_4
+ 2 \mathbf{e}''_5 - 2 \mathbf{e}''_6 
+ 3 \mathbf{e}''_7 - \mathbf{e}''_8 - \mathbf{e}''_9 - \mathbf{e}''_{10}
\right) 
\nonumber \\
&& 
\alpha_5 = \mathbf{e}''_5 - \mathbf{e}''_6, \quad
\alpha_6 = \mathbf{e}''_7 - \mathbf{e}''_8, \quad
\alpha_7 = \mathbf{e}''_8 - \mathbf{e}''_9, \quad
\alpha_{-\gamma} = \mathbf{e}'_9 - \mathbf{e}'_{10}.
\end{eqnarray}
where we used the still different orthonormal basis $\mathbf{e}''_i$.
We find the correspondence between the orthonormal basis $\mathbf{e}''_i$ and the parameters $L_i$, $M_j$ and $N_k$ as 
\begin{eqnarray}
L_i: \,\, \mathbf{e}''_i, \qquad
N_i: \,\, \mathbf{e}''_{i+4}, \qquad
M_i: \,\, \mathbf{e}''_{i+6}.
\end{eqnarray}

The relation between these parameters is obtained by equating the 
parameters corresponding to the corresponding simple roots.
For example, $\alpha_1$ corresponds to $m'_1 - m'_2$ in the $SO(12) \times SU(2) $ parameters 
and $m_1 - m_2$ in the $SU(8)$ parameters, respectively, 
and thus we equate these two. 
In this way, we find 7 independent equations in total and the relation between them are obtained by solving them.

The parameters used in the Seiberg-Witten curve is actually the exponentiated values.
Therefore, we define $\widetilde{m}_i = \exp (-\beta m_i)$,  $q = \exp( - \beta m'_0 )$ and so on,
where $\beta$ is the circumference of the compactified circle. 
Then, the relation between $SU(8)$ parameters $\widetilde{m}_i$ and
 the $SO(12) \times SU(2)$ parameters $\widetilde{m}'_i$, $q$ are 
\begin{eqnarray}
&&\widetilde{m}_i = \frac{\widetilde{m}'_i}{\prod_{i=1}^6 \widetilde{m}'_i{}^{\frac{1}{4}}}
\quad
(i=1,\cdots 6)
\nonumber \\
&&\widetilde{m}_7 = q^{\frac{1}{2}} \prod_{i=1}^6 \widetilde{m}'_i{}^{\frac{1}{4}},
\qquad
\widetilde{m}_8 = q^{-\frac{1}{2}} \prod_{i=1}^6 \widetilde{m}'_i{}^{\frac{1}{4}}.
\end{eqnarray}
The relation between $SU(8)$ parameters $\widetilde{m}_i$ and
the $SU(4) \times SU(4) \times SU(2)$ parameters $L_i$, $M_i$, $N_i$ are
\begin{eqnarray}
\widetilde{m}_i = {N}_2{}^{\frac{1}{2}} {L}_i, \qquad 
\widetilde{m}_{i+4} = {N}_1{}^{\frac{1}{2}} {M}_i \qquad
(i=1,2,3,4).
\end{eqnarray}
This is exactly what we have already obtained in \eqref{E7paramrel}.
%%%%%%%%%%%%%%%%

\section{Maximal Compact Subgroups via Hanany-Witten transitions}
Here we discuss various toric like diagrams for $5\le N_f\le 7$ flavors showing maximal compact subgroups for $E_{N_f+1}$  via the Hanany-Witten transition which we mentioned in the main text\footnote{The authors thank Amihay Hanany suggesting to explore all possible subgroups.}. Since we use $(p,q)$ 5-brane web diagram, maximal compact subgroups of $E_{N_f+1}$ are restricted to those of type $A_N$. As there are many ways of getting to diagram showing the subgroup symmetry, we simply list a representative toric-like diagram of type $A_N$ below.

\subsection{$N_f=5$ or $E_6$ case}
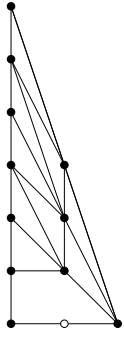
\begin{figure}[t]
\centering
\begin{adjustbox}{width=0.10\textwidth}
\begin{tikzpicture}
  [inner sep=0.5mm,
 dot/.style={fill=black,draw,circle,minimum size=1pt},
 whitedot/.style={fill=white,draw,circle,minimum size=1pt}]
    \draw (0,0) -- (2,0) -- (0,6) -- (0,0);
    \draw (0,1) -- (1,1) -- (1,3);
    \draw (2,0) -- (0,2);     \draw (2,0) -- (1,2);

    \draw (2,0) -- (1,3) -- (0,6);
    %\draw (1,3) -- (0,7);     
    \draw (1,3) -- (0,5);
    \draw (1,2) -- (0,5);     \draw (1,2) -- (0,4);    \draw (1,2) -- (0,3);
    \draw (1,1) -- (0,3);

    \node[dot] at (0,0) {}; 
    \node[whitedot] at (1,0) {}; 
    \node[dot] at (2,0) {};
    \node[dot] at (0,1) {}; 
    \node[dot] at (1,1) {};
    \node[dot] at (0,2) {}; 
    \node[dot] at (1,2) {};
    \node[dot] at (0,3) {}; 
    \node[dot] at (1,3) {};
    \node[dot] at (0,4) {}; 
    %\node[whitedot] at (1,4) {};
    \node[dot] at (0,5) {};
    \node[dot] at (0,6) {};
    %\node[dot] at (0,7) {};\node[dot] at (0,8) {};

%    \draw[->] (-0.3,-0.3) -- (-0.3,0.7); \node[] at (-0.7,0.4) [label=above:$w$] {};
%    \draw[->] (-0.3,-0.3) -- (0.7,-0.3); \node[] at (0.4,-0.9) [label=above:$T$] {};
\end{tikzpicture}
\end{adjustbox}
\caption{A toric-like diagram for the $SU(2)$ theory with $N_f=5$ flavors of  $SU(6)\times SU(2)\subset E_6$ symmetry}\label{SU6SU2}
\end{figure}

The $N_f=5$ case has two maximal compact subgroups of type $A_N$, which are
\begin{align}
E_6 \supset &SU(3) \times SU(3)\times SU(3),\\
E_6 \supset &SU(6) \times SU(2).
\end{align}
As already discussed, the first one is dictated from the $T_3$ diagram as in Figure \ref{webe6}. A toric-like diagram for the second maximal subgroup is given in Figure \ref{SU6SU2}.

\subsection{$N_f=6$  or $E_7$ case}
The $N_f=6$ case has three maximal compact subgroups of type $A_N$, which are
\begin{align}
E_7 \supset& SU(4) \times SU(4)\times SU(2),\\
E_7 \supset& SU(8),\\
E_7 \supset& SU(6) \times SU(3),
\end{align}
whose toric-like diagrams are given in Figure \ref{webe7}, in Figure \ref{dote7su8} and in Figure \ref{SU6SU3}, respectively. %With $SL(2,\mathbb{Z})$, the last one can be depicted as Figure \ref{SU6SU3-1}.

\subsection{$N_f=7$  or $E_8$ case}
The $N_f=8$ case has four maximal compact subgroups of type $A_N$, which are
\begin{align}
E_8 \supset& SU(6) \times SU(3)\times SU(2),\\
E_8 \supset& SU(9),\\
E_8 \supset& SU(8) \times SU(2),\label{e8tosu8su2}\\
E_8 \supset& SU(5) \times SU(5).\label{e8tosu5su5}
\end{align}
The first two are given in Figure \ref{T6e8toric} and in Figure \ref{e8tridots}, respectively. 
The toric-like diagram for \eqref{e8tosu8su2} is given in Figure \ref{SU8SU2}:
\begin{figure}[H]
\centering
\begin{adjustbox}{width=0.2\textwidth}
\begin{tikzpicture}
 [inner sep=0.5mm,
 dot/.style={fill=black,draw,circle,minimum size=1pt},
 whitedot/.style={fill=white,draw,circle,minimum size=1pt}]
    \draw (0,0)  -- (0,-1)  -- (4,3)  -- (0,7) -- (0,6);
    \draw (0,0)  -- (3,3)  -- (0,6);
    \draw (0,1)  -- (2,3)  -- (2,4) -- (0,4);
    \draw (0,2)  -- (1,3)  -- (1,5);
    \draw (0,4)  -- (1,3);
    \draw (0,5)  -- (2,5) -- (2,4);
    \draw (3,3)  -- (0,3);
    \draw (2,3)  -- (0,5);
    \draw (0,0)  -- (0,6);
    \draw (3,3)  -- (4,3);

    \node[whitedot] at (1,1) {};
    \node[whitedot] at (2,2) {};
    \node[whitedot] at (1,2) {};
    \node[whitedot] at (3,4) {};
    \node[whitedot] at (1,6) {};
    \node[whitedot] at (1,0) {};
    \node[whitedot] at (2,1) {};
    \node[whitedot] at (3,2) {};

    \node[dot] at (0,0) {};
    \node[dot] at (3,3) {};
    \node[dot] at (0,1) {};
    \node[dot] at (2,3) {};
    \node[dot] at (0,2) {};
    \node[dot] at (1,3) {};
    \node[dot] at (2,4) {};
    \node[dot] at (0,3) {};
    \node[dot] at (1,4) {};
    \node[dot] at (0,4) {};
    \node[dot] at (1,5) {};
    \node[dot] at (0,5) {};
    \node[dot] at (0,6) {};
    \node[dot] at (4,3) {};
    \node[dot] at (2,5) {};
    \node[dot] at (0,-1) {};
    \node[dot] at (0,7) {};
\end{tikzpicture}
\end{adjustbox}
\caption{A toric-like diagram of $SU(2)$ theory with $N_f=7$ flavors of $SU(8)\times SU(2)\subset E_8$ symmetry}\label{SU8SU2}
\end{figure}
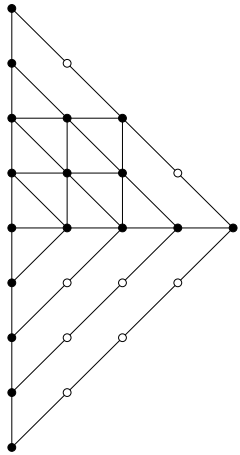
\noindent The toric-like diagram for \eqref{e8tosu5su5} is given in Figure \ref{SU5SU5}:
\begin{figure}[H]
\centering
\begin{adjustbox}{width=0.2\textwidth}
\begin{tikzpicture}
 [inner sep=0.5mm,
 dot/.style={fill=black,draw,circle,minimum size=1pt},
 whitedot/.style={fill=white,draw,circle,minimum size=1pt}]
    \draw (0,-1)  -- (-1,-2)  -- (-1,8)  -- (0,7);
    \draw (0,-1)  -- (4,3)  -- (0,7) -- (0,6);
    \draw (2,4)  -- (0,6);
    \draw (2,3)  -- (2,4) -- (0,4);
    \draw (1,4)  -- (1,5);
    %\draw (0,4)  -- (1,3);
    \draw  (2,5) -- (2,4);
    %\draw (3,3)  -- (0,3);
    \draw (2,3)  -- (0,5);
    \draw (-1, 2) -- (0,3)  -- (0,6);
    \draw (2,3)  -- (4,3);
    \draw (-1,4)  -- (0,4);
    \draw (2,4)  -- (3,4);
    \draw (0,5) -- (-1,6)  -- (1,6) -- (1,5);
    \draw (0,5) -- (2,5);
	\draw (0,3) -- (2,3);
	\draw (-1,0) -- (2,3);

    \node[whitedot] at (1,1) {};
    \node[whitedot] at (2,2) {};
    \node[whitedot] at (1,2) {};
    \node[whitedot] at (1,0) {};
    \node[whitedot] at (2,1) {};
    \node[whitedot] at (3,2) {};
    \node[whitedot] at (-1,-1) {};
    \node[whitedot] at (-1,1) {};
    \node[whitedot] at (-1,3) {};
    \node[whitedot] at (-1,5) {};
    \node[whitedot] at (-1,7) {};
    \node[whitedot] at (0,0) {};
    \node[whitedot] at (0,1) {};
    \node[whitedot] at (0,2) {};
    \node[whitedot] at (0,-1) {};
    \node[whitedot] at (1,3) {};
    \node[whitedot] at (3,3) {};

    \node[dot] at (2,3) {};
    \node[dot] at (2,4) {};
    \node[dot] at (0,3) {};
    \node[dot] at (1,4) {};
    \node[dot] at (0,4) {};
    \node[dot] at (1,5) {};
    \node[dot] at (0,5) {};
    \node[dot] at (0,6) {};
    \node[dot] at (4,3) {};
    \node[dot] at (2,5) {};
    \node[dot] at (0,7) {};
    \node[dot] at (1,6) {};
    \node[dot] at (3,4) {};
    \node[dot] at (-1,-2) {};
    \node[dot] at (-1,0) {};
    \node[dot] at (-1,2) {};
    \node[dot] at (-1,4) {};
    \node[dot] at (-1,6) {};
    \node[dot] at (-1,8) {};

\end{tikzpicture}
\end{adjustbox}
\caption{A toric-like diagram of $SU(2)$ theory with $N_f=7$ flavors of $SU(5)\times SU(5)\subset E_8$ symmetry}\label{SU5SU5}.
\end{figure}
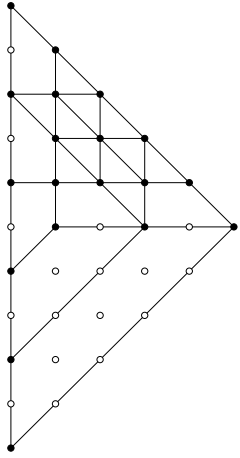

%We note that purely based on the toric-like diagram, we observe that this web (or toric-like) diagram can be made by gluing two differently tune $T_4$ web (or toric-like) diagrams one of which has the global symmetry $SU(4)\times SU(4)\times SU(2)$ with one Coulomb modulus, and the other of which has the global symmetry $SU(4)\times SU(4)$ without any Coulomb modulus (free theory). By gluing $SU(4)$ symmetry of each, at least graphical level, it looks possible to make a diagram of $SU(8)\times SU(2)$ subgroup of $E_8$. It should be also stressed that by gluing it requires the symmetry enhancement from $SU(4)\times SU(4) \subset SU(8)$ to be able to relate to the $E_8$ subgroup, although we do not have a clear understanding of this phenomenon. 

% \bibliography{5dSW}   
% \bibliographystyle{utphys}
% \end{document}
%%%%%%%%%%%%%%%%%%%%%%%%%%%%%%%%%%%%%%%%%%%%%%%
%%%%%%%%%%%%%%%%%%%%%%%%%%%%%%%%%%%%%%%%%%%%%%%
%%%%%%%%%%%%%%%%%%%%%%%%%%%%%%%%%%%%%%%%%%%%%%%
%%%%%%%%%%%%%%%%%%%%%%%%%%%%%%%%%%%%%%%%%%%%%%%
%%%%%%%%%%%%%%%%%%%%%%%%%%%%%%%%%%%%%%%%%%%%%%%
%%%%%%%%%%%%%%%%%%%%%%%%%%%%%%%%%%%%%%%%%%%%%%%
\providecommand{\href}[2]{#2}\begingroup\raggedright\endgroup

\end{document}